\documentclass[aps,pra,twocolumn,showpacs,superscriptaddress]{revtex4-1}

\usepackage{graphicx}
\usepackage{amsmath}  
\usepackage{amsthm}  
\usepackage{amssymb}  
\usepackage{color}
\usepackage{longtable}
\usepackage{multirow}
\usepackage{mathrsfs}
\usepackage{braket}
\usepackage{xfrac}
\usepackage[T1]{fontenc}

\let\savedegree\corresponds
\let\corresponds\relax
\usepackage{mathabx}
\let\corresponds\savedegree

\usepackage{float}
\usepackage{bm}
\usepackage[percent]{overpic}
\usepackage{mathrsfs}
\usepackage{mathcomp}
\usepackage[table,dvipsnames]{xcolor}

\begin{document}
\title{A combined-probability space and (un)certainty relations for a finite-level quantum system}

\author{Arun Sehrawat}
\email[email: ]{aruns@iisermohali.ac.in}
\affiliation{Department of Physical Sciences, 
	Indian Institute of Science Education \& Research (IISER) Mohali, 
	Sector 81 SAS Nagar, Manauli PO 140306, 
	Punjab, India}


\begin{abstract}
The Born rule provides a probability vector (distribution) with a quantum state for a measurement setting.
For two settings, we have 
a pair of vectors from the same quantum state. 
Each pair forms a combined-probability vector that obeys certain quantum constraints, which are triangle inequalities in our case.
Such a restricted set of combined vectors, titled combined-probability space, is presented here for a $d$-level quantum system (qudit). The combined space turns out a compact convex subset of a Euclidean space, and all its extreme points come from a family of parametric curves. 
Considering a suitable concave function on the combined space to estimate the uncertainty, we deliver an uncertainty relation by finding its global minimum at the curves for a qudit.
If one chooses an appropriate concave (or convex) function, then there is no need to search for the absolute minimum (maximum) on the whole space, it will be at the parametric curves.
So these curves are quite useful for establishing an uncertainty (or a certainty) relation for a general pair of settings. In the paper, we also demonstrate that many known tight (un)certainty relations for a qubit can be obtained with the triangle inequalities.
\end{abstract}


\maketitle

\section{Introduction}\label{sec:Intro}

Every setting for a measurement on a quantum system can be completely
specified by an orthonormal basis of the system's Hilbert space.
Identical systems can be independently prepared in a (pure) state $\rho$ such that, every time, we get a definite outcome when a system is measured in a setting $a$.
If we change $a$ to a physically distinct setting $b$, then we
observe---sometimes one and sometimes other---multiple outcomes.
In other words, there the probability is one for an outcome in $a$-setting, whereas none of the probabilities is one in $b$-setting.
Of course, in any setting, all the probabilities are nonnegative numbers that sum up to one.
Apart from that, the probability vectors (distributions) $\vec{p}$ and $\vec{q}$---associated with the two settings $a$ and $b$, respectively---must follow certain constraints, called \emph{quantum constraints} (QCs), together.

Historically, such QCs are expressed in terms of uncertainty relations (URs) by taking Hermitian operators rather than orthonormal bases.
An UR is an inequality, ${\mathsf{c}(a,b,\rho)\leq\mathsf{u}(a,b,\rho)}$,
between two real-valued functions: uncertainty measure $\mathsf{u}$ and its lower bound $\mathsf{c}$.
In 1927, Heisenberg introduced the first UR \cite{Heisenberg27,Wheeler83} 
(derived by Weyl in \cite{Weyl32}) 
for the position and momentum operators. 
Different aspects of his seminal work are reviewed in \cite{Busch07}.
Robertson \cite{Robertson29} generalized the Heisenberg's relation for an arbitrary pair of operators by employing the standard deviation as a measure of uncertainty.
In Robertson's UR, the lower bound $\mathsf{c}$ is a function of state $\rho$. Deutsch criticized it and introduced a new UR \cite{Deutsch83} for a finite-dimensional state space by taking entropy as a measure of uncertainty.
He achieved a state independent ${\mathsf{c}(a,b)}$.
Later, a better lower bound was conjectured by Kraus \cite{Kraus87} and then proved by Maassen and
Uffink \cite{Maassen88}.
Such URs are---known as entropy URs---reviewed in \cite{Wehner10,Bialynicki11,Coles17}.

Throughout the article, we are considering $d$-level quantum systems (qudits) and projective measurements.
Our primary objective is to study a set of combined-probability vectors ${(\vec{p},\vec{q}\,)}$, called \emph{combined-probability space}, where every vector respects certain, if not all, QCs.  
Here the elemental QCs are the triangle inequalities (TIs) between
\emph{quantum angles}, and the (un)certainty relations emerge from them.
As an angle between a pair of kets---called quantum angle---is a metric over the set of all pure states \cite{Wootters81}, we own TIs.
Landau and Pollak obtained a single TI \cite{Landau61} of this kind for continuous-time signals and provided a classical UR
(see also Sec.~8 in \cite{Folland97}).

In Sec.~\ref{sec:PS-C}, we present the combined space that is a compact convex subset of the $2d$-dimensional real vector space $\mathbb{R}^{2d}$.
Thanks to the Krein-Milman theorem (see Theorem~${3.3.5}$ and Appendix~A.3 in \cite{Niculescu93}), every compact convex subset of $\mathbb{R}^{2d}$
can be generated by the convex combinations of its extreme points.
As a principal result, we provide a family of \emph{parametric curves} in Sec.~\ref{sec:PS-C}, which represents all the extreme points of the combined space.
In the case of ${d=2}$, all the parametric curves form an ellipse, and the same ellipse also appears in \cite{Lenard72,Larsen90,Kaniewski14} as a special case.

An uncertainty measure ${\mathsf{u}(a,b,\rho)\equiv\mathsf{u}(\vec{p},\vec{q}\,)}$ should be a concave function on the combined-probability space, argued in 
the beginning of Sec.~\ref{sec:UM-UR}.
The concavity of $\mathsf{u}$ ensures that its global minimum $\mathsf{c}$ will occur at the parametric curves (extreme points) of the space (see Theorem~${3.4.7}$ and Appendix~A.3 in \cite{Niculescu93}).
Hence, one can exploit these curves to 
obtain an UR, rather easily, for her or his liking of $\mathsf{u}$ and, of course, for general measurement settings $a$ and $b$.

In Sec.~\ref{sec:UM-UR}, we choose a concave, thus uncertainty, measure ${\mathfrak{u}(\vec{p},\vec{q}\,)}$.
A significance of our choice
lies in the fact that $\mathfrak{u}$ is again a concave function on every parametric curve (that is, as a function of the parameter). Therefore its absolute minimum $\mathfrak{c}$ will occur nowhere but at the endpoint(s) of these curves.
A simple three-step procedure is delivered to find the lower bound ${\mathfrak{c}\leq\mathfrak{u}}$ for an arbitrary pair ${\{a,b\}}$ of settings and for a finite $d$.
One can employ an ordinary computer to run the procedure.
Besides, $\mathfrak{c}$ is presented in analytic forms for ${d=2,3,}$ and in the case of mutually unbiased bases (MUBs) \cite{Durt10}. 
References~\cite{Kraus87,Larsen90,Ivanovic92,Sanchez-Ruiz95,Ballester07,Wu09,Mandayam10} contains URs particularly for MUBs.
At the end of Sec.~\ref{sec:UM-UR}, we provide another uncertainty measure that is also concave on all the parametric curves, so the whole analysis given before for $\mathfrak{u}$ can be straightforwardly applied to this measure.

If a suitable concave function can be a measure of the uncertainty, then an appropriate convex function will be a measure of certainty.
In Sec.~\ref{sec:other-UM}, we pick some other concave and convex functions and exhibit that the tight (un)certainty relations given in \cite{Rastegin12,Larsen90,Busch14,Garrett90,Sanchez-Ruiz98,Ghirardi03,Bosyk12,Vicente05,Zozor13,Deutsch83,Maassen88} for a qubit can be achieved with the TIs that specifies the ellipse.
We conclude the article with Sec.~\ref{sec:Conc-Out}.

The appendices are kept for certain technical details and proofs:
the TIs are derived in Appendix~\ref{sec:Der-q-const}.
It is manifested in Appendix~\ref{sec:Compact-Convex-w} that
the combined space is a compact convex set.
The parametric curves are explicitly obtained in Appendix~\ref{sec:Extreme-Points-w} with the help of 
Appendix~\ref{sec:inter-results}.

\section{Quantum constraints and combined-probability space}\label{sec:PS-C}

In quantum theory, observables are represented by Hermitian operators.
If such an operator is degenerate, then it possesses more than one eigenbases, where some of them can represent physically different measurement setups.
Hence, `measurement in an orthonormal basis' of the underlying Hilbert space is rather well defined than
`a measurement of an operator' (see Chapter~7 in \cite{Peres93}). 
In fact, measurement in a basis $\mathcal{B}_a$ measure all the operators whose eigenbasis is $\mathcal{B}_a$. 
Moreover, Deutsch pointed out that a measure of uncertainty for a discrete observable must not depend on its eigenvalues, but on
its eigenbasis \cite{Deutsch83}.
With all these considerations, we choose orthonormal bases instead of Hermitian operators to specify different projective measurements for a qudit.

We begin with two orthonormal bases  
\begin{equation}
	\label{AB-bases}
\mathcal{B}_a:=\big\{|a_i\rangle\big\}_{i=1}^{d} \quad \mbox{and} \quad  
\mathcal{B}_b:=\big\{|b_j\rangle\big\}_{j=1}^{d}
\end{equation}
of a $d$-dimensional Hilbert space $\mathscr{H}_d$ to depict the two measurement settings $a$ and $b$, respectively.
In this paper, all (un)certainty relations are \emph{preparation} (un)certainty relations that are applicable in the following experimental scheme.
\begin{equation}
\label{expt-situ}
\parbox{0.8\columnwidth}
{
$N$ number of independent qudits are identically prepared in a quantum state $\rho$. Then half of them are measured in the basis $\mathcal{B}_a$ and the rest in $\mathcal{B}_b$, one by one.
 }
\end{equation}
A similar scenario Peres used in his book~\cite{Peres93} at page~93 to interpret the position-momentum UR. 
In proposal~\eqref{expt-situ}, clearly, the two measurements have no influence whatsoever on each other.

Throughout the text, we 
assume $\rho$ is a pure quantum state ${|\psi\rangle\langle \psi|}$ so that we can associate angles~\eqref{alpha-beta} and TIs~\eqref{t-ineq-3} with the state vector ${|\psi\rangle}$.
Although every (un)certainty relation presented in this paper as it is applicable for every qudit's state 
[see the text around \eqref{pq rho}].

The state ${\rho=|\psi\rangle\langle \psi|}$ provides two probability distributions for the two measurement settings [given in~\eqref{AB-bases}] by the Born rule:
\begin{equation}
\label{pq}
p_i=|\langle a_i|\psi\rangle|^2 \quad \mbox{and} \quad  
q_j=|\langle b_j|\psi\rangle|^2
\end{equation}
are the probabilities of getting outcome $a_i$ in the $a$-setting and outcome $b_j$ in the $b$-setting, respectively.   
Next, we present quantum angles:
\begin{equation}
\label{alpha-beta}
\alpha_i=\arccos|\langle a_i|\psi\rangle| 
\quad \mbox{and} \quad  
\beta_j=\arccos|\langle b_j|\psi\rangle|
\end{equation}
are the angles between $|\psi\rangle$ and $|a_i\rangle$ and between $|\psi\rangle$ and $|b_j\rangle$, respectively. 
In the entire article, we consider only the principal values ${[0,\pi]}$ of the (multivalued) $\arccos$ function.
With \eqref{pq} and \eqref{alpha-beta}, one can recognize that the absolute value of the inner product establishes a one-to-one correspondence between the angles---that belong to ${[0,\tfrac{\pi}{2}]}$---and the probabilities---that lie in ${[0,1]}$.

Related to the $a$-setting, every probability vector ${\vec{p}:=(p_1,\cdots,p_d)}$ satisfies 
\begin{eqnarray}
\label{p-const1}
\textstyle\sum\nolimits_{i=1}^{d}p_i&=&1\quad \mbox{and} \\  
\label{p-const2}
0&\leq&p_i\quad \mbox{for all} \quad 1\leq i\leq d\;,
\end{eqnarray}
and the collection of all such vectors constitutes a probability space $\Omega_a$.
Similarly, $\Omega_b$ is---related to the basis $\mathcal{B}_b$---defined be the constraints 
\begin{eqnarray}
\label{q-const1}
\textstyle\sum\nolimits_{j=1}^{d}q_j&=&1\quad \mbox{and} \\  
\label{q-const2}
0&\leq&q_j\quad \mbox{for all} \quad 1\leq j\leq d\;.
\end{eqnarray}
Equations~\eqref{p-const1} and \eqref{q-const1} state that all the probabilities add up to one, and inequalities~\eqref{p-const2} and \eqref{q-const2} tell that probabilities are nonnegative numbers.
Both $\Omega_a$ and $\Omega_b$ are---the standard ${(d-1)}$-simplices---compact convex subsets of the $d$-dimensional real vector space $\mathbb{R}^d$, and their Cartesian product ${\mathbf{\Omega}:=\Omega_a\times\Omega_b}$ is a compact convex subset of $\mathbb{R}^{2d}$ [see Appendix~\ref{sec:Compact-Convex-w}].
Basically, $\mathbf{\Omega}$ is determined by the conditions \eqref{p-const1}--\eqref{q-const2}.

Performing measurement on every qudit using a single setting, say $a$, looks like throwing a $d$-sided dice, every time.
The vector $\vec{p}$ alone is limited by \eqref{p-const1} and \eqref{p-const2} that specify $\Omega_a$, which is also the probability space of a $d$-sided dice.
Whereas the experimental scheme~\eqref{expt-situ}
is not similar to throwing one out of two $d$-sided dices at a time, although $\mathbf{\Omega}$ is the probability space of two dices:
every pure or mixed state of a qudit gives a unique pair ${(\vec{p},\vec{q}\,)\in\mathbf{\Omega}}$ by the Born rule [see \eqref{pq} and \eqref{pq rho}], but \emph{not every} pair ${(\vec{p},\vec{q}\,)\in\mathbf{\Omega}}$ has a quantum state.
For example, if ${|\langle a_i|b_j\rangle|\neq1}$ for some $i,j$,
then one cannot get always the same outcome: $a_i$ in the $a$-setting and $b_j$ in the $b$-setting.
In other words, it is impossible to \emph{prepare} \cite{prep} a quantum system in a state (in this case, there exists no quantum state) that can provide ${(\vec{p},\vec{q}\,)}$, where ${p_i=1=q_j}$, which identifies an extreme point of $\mathbf{\Omega}$.

So, other than \eqref{p-const1}--\eqref{q-const2}, there are certain constraints that are purely quantum mechanical in nature and must be obeyed by $\vec{p}$ and $\vec{q}$ \emph{together}.
In our case, QCs are the TIs given in~\eqref{t-ineq-3}, which
arise naturally from the structure of Hilbert space on which quantum theory is based.
To write the TIs, we need  
\begin{equation}
\label{r}
\qquad \qquad\quad
r_{ij}=|\langle a_i|b_j\rangle|^2\qquad
\big(1\leq i,j\leq d\big)
\end{equation}
that is the probability of getting outcome ${a_i}$ if ${|b_j\rangle\langle b_j|}$ (or ${b_j}$ if ${|a_i\rangle\langle a_i|}$) is our state for the system.
Like $\alpha_i$ and $\beta_j$ in~\eqref{alpha-beta}, 
\begin{equation}
\label{theta}
\theta_{ij}=\arccos|\langle a_i|b_j\rangle|
\end{equation}
is the angle between the pure states ${|a_i\rangle\langle a_i|}$ and ${|b_j\rangle\langle b_j|}$. 
In the subscripts of $r_{ij}$ and $\theta_{ij}$, from left, the first 
and second indices are reserved for $\mathcal{B}_a$ and $\mathcal{B}_b$, respectively. 
Therefore, note that ${r_{ji}=|\langle a_j|b_i\rangle|^2}$ is different from $r_{ij}$, and likewise for $\theta$.

After choosing the measurement settings, $\mathcal{B}_a$ and $\mathcal{B}_b$ in~\eqref{AB-bases}, the entries in 
\begin{equation}
\label{r-theta-matrices}
R:=\begin{pmatrix}
r_{11} & \cdots & r_{1d} \\
\vdots  & \ddots & \vdots  \\
r_{d1} & \cdots & r_{dd} 
\end{pmatrix}
\quad \mbox{and} \quad  
\varTheta:=\begin{pmatrix}
\theta_{11} & \cdots & \theta_{1d} \\
\vdots  & \ddots & \vdots  \\
\theta_{d1} & \cdots & \theta_{dd} 
\end{pmatrix}
\end{equation} 
get fixed by~\eqref{r} and \eqref{theta}. 
Each entry in $R$ and in $\varTheta$ belong to ${[0,1]}$ and ${[0,\tfrac{\pi}{2}]}$, respectively. Sum of all the entries in each row and every column of $R$ is one, thus it is a doubly stochastic matrix.
If the two measurement settings described by \eqref{AB-bases} are physically the same, then $R$
will be a permutation matrix.
For every state vector ${|\psi\rangle\in\mathscr{H}_d}$, there are three TIs 
\begin{equation}
\label{t-ineq-3}
|\theta_{ij}-\beta_j|\,\leq\,\alpha_i\,\leq\,\theta_{ij}+\beta_j
\end{equation}
attached to each entry in $\varTheta$.
These TIs [see \eqref{QC_beta-alpha}] are derived in Appendix~\ref{sec:Der-q-const}.

For simplicity, out the three TIs~\eqref{t-ineq-3}, here we choose only one
\begin{equation}
\label{t-ineq}
\theta_{ij}\leq\alpha_{i}+\beta_j\quad\mbox{for every} \quad 1\leq i,j\leq d\,.
\end{equation}
Angles $\alpha_i$ and $\beta_j$ vary, whereas $\theta_{ij}$ is fixed, as we change the state vector ${|\psi\rangle}$.
The kets that saturates TI~\eqref{t-ineq} for certain $i,j$ lie in the linear span of ${\{|a_i\rangle,|b_j\rangle\}}$
[consider \eqref{psi-eq-a} and \eqref{psi-eq-b} with ${0\leq\beta\leq\theta}$ from Appendix~\ref{sec:Der-q-const}].
In the triangle \emph{equality} (TE) ${\theta_{ij}=\alpha_{i}+\beta_j}$,
$\alpha_i$ and $\beta_j$ are reminiscent of complementary angles from planar geometry, and ${0\leq\alpha_i,\beta_j\leq\theta_{ij}}$. 
Identifying $f$, $D$, and $B$ in \cite{Landau61} by our ${|\psi\rangle}$, 
${|a\rangle\langle a|}$, and ${|b\rangle\langle b|}$, respectively,
one can see that the TI ${\theta\leq\alpha+\beta}$
is obtained by Landau and Pollak for continuous-time signals
(see also Sec.~8 in \cite{Folland97}).
They also plotted elliptic curves (for different $\theta$s) one of this kind is shown in Fig.~\ref{fig:u,pi/6} between the point $E_1$ and $E_2$ (see also \cite{Lenard72}).
The results in \cite{Landau61, Lenard72} are more general than here, but they are only for a pair of projectors.
Whereas, we take every possible pair ${|a_i\rangle\langle a_i|}$ and ${|b_j\rangle\langle b_j|}$ and present three TIs [see \eqref{t-ineq-3}], not just one, for each pair.

The cosine function is strictly decreasing on ${[0,\pi]}$, so applying it on both sides of TI~\eqref{t-ineq} and using \eqref{pq}, \eqref{alpha-beta}, \eqref{r}, and \eqref{theta}, we attain 
\begin{equation}
\label{rpq}
\sqrt{p_{i}\,q_{j}}\;\leq\;\sqrt{r_{ij}}+\textstyle\sqrt{(1-p_{i})(1-q_{j})} 
\end{equation}
after a rearrangement of terms.
As both sides in \eqref{rpq} are nonnegative functions of the probabilities,
squaring and further simplification lead to
\begin{equation}
\label{rpq-2}
p_i+q_j\;\leq\,r_{ij}+1+2\sqrt{r_{ij}(1-p_i)(1-q_j)}
\end{equation}
for every ${1\leq i,j\leq d}$.

All those pairs ${(\vec{p},\vec{q}\,)\in\mathbf{\Omega}}$
that obey QC~\eqref{rpq-2} for every ${1\leq i,j\leq d}$
build the combined-probability space $\boldsymbol{\omega}$ for the two measurement bases in \eqref{AB-bases}.
In the case of ${d>2}$, even if we consider all TIs given in \eqref{t-ineq-3} for each ${1\leq i,j\leq d}$, they do not capture the full QCs for a general pair of settings.
Therefore, one can still find some ${(\vec{p},\vec{q}\,)\in\boldsymbol{\omega}}$ that corresponds to no quantum state.
Nevertheless, our analysis relies on the following fact:
\emph{every ${(\vec{p},\vec{q}\,)}$ that does not belong to $\boldsymbol{\omega}$ cannot be obtained from a quantum state, thus it is discarded.}
To investigate a space $\boldsymbol{\omega}_{{\textsc{q}}}$---that contains all those, and only those, pairs ${(\vec{p},\vec{q}\,)}$ that originate from the quantum states---is not the aim of this paper. 
However, it is not tough to realize that
${\boldsymbol{\omega}_{{\textsc{q}}}=\boldsymbol{\omega}}$ for ${d=2}$; in general, ${\boldsymbol{\omega}_{{\textsc{q}}}\subseteq\boldsymbol{\omega}}$.

Note that $\boldsymbol{\omega}$ is a \emph{proper} subset of $\mathbf{\Omega}$.
To prove this one can show: only one out of the two extreme points---specified by ${p_i=1=q_j}$ and ${p_i=1=q_{l}}$, where ${j\neq l}$---of $\mathbf{\Omega}$ can belong to $\boldsymbol{\omega}$.
Recall that if and only if ${r_{ij}=1}$ then the point described by ${p_i=1=q_j}$ belongs to $\boldsymbol{\omega}$, otherwise ${\theta_{ij}\leq\alpha_{i}+\beta_j}$ will be violated.
Secondly, if ${r_{ij}=1}$ then ${r_{il}=0}$, and ${\theta_{il}\leq\alpha_{i}+\beta_l}$ cannot be obeyed by the other point; hence that stays outside of $\boldsymbol{\omega}$.

The space $\boldsymbol{\omega}$ is---held by the conditions~\eqref{p-const1}--\eqref{q-const2} and \eqref{rpq-2}---a compact and convex subset of $\mathbb{R}^{2d}$ 
[for a proof, see Appendix~\ref{sec:Compact-Convex-w}].
Every point of such a set can be written as a convex combination of its extreme points due to the Krein-Milman theorem (see Theorem~${3.3.5}$ and Appendix~A.3 in \cite{Niculescu93}).
We begin our journey from an interior point of $\boldsymbol{\omega}$ in Appendix~\ref{sec:interior-w} and arrive at its extreme points at the end of Appendix~\ref{sec:extreme-w}.
There it is concluded that the set of all extreme points of $\boldsymbol{\omega}$ comes from a family of parametric curves.

One can skip all those technical details and start
constructing the parametric curves straight from the conclusion~\eqref{Ext-pts}:
the first step is to pick a set of $m$ angles from a single column or row of the matrix $\varTheta$ given in \eqref{r-theta-matrices}. 
Such a set is called $m$-set, and ${1\leq m\leq d-1}$.
For instance, we pick the top $m$ angles ${\{\theta_{i1}\}_{i=1}^m}$ from the first column.
Then we associate $m$ TEs with the $m$-set as
\begin{equation}
\label{m-t-eq}
\alpha_i=\theta_{i1}-\beta_1 \quad \mbox{for all}\quad i=1,\cdots,m
\end{equation}
by taking $\beta_1$, where the subscript~1 reflects the selected column.

Next, with~\eqref{pq} and \eqref{alpha-beta}, we assign ${m+1}$ probabilities to the angles: ${p_i={\cos\alpha_i}^2}$ and ${q_1={\cos\beta_1}^2}$.
They create the probability vectors
\begin{eqnarray}
\label{p-curves}
    \vec{p}{\scriptstyle(\beta_1)}&=&
    \big({\cos\alpha_1}^2,\cdots,{\cos\alpha_m}^2,
    \mathbf{0}\,,\,p_s\,,\,\mathbf{0}\big)\,,\qquad \\
\label{q-curves}
    \vec{q}{\scriptstyle(\beta_1)}&=&
    \big({\cos\beta_1}^2,\mathbf{0}\,,\,q_t\,,\,\mathbf{0}\big)\,,
    \qquad \mbox{where} \\
\label{p_s}
    p_{s}&=&1-{\textstyle\sum\nolimits_{i=1}^{m}{\cos\alpha_i}^2}
   \quad \ \; (m+1\leq s\leq d)\,,\qquad\qquad\\
\label{q_l}
    q_{t}&=&1-{\cos\beta_1}^2 \qquad\qquad\quad (2\leq t\leq d)\,,\quad 
    \mbox{and}\\
\label{bf-0-}
    \mathbf{0}&\equiv& 0,\cdots,0\,.
\end{eqnarray} 
One can observe that ${\big(\vec{p}{\scriptstyle(\beta_1)}\,,\,\vec{q}{\scriptstyle(\beta_1)}\big)}$ serves as a vector-valued function of a single real parameter $\beta_1$, thus it exhibits a parametric curve.
Since the curve is associated with an $m$-set and all its points obey $m$ TEs~\eqref{m-t-eq}, we call it an $m$-parametric curve.

A part of the curve, identified by the upper and lower limits ${\beta'\leq\beta_1\leq\beta''}$, lies in $\boldsymbol{\omega}$ and represents its extreme points because ${\big(\vec{p}{\scriptstyle(\beta_1)}\,,\,\vec{q}{\scriptstyle(\beta_1)}\big)}$ cannot be written into a convex combination of other points of $\boldsymbol{\omega}$. 
In Appendix~\ref{sec:limit-beta-1}, we realize that the two limits are fixed by
\begin{eqnarray}
\label{beta'-1}	
p_s{\scriptstyle(\beta')}={\cos(\theta_{s1}-\beta')}^2 
&\quad \mbox{ when }\quad &  1\leq m\leq d-1\,,\qquad\\
\label{beta'' 1=m}
\beta''=\tfrac{\theta_{11}-\theta_{1t}}{2}+\tfrac{\pi}{4} 
&\quad \mbox{ when }\quad &  1=m\,,\quad\mbox{and}\qquad\\
\label{beta'' 1<m}
p_s{\scriptstyle(\beta'')}=0\
&\quad \mbox{ when }\quad & 1<m\leq d-1
\end{eqnarray} 
[see \eqref{lim-bete}].
Equations~\eqref{beta'-1} and \eqref{beta'' 1<m} are like Eq.~\eqref{eq-beta-1}, whose 
roots are stated in \eqref{eq-beta-roots}.
Always the root with +~sign delivers the correct limit [for justifications, see the last paragraph in Appendix~\ref{sec:limit-beta-1}].

If one chooses an $m$-set from a row of $\varTheta$, say 
${\{\theta_{1j}\}_{j=1}^m}$, then the $m$-parametric curve is constructed as
\begin{eqnarray}
\label{m-t-eq row}
\beta_j&=&\theta_{1j}-\alpha_1 \quad \mbox{for all}\quad 
j=1,\cdots,m\,,\\
\label{p-curves row}
\vec{p}{\scriptstyle(\alpha_1)}&=&
\big({\cos\alpha_1}^2,\mathbf{0}\,,\,p_s\,,\,\mathbf{0}\big)\,,\qquad \\
\label{q-curves row}
\vec{q}{\scriptstyle(\alpha_1)}&=&
\big({\cos\beta_1}^2,\cdots,{\cos\beta_m}^2\,,\mathbf{0}\,,\,q_t\,,\,\mathbf{0}\big)\,,
\qquad\quad \\
\label{p_s row}
p_{s}&=&1-{\cos\alpha_1}^2
 \qquad\qquad\quad (2\leq s\leq d)\,,\quad\mbox{and} \\
\label{q_l row}
q_{t}&=&1-{\textstyle\sum\nolimits_{j=1}^{m}{\cos\beta_j}^2}
\quad \ \; (m+1\leq t\leq d)\,.\qquad\qquad
\end{eqnarray} 
Now the parameter is ${\alpha_1\in[\alpha',\alpha'']}$, and the limits are determined by
\begin{eqnarray}
\label{alpha'-1}	
q_t{\scriptstyle(\alpha')}={\cos(\theta_{1t}-\alpha')}^2 
&\quad \mbox{ when }\quad &  1\leq m\leq d-1\,,\qquad\\
\label{alpha'' 1=m}
\alpha''=\tfrac{\theta_{11}-\theta_{s1}}{2}+\tfrac{\pi}{4} 
&\quad \mbox{ when }\quad &  1=m\,,\quad\mbox{and}\qquad\\
\label{alpha'' 1<m}
q_t{\scriptstyle(\alpha'')}=0\
&\quad \mbox{ when }\quad &  1<m\leq d-1\,.
\end{eqnarray} 
One can check that, for ${m=1}$,
both \eqref{m-t-eq}--\eqref{beta'' 1=m} and
\eqref{m-t-eq row}--\eqref{alpha'' 1=m} describe the same thing, provided $s$ and $t$ are identical in both the cases.
So an $m$-parametric curve is identified by an $m$-set and the positions of $p_s$ and $q_t$ (that is, $s$ and $t$) in $\vec{p}$ and $\vec{q}$, respectively.

Let us count the total number of curves such as describe by \eqref{m-t-eq}--\eqref{q_l}.
One can harvest $\tfrac{d!}{m!(d-m)!}$ distinct $m$-sets from a single column of $\varTheta$, and there are total $d$ columns. 
The probability $p_s$ can take ${d-m}$ separate places in $\vec{p}$ of~\eqref{p-curves} for distinct $s$, and $q_t$ can take ${d-1}$ separate places in $\vec{q}$ of~\eqref{q-curves} for distinct $t$.
Thus we have $(d-m)(d-1)$ individual $m$-parametric curves with a
single $m$-set.
Since ${1\leq m\leq d-1}$, we collect
\begin{equation}
\label{total-curves col/row}
d\,\textstyle\sum\nolimits_{m=1}^{d-1}\tfrac{d!}{m!(d-m)!}(d-1)(d-m)
\end{equation}
number of curves, where each $m$-set is made of angles from a column of $\varTheta$.

We secure the same number if we consider rows, rather than columns, to build an $m$-set and then a curve such as given 
by \eqref{m-t-eq row}--\eqref{q_l row}.
For ${m=1}$, every $m$-set is a part of a row as well as a part of a column.
So, to avoid double counting errors, we take the cases ${m=1}$ and ${m>1}$ separately.
In total, there are 
\begin{eqnarray}
\label{total-curves}
d^2(d-1)^2&&\,+\,2d\,\textstyle\sum\nolimits_{m=2}^{d-1}\tfrac{d!}{m!(d-m)!}(d-1)(d-m)\quad\quad\nonumber\\
\quad&&=d^2(d-1)[2^d-(d+1)]
\end{eqnarray}
number of parametric curves for a qudit.

If one adopts a suitable concave function ${\mathsf{u}(\vec{p},\vec{q}\,)}$ on the combined space $\boldsymbol{\omega}$ to estimate the uncertainty, then its 
absolute minimum will occur \emph{only} at the parametric curves (see Theorem~${3.4.7}$ and Appendix~A.3 in \cite{Niculescu93}).
So ultimately one needs to find absolute minima of, at most, $d^2(d-1)[2^d-(d+1)]$ functions, each of a single variable [for example, see~\eqref{u-beta}]. 
Then the smallest minimum will be the lower bound ${\mathsf{c}\leq\mathsf{u}}$ in an UR.
This task can be easily completed with a regular computer.
In the next two sections, we discuss certain concave as well as convex functions on $\boldsymbol{\omega}$.

\section{Uncertainty measures and relations}\label{sec:UM-UR}

If $u$ quantifies the uncertainty---about the outcomes $a_i$ when a qudit is measured in the basis $\mathcal{B}_a$ of \eqref{AB-bases}---then $u$ should be a concave function of ${\vec{p}\in\Omega_a}$. 
It is because mixing probability distributions, $\vec{p}\,'$ and $\vec{p}\,''$ as ${\lambda\,\vec{p}\,'+(1-\lambda)\vec{p}\,''=\vec{p}}$ with ${\lambda\in[0,1]}$,
can only increase uncertainty
${\lambda\,u(\vec{p}\,')+(1-\lambda)u(\vec{p}\,'')\leq u(\vec{p}\,)}$
(see Chapter~9 in \cite{Peres93}).
In this regard, every mixed state, say 
${\lambda|\psi'\rangle\langle\psi'|+
	(1-\lambda)|\psi''\rangle\langle\psi''|=\rho_{\text{mix}}}$,
has more uncertainty.

So, here, we adopt a real-valued smooth concave function
\begin{eqnarray}
	\label{ua}
	u(\vec{p}\,):=\textstyle\sum\nolimits_{i=1}^{d}\sqrt{p_i}
\end{eqnarray}
as an uncertainty measure.
It is associated with the Tsallis entropy \cite{Tsallis88} ${S_{\sfrac{1}{2}}(\vec{p}\,)=2K(u(\vec{p}\,)-1) }$, where $K$ 
the Boltzmann constant.
To prove ${u(\vec{p}\,)}$ is a concave function on $\Omega_a$, it is sufficient to demonstrate that the ${(d-1)\times(d-1)}$ Hessian matrix---that is a symmetric matrix of second-order partial derivatives of $u$---is a negative semidefinite matrix at every point in $\Omega_a$ (see Theorem~${4.5}$ in \cite{Rockafellar70}).
At an interior point (where all ${p_i>0}$) of $\Omega_a$, the entry in the $k$th row and $l$th column ${(1\leq l,k\leq d-1)}$ in the Hessian matrix is 
\begin{equation}
\label{h_kl}
\frac{\partial^2 u}{\partial p_k\partial p_l}
=-\frac{1}{4}\left(\frac{1}{p_l^{3/2}}\delta_{lk}+
\frac{1}{p_d^{3/2}}\right)=\frac{\partial^2 u}{\partial p_l\partial p_k}\,,
\end{equation}
where ${p_d=1-\sum\nolimits_{i=1}^{d-1}p_i}$ and $\delta_{lk}$ is the Kronecker delta function.
These entries indeed provide a negative definite matrix, thus ${u(\vec{p}\,)}$ is strictly concave in the interior of $\Omega_a$. At a boundary point (where one or more ${p_i=0}$), all the partial derivatives in a certain row(s) and column(s) of the Hessian matrix become zero, thus the matrix turns out to be a negative semidefinite and ${u(\vec{p}\,)}$ to be a concave function.
By the way, ${u(\vec{p}\,)}$ can be employed for the entanglement detection (see Remark~2 in \cite{Sehrawat16}).

If the state vector ${|\psi\rangle}$ is an equal superposition of all the kets in $\mathcal{B}_a$ or the state is completely mixed, then all the outcomes $a_i$ will be equally probable: ${p_i=\tfrac{1}{d}}$ for every ${1\leq i\leq d}$ is the center of $\Omega_a$, where ${u(\vec{p}\,)}$ reaches its maximum value $\sqrt{d}$. 
Whereas, only in the case of a definite outcome---that is when ${|\psi\rangle\langle\psi|=|a_i\rangle\langle a_i|}$, and then ${p_i=1}$ for a particular $i$---we have the minimum uncertainty ${u(\vec{p}\,)=1}$ as it should be. Note that ${p_i=1}$ characterizes  an extreme point of $\Omega_a$.

To establish a measure of combined uncertainty 
for the experimental proposal~\eqref{expt-situ}, we take the same function,
\begin{eqnarray}
\label{ub}
 u(\vec{q}\,)=\textstyle\sum\nolimits_{j=1}^{d}\sqrt{q_j}\,,
\end{eqnarray}
for the $b$-setting.
Like ${u(\vec{p}\,)}$ of \eqref{ua}, ${u(\vec{q}\,)}$ is a concave function on $\Omega_b$ with the range ${[1,\sqrt{d}\,]}$.
Now we define our combined uncertainty measure 
\begin{equation}
	\label{u-pq}	\mathfrak{u}(\vec{p},\vec{q}\,):=u(\vec{p}\,)+u(\vec{q}\,)=
	\textstyle\sum\nolimits_{l=1}^{d}\big(\sqrt{p_l}+\sqrt{q_l}\,\big)
\end{equation}
on the convex set $\boldsymbol{\omega}$, rather than $\mathbf{\Omega}$.
Sum of two concave functions is concave, so ${\mathfrak{u}}$ is also a concave function.

A mixed quantum state is a convex combination of pure states, the probabilities 
\begin{equation}
\label{pq rho}
p_i=\text{tr}\,\bm(\varrho\,|a_i\rangle\langle a_i|\bm) \quad \mbox{and} \quad  
q_j=\text{tr}\,\bm(\varrho\,|b_j\rangle\langle b_j|\bm)
\end{equation}
are linear functions of the state $\varrho$ (${0\leq\varrho}$, ${\text{tr}(\varrho)=1}$), and $\boldsymbol{\omega}$ is a compact and convex set.
As a result, every $(\vec{p},\vec{q}\,)$ associated with any (pure or mixed) quantum state lies in $\boldsymbol{\omega}$.
And, because ${\mathfrak{u}}$ is a concave function on $\boldsymbol{\omega}$, our UR given in \eqref{u-range} applies to every state for a qudit.
This is also true in the case of other (un)certainty relations presented in Sec.~\ref{sec:other-UM}, because mostly there also we have either a concave or a convex function.
In \eqref{ur-pro,d=2,UR} and \eqref{max-min entropy UR}, the functions are neither concave nor convex on $\boldsymbol{\omega}$, but the relations are followed by every qubit's state.
By the way, one can check that if 
${\varrho=|\psi\rangle\langle\psi|}$ then 
the Born rule~\eqref{pq rho} reduces to \eqref{pq}.

The range of ${\mathfrak{u}(\vec{p},\vec{q}\,)}$ and our UR are presented as
\begin{eqnarray}
\label{u-range}
&&2\,\leq\, \mathfrak{c}\,\leq\, \mathfrak{u}(\vec{p},\vec{q}\,)
\,\leq\, \,2\sqrt{d}\,,\quad\mbox{where}\\
\label{c}
&&\mathfrak{c}:=\operatorname*{min}_{
	(\vec{p},\vec{q}\,)\,\in\,\boldsymbol{\omega}\,}\mathfrak{u}(\vec{p},\vec{q}\,)
\end{eqnarray} 
is the global minimum that
will occur at the $m$-parametric curves [given in Sec.~\ref{sec:PS-C}].
Whereas, $\mathfrak{u}$ gains its absolute maximum ${2\sqrt{d}}$
only at the point identified by 
${p_i=\tfrac{1}{d}=q_j}$ for all ${1\leq i,j\leq d}$. It is called the \emph{center} of $\boldsymbol{\omega}$, which represents the uniform distribution for both the settings. 
Now recall from Sec.~\ref{sec:PS-C} that an extreme point of $\mathbf{\Omega}$, describe by ${p_i=1=q_j}$, belongs to $\boldsymbol{\omega}$ if and only if
${|a_i\rangle\langle a_i|=|b_j\rangle\langle b_j|}$.
Only in such a situation---that does not necessarily require both the bases $\mathcal{B}_a$ and $\mathcal{B}_b$ to be the same in any way---we have the \emph{trivial} lower bound ${\mathfrak{c}=2}$ and thus the UR ${2\leq\mathfrak{u}}$.
A similar statement is made by Deutsch in \cite{Deutsch83}.
For ${d=2}$, the trivial case is possible if and only if the two measurement settings are (physically) the same.
A nontrivial lower bound ${\mathfrak{c}>2}$ materializes when the settings are completely different, that is when ${r_{ij}<1}$ for every ${1\leq i,j\leq d}$. 
So the following analysis is obviously for the nontrivial cases.

To find the lower bound~\eqref{c} and to establish the UR ${\mathfrak{c}\leq\mathfrak{u}}$, we write the functional form 
\begin{equation}
 \label{u-beta}
 \mathfrak{u}(\beta_1)=\textstyle\sum\nolimits_{i=1}^{m}\cos\alpha_i+
 \sqrt{p_s}+\cos\beta_1+\sin\beta_1\,,
\end{equation}
which ${\mathfrak{u}(\vec{p},\vec{q}\,)}$ of \eqref{u-pq} acquires on an $m$-parametric curve specified by \eqref{m-t-eq}--\eqref{bf-0-}.
To show that ${\mathfrak{u}}$ of \eqref{u-beta} is a concave function of $\beta_1$, we present
 \begin{eqnarray}
 \label{d-u/d-beta}
 &&\frac{\partial^2\,\mathfrak{u}}{{\partial\beta_1}^2}=
 -\left[{\textstyle\sum\nolimits_{i=1}^{m}}\cos\alpha_i+
       \cos\beta_1+\sin\beta_1\right]+
       \frac{\partial^2\,\sqrt{p_s}}{{\partial\beta_1}^2}\,,\quad\ \ \\
\label{d-sqrtps/d-beta}
&&\frac{\partial^2\sqrt{p_s}}{{\partial\beta_1}^2}=
  -\,\frac{1}{4\,p_s^{3/2}}
  \left(\frac{\partial\,p_s}{\partial\beta_1}\right)^2
  +\frac{1}{2\sqrt{p_s}}\,\frac{\partial^2\,p_s}{{\partial\beta_1}^2}\,,
  \quad\mbox{and}\\
\label{d-ps/d-beta}  
&&\frac{\partial^2\,p_s}{{\partial\beta_1}^2}=
  -2\left[\,2\,p_s+(m-2)\,\right].
 \end{eqnarray}  
With these derivatives, one can clearly see ${\frac{\partial^2\,\mathfrak{u}}{{\partial\beta_1}^2}<0}$ for 
${1< m\leq d-1}$. Whereas, for ${m=1}$, one can directly realize ${\frac{\partial^2\,\mathfrak{u}}{{\partial\beta_1}^2}=
	-\mathfrak{u}<0}$.
This proves that $\mathfrak{u}$ is a (strictly) concave function on every parametric curve.
Therefore, its global minimum $\mathfrak{c}$ will always be at the endpoints of the curves.
Endpoints of an $m$-parametric curve are identified by the two limits on a parameter [see \eqref{beta'-1}--\eqref{beta'' 1<m} as well as \eqref{alpha'-1}--\eqref{alpha'' 1<m}].

It is manifested in Appendix~\ref{sec:limit-beta-1} that, to compute a limit, we always have to solve an equation such as \eqref{eq-beta-1}; which carries \textsc{m} number of angles from a column or a row of $\varTheta$ [given in \eqref{r-theta-matrices}].
Note that we use small letter `$m$' ${(1\leq m\leq d-1)}$ when we construct a parametric curve with an $m$-set [see Sec.~\ref{sec:PS-C}] and use
capital letter `\textsc{m}' ${(2\leq \textsc{m}\leq d)}$ when we compute a limit with an \textsc{m}-set.
Essentially, one needs to follow a three-step procedure 
to compute a limit and then the value of $\mathfrak{u}$ [defined in \eqref{u-pq}, see also \eqref{u-beta}] at the corresponding endpoint of a curve:
\begin{equation}
\label{3-steps}
\parbox{0.85\columnwidth}
{%
	1. Pick an $\textsc{m}$-set from a column or a row of~$\varTheta$, say ${\{\theta_{\scriptstyle 1},
		\cdots,\theta_{\scriptstyle\textsc{m}}\}}$, here only one index of $\theta$ is shown.\\
	2.~Solve ${{\textstyle\sum\nolimits_{l=1}^{\textsc{m}}
			{\cos(\theta_{l}-\chi)}^2}=1}$ for $\chi$ that represents a limit.\\
	3. Compute ${c_\textsc{m}:={\textstyle\sum\nolimits_{l=1}^{\textsc{m}}\cos(\theta_{l}-\chi)}+\cos\chi+\sin\chi}$ that is the value of $\mathfrak{u}$ at an endpoint.  
}
\end{equation}
The equation in Step~2 is like Eq.~\eqref{eq-beta-1} that is solved in Appendix~\ref{sec:limit-beta-1}, and every time we take the solution~\eqref{eq-beta-roots} with +~sign.
One can observe that $\chi$ and therefore $c_\textsc{m}$
are solely determined by the $\textsc{m}$-set picked in Step~1.

After repeating the three-step procedure for every $\textsc{m}$-set and for each ${2\leq \textsc{m}\leq d}$, we collect a set of values ${\{c_\textsc{m}\}}$ for all the endpoints.
Then, the smallest value in this set will be $\mathfrak{c}$ [defined by
\eqref{c}], and thus we own our UR $\mathfrak{c}\leq\mathfrak{u}$ [presented in~\eqref{u-range}].
Since every $c_\textsc{m}$ is determined by the entries in $\varTheta$-matrix, the
lower bound $\mathfrak{c}$---depends only on the measurement bases in~\eqref{AB-bases}---is \emph{independent} of a quantum state.
Besides, to compute $\mathfrak{c}$, we can employ an ordinary computer, which repeats the three steps of \eqref{3-steps} by taking
\begin{equation}
\label{total-endpoints}
2d\,\textstyle\sum\nolimits_{\textsc{m}=2}^{d}\tfrac{d!}{\textsc{m}!(d-\textsc{m})!}=
2d\,[\,2^{d}-(d+1)]
\end{equation}
number of $\textsc{m}$-sets one by one. In fact, ${2d\,[\,2^{d}-(d+1)]}$ is the total number of endpoints for a qudit.

Although we have the solution~\eqref{eq-beta-roots} for Step~2, it is easy to calculate 
$\chi$ and $c_\textsc{m}$ for ${\textsc{m}=2,d}$.
For a 2-set ${\{\theta_1,\theta_2\}}$, one can directly realize
\begin{eqnarray}
\label{chi_textsc{m}=2}
\chi&=&\tfrac{\theta_1+\theta_2}{2}-\tfrac{\pi}{4}\,,
\quad\mbox{and then}\\
\label{c2_for_textsc{m}=2}
c_2(\theta_1,\theta_2)&=&
\sqrt{2}\left[\cos\big(\tfrac{\theta_1-\theta_2}{2}\big)+\sin\big(\tfrac{\theta_1+\theta_2}{2}\big)\right]\\
&=&
\tfrac{1}{\sqrt{2}}{\scriptstyle\big(\sqrt{1+\sqrt{r_1}}+\sqrt{1-\sqrt{r_1}}\big)\big(\sqrt{1+\sqrt{r_2}}+\sqrt{1-\sqrt{r_2}}\big)}.\qquad
\end{eqnarray}
Every endpoint of a ${m=1}$ parametric curve is determined by a set of ${\textsc{m}=2}$ angles [see \eqref{beta'-1}, \eqref{beta'' 1=m}, \eqref{alpha'-1}, and \eqref{alpha'' 1=m}].
For a $d$-set ${\{\theta_1,\cdots,\theta_d\}}$, that is an entire column or row of $\varTheta$, we have the total probability ${{\textstyle\sum\nolimits_{l=1}^{d}{\cos\theta_{l}}^2}=1}$.
Therefore, we obtain the solution
\begin{eqnarray}
\label{chi_textsc{m}=d}
\chi&=&0\,,
\quad\mbox{and then}\\
\label{cd_for_textsc{m}=d}
c_d(\theta_1,\cdots,\theta_d)&=&
{\textstyle\sum\nolimits_{l=1}^{d}\cos\theta_{l}}+1=
{\textstyle\sum\nolimits_{l=1}^{d}\sqrt{r_{l}}}+1\,.\qquad
\end{eqnarray}

For general measurement settings, it is---easy to compute but---difficult to express $\mathfrak{c}$ in an analytic form.
Nevertheless, we present it for ${d=2,3,}$ and when the measurement bases in~\eqref{AB-bases} are MUBs \cite{Durt10}.

In the case of a qubit, ${d=2}$, a (un)certainty relation can be stated with the three probabilities $p_1$, $q_1$, and $r_{11}$,
hence we drop their subscripts here and in the next section.
Furthermore, all the TIs~\eqref{t-ineq} can now be put together as
\begin{equation}
	\label{d2-cons}
	\theta\leq\alpha+\beta\leq\pi-\theta\quad\mbox{and}\quad
	|\alpha-\beta|\leq\theta\,,
\end{equation}
where $\alpha$, $\beta$, and $\theta$ are associated with $p$, $q$, and $r$, respectively [through \eqref{pq}, \eqref{alpha-beta}, \eqref{r}, and \eqref{theta}].
Here only ${m=1}$ parametric curves exist, which are 
four in total [see with~\eqref{total-curves}].
To draw an endpoint of a curve, we can use either \eqref{chi_textsc{m}=2} or \eqref{chi_textsc{m}=d}; both are equal (because ${\theta_1+\theta_2=\tfrac{\pi}{2}}$).
There are only four [see \eqref{total-endpoints}] endpoints
${E_1,\cdots,E_4}$.
Next, one can realize that \eqref{c2_for_textsc{m}=2} and \eqref{cd_for_textsc{m}=d} are also the same for a qubit.
Furthermore, $c_d$ is even identical for every ${\textsc{m}=2}$ set.
It implies that
our combined uncertainty function~\eqref{u-pq} takes the same value at all the four endpoints, thus $\mathfrak{c}=c_d=c_2$ and
\begin{equation}
\label{d2-UR}
\underbrace{\sqrt{r}+\sqrt{1-r}+1}_{\textstyle\mathfrak{c}(r)}\,\leq\,
\underbrace{\sqrt{p}+\sqrt{1-p}+\sqrt{q}+\sqrt{1-q}}_{\textstyle\mathfrak{u}(p,q)}
\end{equation}
is an UR for ${d=2}$.
It is also given in \cite{Rastegin12}.

Together all the parametric curves---that represent all the extreme points of the combined-probability space $\boldsymbol{\omega}$---can be expressed by an ellipse
\begin{equation}
\label{ellipse}
(p{\scriptstyle(\vartheta)},q{\scriptstyle(\vartheta)})=
\big({\cos(\theta-\vartheta)}^2,{\cos\vartheta}^2\,\big)
\ \mbox{with}\ \vartheta\in[0,\pi)
\end{equation}
in the case of a qubit.
As a special case, the same ellipse also appears in \cite{Lenard72,Larsen90,Kaniewski14} through different routes \cite{diff-routes},
although our approach is closer to \cite{Lenard72}.
One can observe that the ellipse turns into a circle for ${\theta=\tfrac{\pi}{4}}$ and into certain line segments for ${\theta=0,\tfrac{\pi}{2}}$.
In Fig.~\ref{fig:u,pi/6}, we present a contour plot of ${\mathfrak{u}(p,q)}$ on $\boldsymbol{\omega}$ by taking ${r=\tfrac{3}{4}}$.
So ${\theta=\tfrac{\pi}{6}}$, and
one can see that $\boldsymbol{\omega}$ is bounded by
the ellipse~\eqref{ellipse}.
Furthermore, by putting ${\vartheta=0,\theta,\tfrac{\pi}{2},\tfrac{\pi}{2}+\theta}$
in ${(p{\scriptstyle(\vartheta)},q{\scriptstyle(\vartheta)})}$, we can have the four endpoints ${E_1,\cdots,E_4}$, respectively.

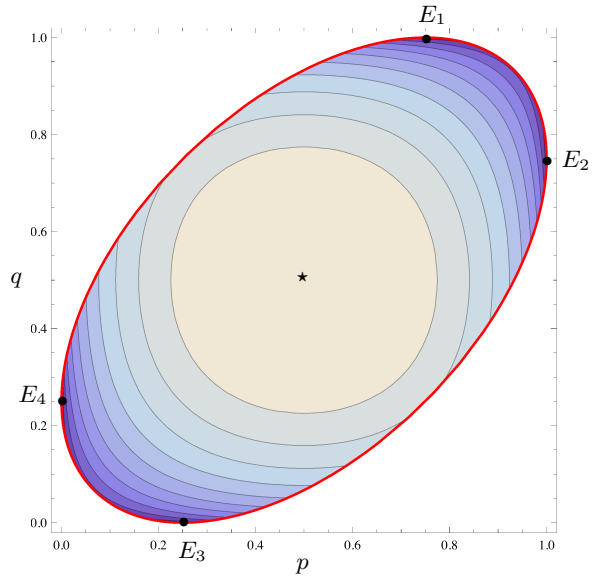
\begin{figure}
	\centering
	\begin{overpic}[width=0.4\textwidth 
		]{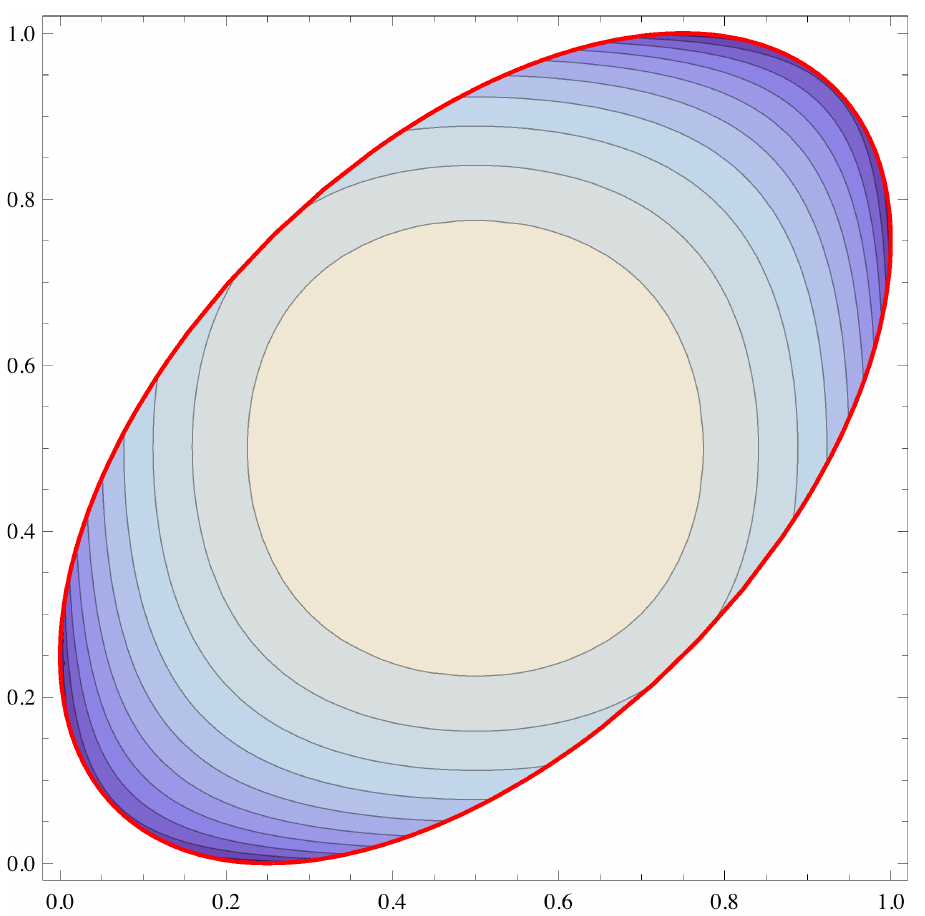}
		\put (50,-3) {$p$}
		\put (-3,50) {$q$}
		\put (72.5,98.5) {$E_1$}	\put (73,94.2) {$\bullet$}
		\put (95.4,71.5) {$\bullet\ E_2$}
		\put (28,-1) {$E_3$}	\put (28,4.5) {$\bullet$}
		\put (-1.5,28) {$E_4$}	\put (5.5,27) {$\bullet$}
		\put (50,50) {$\star$}
	\end{overpic} 
	\caption{For ${d=2}$ and ${r=\tfrac{3}{4}}$, contour plot of ${\mathfrak{u}(p,q)}$ on $\boldsymbol{\omega}$, where a darker shade represents a smaller value of $\mathfrak{u}$.
		The square-shaped and elliptical regions are $\mathbf{\Omega}$
		and $\boldsymbol{\omega}$, respectively.
		Note that ${\boldsymbol{\omega}\subset\mathbf{\Omega}\subset\mathbb{R}^{4}}$ and the unseen coordinates are ${p_2=1-p}$ and ${q_2=1-q}$ for each point. 
		For every ${r\in[0,1]}$, ${\mathfrak{u}}$ hits its global minimum ${\mathfrak{c}}$ [given in \eqref{d2-UR}] on $\boldsymbol{\omega}$ at all the four points ${E_1,\cdots,E_4}$, which are marked by the bullets ${(\bullet)}$.
		And, ${\mathfrak{u}}$ achieves its global maximum ${2\sqrt{2}}$ [stated in \eqref{u-range}] always at the center, ${p=\tfrac{1}{2}=q}$ indicated by the star ${(\star)}$, of $\boldsymbol{\omega}$.}
	\label{fig:u,pi/6}  
\end{figure}

In the case of ${d=2}$, there always exist a quantum state for each point in $\boldsymbol{\omega}$, thus ${\boldsymbol{\omega}=\boldsymbol{\omega}_{\textsc{q}}}$.
For instance, the
kets such as \eqref{psi-eq-a} and \eqref{psi-eq-b} correspond to 
points on the ellipse~\eqref{ellipse} by the Born rule~\eqref{pq}. 
In particular, the kets of basis $\mathcal{B}_a$ correspond to the points ${\{E_2,E_4\}}$, and the kets of $\mathcal{B}_b$ are related with ${\{E_1,E_3\}}$.
So the lower bound $\mathfrak{c}(r)$ in the UR~\eqref{d2-UR} is achieved---hence, it is a \emph{tight} UR---only by those state vectors $|\psi\rangle$ that (up to a phase factor) belong to one of the bases in~\eqref{AB-bases}.
The lower bound will be the largest ${\sqrt{2}+1}$ when, ${r=\tfrac{1}{2}}$, the measurement bases are MUBs [see also \eqref{c-MUB}].

An UR is called tight if there exists a quantum state that saturates the UR.
In the case of a qubit, all the relations mentioned in this and the next section are tight because 
${\boldsymbol{\omega}=\boldsymbol{\omega}_{\textsc{q}}}$.
For ${d\geq3}$, ${\boldsymbol{\omega}_{\textsc{q}}\subseteq\boldsymbol{\omega}}$, 
hence our UR ${\mathfrak{c}\leq\mathfrak{u}}$ is not tight in general.

In the case of ${d=3}$ (qutrit), there are only two kinds of parametric curves (for ${m=1,2}$), and two types of endpoints (for ${\textsc{m}=2,3}$).
So \eqref{chi_textsc{m}=2} and \eqref{chi_textsc{m}=d} can specify any endpoint for a qutrit.
To compute the lower bound $\mathfrak{c}$, we have to evaluate the function $c_2$ of \eqref{c2_for_textsc{m}=2} for every 2-set and $c_d$ of \eqref{cd_for_textsc{m}=d} every $d$-set drawn from the $\varTheta$-matrix.
For ${d=3}$, there are 18 {2-sets} and 6 $d$-sets [see the total in~\eqref{total-endpoints}].
Then, the smallest out of the ${18+6=24}$ values will be our $\mathfrak{c}$. 
Now let us consider a pair of MUBs \cite{Durt10} for a finite dimension $d$.

If the two bases given in~\eqref{AB-bases} are such that
${r_{ij}=\tfrac{1}{d}}$ for every ${1\leq i,j\leq d}$ [for $r_{ij}$, see \eqref{r}], then they 
are called MUBs and the measurement settings $a$ and $b$ are designated as \emph{complementary} \cite{Kraus87}.
In the case of MUBs, ${\theta_{ij}=\arccos\tfrac{1}{\sqrt{d}}}$ for every $i,j$, so one can straightforward realize
\begin{eqnarray}
\label{beta_end-MUB}
\chi&=&\arccos\tfrac{1}{\sqrt{d}}-
\arccos\tfrac{1}{\sqrt{\textsc{m}}}\,,
\quad\mbox{and}\\
\label{u_beta_end-MUB}
c_\textsc{m}&=&
\sqrt{\textsc{m}}+\tfrac{1+\sqrt{(d-1)(\textsc{m}-1)}+\sqrt{d-1}\;-\sqrt{\textsc{m}-1}}{\sqrt{d\,\textsc{m}}}
\end{eqnarray}
in Step~2 and 3 of the three-step procedure~\eqref{3-steps}.
One can acknowledge that here $\chi$ and $c_\textsc{m}$
depend on ${\textsc{m}=2,\cdots,d}$, not on a particular $\textsc{m}$-set, because every $\theta$ is the same.
Furthermore, $\chi$ decreases, whereas $c_\textsc{m}$ increases, with
$\textsc{m}$.
Hence the lower bound is
\begin{equation}
\label{c-MUB}
\mathfrak{c}^{(d)}_{\textsc{mub}}=c_2=
\sqrt{2}\left(1+\tfrac{\sqrt{d-1}}{\sqrt{d}}         
\,\right)\,, 
\end{equation}
which does not deliver a tight UR when ${d>2}$, whereas tight URs  \cite{Kraus87,Maassen88,Ballester07} are
known for MUBs in a finite $d$. 
We close this section with the following remarks.

{\textbf{Remark~1:}} By the Born rule~\eqref{pq}, ${|\psi\rangle=|a_i\rangle}$ provides an extreme point, given by ${p_i=1}$ and ${\vec{q}=(r_{i1},\cdots,r_{id})}$, of $\boldsymbol{\omega}$
[see \eqref{ext-pt_m=1,0=p p} and \eqref{ext-pt_m=1,0=p q} in Appendix~\ref{sec:extreme-w}]. 
At this point the combined uncertainty function~\eqref{u-pq} has the value
${1+\textstyle\sum\nolimits_{j=1}^{d}\sqrt{r_{ij}}\,}$ [see also \eqref{cd_for_textsc{m}=d}].
Likewise, ${|\psi\rangle=|b_j\rangle}$ gives the 
combined uncertainty ${1+\textstyle\sum\nolimits_{i=1}^{d}\sqrt{r_{ij}}\,}$.
Now we take the minimum value
\begin{eqnarray}
\label{c-basis}
\mathfrak{c}_{\text{bases}}&:=&
\min\{\,\mathfrak{u}_a\,,\,\mathfrak{u}_b\,\}\,,\quad\mbox{where}\\
\label{u_a}	
\mathfrak{u}_a&:=&\operatorname*{min}_{1\leq i\leq d}\,
\big\{1+\textstyle\sum\nolimits_{j=1}^{d}\sqrt{r_{ij}}\,\big\}
\quad\mbox{and}\\
\label{u_b}
\mathfrak{u}_b&:=&\operatorname*{min}_{1\leq j\leq d}\,
\big\{1+\textstyle\sum\nolimits_{i=1}^{d}\sqrt{r_{ij}}\,\big\}\,.
\end{eqnarray} 
Next, one can easily establish
\begin{eqnarray}
\label{c<cQ<cmax}
&&2\leq\mathfrak{c}\leq\mathfrak{c}_{\textsc{q}}\leq\mathfrak{c}_{\text{bases}}\leq1+\sqrt{d}\,,
\quad\mbox{where}\\	
\label{c_Q}
&&\mathfrak{c}_{\textsc{q}}:=\operatorname*{min}_{|\psi\rangle\,\in\,\mathscr{H}_d}\,
\mathfrak{u}(\vec{p},\vec{q}\,)\,.
\end{eqnarray} 
The first inequality in~\eqref{c<cQ<cmax} comes from \eqref{u-range}.
The last inequality is due to ${\textstyle\sum\nolimits_{i=1}^{d}\sqrt{r_{ij}}\leq\sqrt{d}}$ and the
similar relation where the summation is over index $j$ instead of $i$.
$\mathfrak{c}_{\textsc{q}}$ is the \emph{largest} lower bound that defines the \emph{tight} UR ${\mathfrak{c}_{\textsc{q}}\leq\mathfrak{u}(\vec{p},\vec{q}\,)}$.
For $d=2$, our lower bound
${\mathfrak{c}=\mathfrak{c}_{\textsc{q}}=\mathfrak{c}_{\text{bases}}}$, and the UR~\eqref{d2-UR} is tight.
Whereas, if the two bases in~\eqref{AB-bases} share a ket then $\mathfrak{c}$
turns out to be the trivial bound: ${2=\mathfrak{c}=\mathfrak{c}_{\textsc{q}}=
	\mathfrak{c}_{\text{bases}}}$.
One can use~\eqref{c<cQ<cmax} to avoid errors while calculating $\mathfrak{c}$.

\textbf{Remark~2:} The function ${H_{\sfrac{1}{2}}(\vec{p}\,)=2\log u(\vec{p}\,)}$ is the R\'{e}nyi entropy \cite{Renyi61} of order $\tfrac{1}{2}$. Using \eqref{h_kl}, one can realize that ${H_{\sfrac{1}{2}}(\vec{p}\,)}$ is a concave function on $\Omega_a$, hence
the sum
\begin{equation}
\label{hmax-pq}
H_{\sfrac{1}{2}}(\vec{p}\,)+H_{\sfrac{1}{2}}(\vec{q}\,)=
2\log\big[ u(\vec{p}\,)u(\vec{q}\,)\big]
\end{equation}
is concave on $\boldsymbol{\omega}$.
Taking \eqref{d-u/d-beta}--\eqref{d-ps/d-beta}, one can confirm that the sum is also concave on each of the parametric curves, therefore its absolute minimum will be on the endpoints.
By repeating the three-step procedure~\eqref{3-steps}---where in
the third step now we need to compute
\begin{equation}
\label{hM}
h_{\textsc{m}}:=
2\log\big[\bm({\textstyle\sum\nolimits_{l=1}^{\textsc{m}}\cos(\theta_{l}-\chi)}\bm)\bm(\cos\chi+\sin\chi\bm)\big]
\end{equation}
instead of $c_{\textsc{m}}$---for every $\textsc{m}$-set,
we can own an UR based on the combined entropy~\eqref{hmax-pq} for any pair of measurement settings.
Analogues to \eqref{c2_for_textsc{m}=2}, \eqref{cd_for_textsc{m}=d}, and \eqref{u_beta_end-MUB}, here we have
\begin{eqnarray}
\label{h2_for_textsc{m}=2}
h_2(\theta_1,\theta_2)&=&
2\log\big[2\,\cos\big(\tfrac{\theta_1-\theta_2}{2}\big)\sin\big(\tfrac{\theta_1+\theta_2}{2}\big)\big]\\
&=&
2\log\big[\sqrt{1-r_1}+\sqrt{1-r_2}\,\big]\,,\nonumber
\\
\label{hd_for_textsc{m}=d}
h_d(\theta_1,\cdots,\theta_d)&=&
2\log{\textstyle\sum\nolimits_{l=1}^{d}\cos\theta_{l}}=
2\log{\textstyle\sum\nolimits_{l=1}^{d}\sqrt{r_{l}}}\,,
\qquad\\
\mbox{and}\quad
h_\textsc{m}&=&2\log\left[
\tfrac{1+\sqrt{(d-1)(\textsc{m}-1)}+\sqrt{d-1}\;-\sqrt{\textsc{m}-1}}{\sqrt{d}}\right],\qquad
\end{eqnarray}
respectively, with these one can directly get URs for qubit, qutrit, and for a pair of MUBs just like above. 
For a qubit, we express the corresponding tight UR (also obtained in \cite{Rastegin12})
\begin{equation}
\label{d2-ProUR}
\sqrt{r}+\sqrt{1-r}\,\leq\,\big(\sqrt{p}+\sqrt{1-p}\,\big)\big(\sqrt{q}+\sqrt{1-q}\,\big)
\end{equation}
in terms of the product ${u(p)u(q)}$.
In this case, the product turns out not only a concave function on $\boldsymbol{\omega}$ but also on each of the four parametric curves.
And, its absolute minimum---given in left-hand side of \eqref{d2-ProUR}---occurs 
at all the four endpoints ${E_1,\cdots,E_4}$, and the absolute maximum $2$ at the center [denoted by ${\star}$ in Fig.~\ref{fig:u,pi/6}] of $\boldsymbol{\omega}$.

\section{Other (un)certainty measures and relations}\label{sec:other-UM}

The negative of a concave function is a convex function,
hence a suitable convex function can be taken as a measure of certainty, rather than uncertainty.
Here we present other popular measures of (un)certainty and obtain the associated (un)certainty relations for ${d=2}$ by finding the absolute minimum (for concave) and maximum (for convex) on the ellipse~\eqref{ellipse}.
We want to emphasize that all the relations given in this paper for a qubit are already known, thanks to \cite{Larsen90,Busch14,Garrett90,Sanchez-Ruiz98,Ghirardi03,Bosyk12,Vicente05,Zozor13,Deutsch83,Maassen88,Rastegin12},
through different methods.
The following analysis merely shows that they all can be obtained from
the TIs~\eqref{d2-cons} that characterize the ellipse.
Recall that one can have the same ellipse from \cite{Lenard72,Larsen90,Kaniewski14}.

One can always construct Hermitian operators, for example
\begin{equation}
\label{AB}
A=\textstyle\sum\nolimits_{i=1}^{d}a_i|a_i\rangle\langle a_i|
\quad\mbox{and}\quad
B=\textstyle\sum\nolimits_{j=1}^{d}b_j|b_j\rangle\langle b_j|\,,
\end{equation}
by assigning real numbers to the measurement outcomes $a_i$ and $b_j$
for the two settings specified by~\eqref{AB-bases}.
Then ${\textbf{a}:=\{a_i\}_{i=1}^{d}}$ and ${\textbf{b}:=\{b_j\}_{j=1}^{d}}$ are the sets of eigenvalues of $A$ and $B$, respectively.
With \eqref{pq} and \eqref{AB}, one can perceive that the squared standard deviations
\begin{eqnarray}
	\label{Std-A}
{\Delta(A,\rho)}^2&=&\langle\psi| A^2|\psi\rangle-{\langle\psi| A\,|\psi\rangle}^2
	\qquad\nonumber\\
	&=& \textstyle\sum\nolimits_{i=1}^{d}{a_i}^2\,p_i-
	\big(\textstyle\sum\nolimits_{i=1}^{d}a_i\,p_i\big)^2= {\Delta(\textbf{a},\vec{p}\,)}^2\,,\qquad\\
	\label{Std-B}
	{\Delta(B,\rho)}^2&=& \textstyle\sum\nolimits_{j=1}^{d}{b_j}^2\,q_j-
	\big(\textstyle\sum\nolimits_{j=1}^{d}b_j\,q_j\big)^2= {\Delta(\textbf{b},\vec{q}\,)}^2
\end{eqnarray}
are functions of the probabilities as well as the eigenvalues.

Taking ${p_d=1-\textstyle\sum\nolimits_{i=1}^{d-1}p_i}$, like the derivatives~\eqref{h_kl}
of ${u(\vec{p}\,)}$, we get
the second-order partial derivatives 
\begin{equation}
	\label{Std-Hessian}
	\frac{\partial^2\,\Delta^2}{\partial p_k\partial p_l}
	=-2(a_k-a_d)(a_l-a_d)=\frac{\partial^2\,\Delta^2}{\partial p_l\partial p_k}
\end{equation}
of the function~\eqref{Std-A} for ${1\leq k,l\leq d-1}$.
One can validate that the Hessian matrix---made of the derivatives \eqref{Std-Hessian}---is a negative semidefinite matrix for any set \textbf{a} of eigenvalues. Thus, ${{\Delta(\textbf{a},\vec{p}\,)}^2}$ is a concave function on $\Omega_a$ (see Theorem~${4.5}$ in \cite{Rockafellar70}).  
Likewise, ${{\Delta(\textbf{b},\rho)}^2}$ is a concave function on $\Omega_b$.
Hence, analogues to ${\mathfrak{u}(\vec{p},\vec{q}\,)}$ of \eqref{u-pq}, the sum 
\begin{equation}
	\label{Std,sum} \boldsymbol{\Delta}^{\text{sq}}(\textbf{a},\vec{p},\textbf{b},\vec{q}\,)
	:={\Delta(\textbf{a},\vec{p}\,)}^{2}+{\Delta(\textbf{b},\vec{q}\,)}^2
\end{equation} 
establishes a concave, thus uncertainty, measure on the combined space $\boldsymbol{\omega}$.
In \cite{Maccone14}, URs are presented by taking a sum such as \eqref{Std,sum}, however, here the approach is different.

In the case of a qubit (${d=2}$), every measurement setting can also be described by a three-component real vector. 
So, we designate the two settings [see~\eqref{AB-bases}] by certain unit vectors $\widehat{a}$ and $\widehat{b}$ and then construct the Hermitian operators ${A=\widehat{a}\cdot\vec{\sigma}}$ and ${B=\widehat{b}\cdot\vec{\sigma}}$ with the dot product, where $\vec{\sigma}$ is the Pauli vector operator.
One can verify that ${A^2=I=B^2}$, therefore the eigenvalues are:
${\textbf{a}=\{\pm1\}=\textbf{b}}$.
Suppose the kets ${|a_1\rangle}$ and ${|b_1\rangle}$ of the two bases [in~\eqref{AB-bases}] are associated with the eigenvalue ${+1}$ of $A$ and $B$, respectively.
Now one can easily derive the relation
\begin{equation}
\label{angle-bt-axis}
	\text{tr}(A^\dagger B)=
	4\;{|\langle a_1|b_1\rangle|}^2-2=
	2\;\widehat{a}\cdot\widehat{b}
\end{equation}
between the three kinds of inner products.
From Sec.~\ref{sec:UM-UR}, let us recall that we only require
three probabilities $p_1$, $q_1$, and $r_{11}$ to
express a (un)certainty relation for ${d=2}$.
So, there is no further need for the subscripts.
With all the above considerations, $\boldsymbol{\Delta}^{\text{sq}}$ of \eqref{Std,sum} turns out to be the function
\begin{equation}
\label{Std,d=2}
\boldsymbol{\Delta}^{\text{sq}}\bm(\pm1,p,\pm1,q\bm)
= 1-(2p-1)^2 + 1-(2q-1)^2
\end{equation}
of $p$ and $q$.

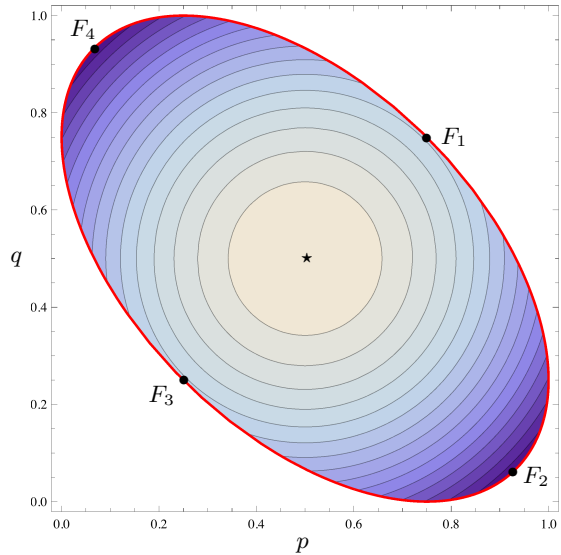
\begin{figure}[h]
	\centering
	\begin{overpic}[width=0.4\textwidth 
		]{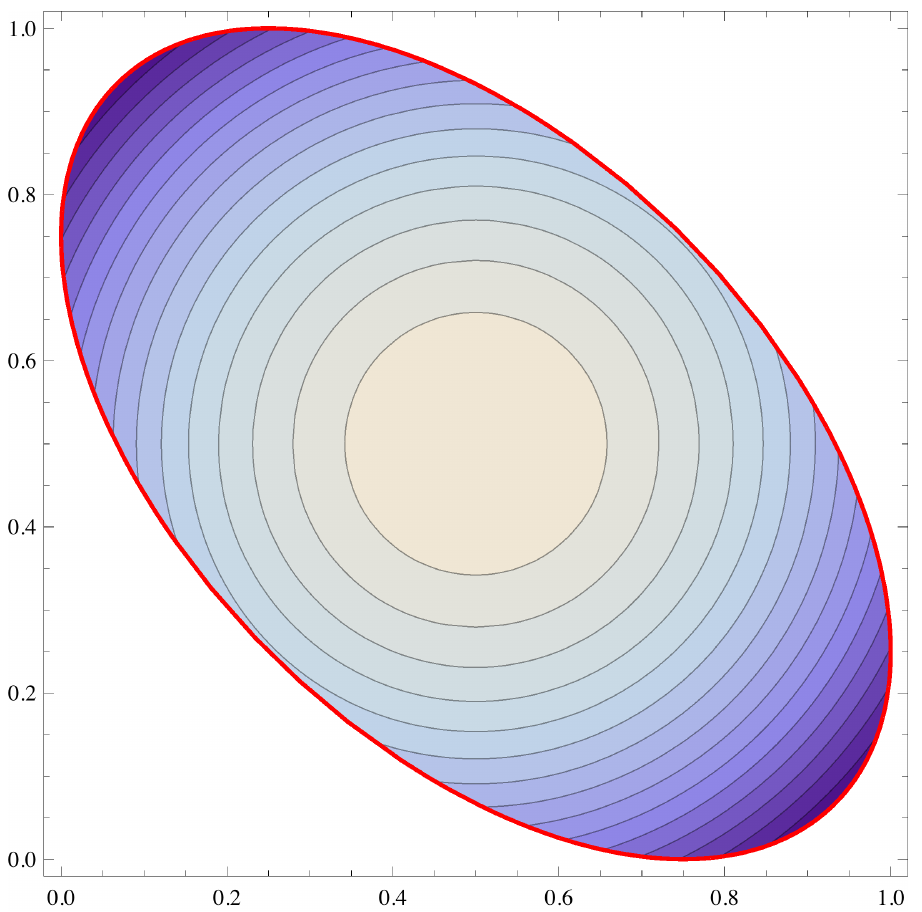}
		\put (50,-3) {$p$}
		\put (-3,50) {$q$}
		\put (73,72) {$\bullet\ F_1$}
		\put (92,9) {$F_2$}	\put (89,10) {$\bullet$}
		\put (22.6,24) {$F_3$}	\put (28,27) {$\bullet$}
		\put (8,92) {$F_4$}	\put (11.5,88.5) {$\bullet$}
		\put (50.8,49.7) {$\star$}
	\end{overpic} 
	\caption{A contour plot of ${\boldsymbol{\Delta}^{\text{sq}}(p,q)}$ of \eqref{Std,d=2} on $\boldsymbol{\omega}$, where a darker shade illustrates a smaller value of $\boldsymbol{\Delta}^{\text{sq}}$. Here ${r=\tfrac{1}{4}}$, therefore ${\boldsymbol{\Delta}^{\text{sq}}}$ reaches its global minimum ${2r}$ [see the UR~\eqref{Std,d=2,UR} and \eqref{min-Std}] at the two points $F_2$ and $F_4$. 
	Whereas, ${\boldsymbol{\Delta}^{\text{sq}}}$ gains its global maximum 2 always at the center, ${p=\tfrac{1}{2}=q}$ denoted by the star ${(\star)}$, of $\boldsymbol{\omega}$. Like Fig.~\ref{fig:u,pi/6}, $\boldsymbol{\omega}$ is the region bounded by the ellipse~\eqref{ellipse}; while ${\theta=\tfrac{\pi}{3}}$ here.}
	\label{fig:std,pi/3}  
\end{figure}

We plot ${\boldsymbol{\Delta}^{\text{sq}}}$ of \eqref{Std,d=2} on ${\boldsymbol{\omega}}$ in Fig.~\ref{fig:std,pi/3} by taking ${r=\tfrac{1}{4}}$.
Since ${\boldsymbol{\Delta}^{\text{sq}}}$ is a concave function on ${\boldsymbol{\omega}}$, its absolute minimum will be at the four parametric curves, which are jointly described by the ellipse~\eqref{ellipse} and by their endpoints ${E_1,\cdots,E_4}$.
To compute the minimum, first, we need to represent ${\boldsymbol{\Delta}^{\text{sq}}}$ as a function of a parameter, like $\mathfrak{u}$ in \eqref{u-beta}, on each curve.
Then, we have to find the critical points of ${\boldsymbol{\Delta}^{\text{sq}}}$.
Here we obtain four critical points ${F_1,\cdots,F_4}$---one on each curve---that are depicted by the bullets ${(\bullet)}$ in Fig.~\ref{fig:std,pi/3}.
By putting 
$\vartheta=\tfrac{\theta}{2},
\tfrac{\theta}{2}+\tfrac{\pi}{4},
\tfrac{\theta}{2}+\tfrac{2\pi}{4},
\tfrac{\theta}{2}+\tfrac{3\pi}{4}$
in ${(p{\scriptstyle(\vartheta)},q{\scriptstyle(\vartheta)})}$
of~\eqref{ellipse}, one can have ${F_1,\cdots,F_4}$, in that order.
Record that the $F$-points are not the endpoints ${E_1,\cdots,E_4}$ that are only shown in Fig.~\ref{fig:u,pi/6}, not in Fig.~\ref{fig:std,pi/3}.

The function ${\boldsymbol{\Delta}^{\text{sq}}}$ of \eqref{Std,d=2} takes the value $2r$ at both the points 
${\{F_2,F_4\}}$ and takes the value ${2(1-r)}$ at ${\{F_1,F_3\}}$.
So the global minimum is 
\begin{equation}
\label{Std,d=2,UR}
\min\big\{2r\,,\,2(1-r)\big\}\,\leq\,\boldsymbol{\Delta}^{\text{sq}}\bm(\pm1,p,\pm1,q\bm)\,,
\end{equation} 
and thus we obtain a tight UR, like~\eqref{d2-UR}. 
One can confirm that the lower bound is
\begin{equation}
\label{min-Std}
\begin{cases}
2r
&\mbox{if}\ \ r\leq\tfrac{1}{2}\quad (\mbox{at } F_2,F_4 
\mbox{ in Fig.~\ref{fig:std,pi/3}})\\
2(1-r)
&\mbox{if}\ \ r\geq\tfrac{1}{2}\quad (\mbox{at } F_1,F_3 
\mbox{ in Fig.~\ref{fig:std,pi/3}})\,.
\end{cases}
\end{equation}

\textbf{Remark~3:} The standard deviation $\Delta\bm(\pm1,p\bm)$ is a concave function of $p$, hence the sum $\Delta\bm(\pm1,p\bm)+\Delta\bm(\pm1,q\bm)$
is a concave function on $\boldsymbol{\omega}$.
As a result, we have another tight uncertainty relation
\begin{equation}
\label{Std,d=2,UR-1}
\sqrt{1-(2r-1)^2}\,\leq\,\Delta\bm(\pm1,p\bm)+\Delta\bm(\pm1,q\bm)\,.
\end{equation}
One can check that the sum reaches its absolute minimum value at all the endpoints ${E_1,\cdots,E_4}$, and has its maximum value $2$ at the center of $\boldsymbol{\omega}$. 
Both the tight URs \eqref{Std,d=2,UR} and \eqref{Std,d=2,UR-1} are known due to \cite{Busch14}.
A quantum state that saturates a tight UR is called \emph{its minimum uncertainty state}.
Since the $E$-points and the $F$-points are not the same, in general,
the set---of minimum uncertainty states---is different for the two URs \eqref{Std,d=2,UR} and \eqref{Std,d=2,UR-1} based on the standard deviation.
Note that we always get the trivial lower bound ${0\leq\Delta(\textbf{a},\vec{p}\,)\Delta(\textbf{b},\vec{q}\,)}$ 
for the product of standard deviations, and this bound can be reached by any ket belongs to either of the bases given in \eqref{AB-bases}.

Next, the Shannon entropy \cite{Shannon48}
\begin{equation}
	\label{Ent-A}
	H(\vec{p}\,)=-\textstyle\sum\nolimits_{i=1}^{d}p_i\log p_i
\end{equation}
is arguably the most famous measure of uncertainty at present. It is superior than the standard deviation ${\Delta(\textbf{a},\vec{p}\,)}$ \cite{Bialynicki11,Coles17} because it only depends on $\vec{p}$, not on
the eigenvalues.
One can show that ${H(\vec{p}\,)\in[0,\log d]}$, and it is a concave function on $\Omega_a$ with the Hassian matrix composed of the second-order derivatives 
\begin{equation}
	\label{Ent-Hessian}
	\frac{\partial^2 H}{\partial p_k\partial p_l}
	=-\left(\frac{1}{p_l}\delta_{lk}+
	\frac{1}{p_d}\right)=\frac{\partial^2 H}{\partial p_l\partial p_k}\,,
\end{equation}
where ${p_d=1-\textstyle\sum\nolimits_{i=1}^{d-1}p_i}$.
Considering the same function for the $b$-setting, that is $H(\vec{q}\,)$, one can formulate a combined uncertainty measure
by the sum ${H(\vec{p}\,)+H(\vec{q}\,)}$ and then produce an entropy
UR \cite{Deutsch83,Kraus87,Maassen88}.
Such URs are reviewed in \cite{Wehner10,Bialynicki11,Coles17}.
For ${d=2}$, the tight entropy UR is achieved in \cite{Garrett90,Ghirardi03} (see also \cite{Sanchez-Ruiz98}), 
and we can directly import all their results here.
In fact, Eq.~(7) in \cite{Garrett90} and Eq.~(2.4) in \cite{Ghirardi03} are ${H(p)+H(q)}$ on the ellipse~\eqref{ellipse}, and they found the absolute minimum of ${H(p)+H(q)}$ on the ellipse.
In \cite{Ghirardi03}, all the results are given in terms of angles between the real unit vectors, which are related to the angles between kets through \eqref{angle-bt-axis}.

We can choose 
\begin{equation}
	\label{u-gamma-p}
	u_\gamma(\vec{p}\,)=\textstyle\sum\nolimits_{i=1}^{d}(p_i)^\gamma
	\quad\mbox{with}\quad 0<\gamma<\infty
\end{equation} 
as another (un)certainty measure, which is closely related to the Tsallis \cite{Tsallis88} and R\'{e}nyi \cite{Renyi61} entropies of order $\gamma$.
One can prove that the Hassian matrix with entries
\begin{equation}
\label{u-t-Hessian}
\frac{\partial^2 u_\gamma}{\partial p_k\partial p_l}
=\gamma(\gamma-1)\left[{p_l}^{\gamma-2}\,\delta_{lk}+{p_d}^{\gamma-2}\right]=
\frac{\partial^2 u_\gamma}{\partial p_l\partial p_k}\,,
\end{equation}
${1\leq k,l\leq d-1}$, is a negative and positive
semidefinite matrix for ${0<\gamma\leq1}$ and ${1\leq\gamma<\infty}$, respectively.
It confirms that ${u_\gamma(\vec{p}\,)}$ is a concave (uncertainty) and convex (certainty) measure when 
${0<\gamma\leq1}$ and ${1\leq\gamma<\infty}$, respectively.
A similar observation is made in \cite{Luis11,Rastegin12}.
In fact, our uncertainty measure ${u(\vec{p}\,)}$ of~\eqref{ua} is ${u_\gamma(\vec{p}\,)}$ with the exponent ${\gamma=\tfrac{1}{2}}$.
Furthermore, the range of ${u_\gamma(\vec{p}\,)}$
is ${[1,d^{1-\gamma}]}$ if ${\gamma\leq1}$ and is ${[d^{1-\gamma},1]}$
if ${1\leq\gamma}$.
When ${\gamma=1}$, ${u_\gamma(\vec{p}\,)=1}$ for every $\vec{p}\in\Omega_a$ due to Eq.~\eqref{p-const1}, thus ${u_1}$ is not a genuine (un)certainty measure.

Like before, one can establish a (un)certainty relation with the sum $u_\gamma(\vec{p}\,)+u_\gamma(\vec{q}\,)$.
For $\gamma=2$, in the case of $d=2$, we obtain
\begin{eqnarray}
	\label{u2,d=2} 
	u_2(p)+u_2(q)&=&
	2-\tfrac{1}{2}\boldsymbol{\Delta}^{\text{sq}}\bm(\pm1,p,\pm1,q\bm)\,,
	\ \mbox{and then}\qquad\ \ \\ 
	\label{u2,d=2,UR}             
	u_2(p)+u_2(q)&\leq& \underbrace{2-\min\bm\{\,r\,,\,1-r\,\bm\}}
	_{\max\{2-r,1+r\}}          
\end{eqnarray}
as a tight certainty relation; which is also given in \cite{Larsen90}
for ${\tfrac{1}{2}\leq r}$.
Due to~\eqref{u2,d=2}, one can immediately derive \eqref{u2,d=2,UR}
from the UR~\eqref{Std,d=2,UR}.
Where $\boldsymbol{\Delta}^{\text{sq}}$ of \eqref{Std,d=2} reaches its absolute minimum (uncertainty) on $\boldsymbol{\omega}$, there the function \eqref{u2,d=2} achieves its global maximum (certainty)
\begin{equation}
	\label{max-u2}
	\max\{2-r,1+r\}=
	\begin{cases}
		2-r\ \mbox{if}\ r\leq\tfrac{1}{2}\ \ (\mbox{at } F_2,F_4 \mbox{ in Fig.~\ref{fig:std,pi/3}})\\
		1+r\ \mbox{if}\ r\geq\tfrac{1}{2}\ \ (\mbox{at } F_1,F_3 \mbox{ in Fig.~\ref{fig:std,pi/3}})\,.
	\end{cases}
\end{equation}
The certainty measure~\eqref{u2,d=2} hits its absolute minimum 1 at the center of $\boldsymbol{\omega}$ [depicted by the star ${(\star)}$ in Figs.~\ref{fig:u,pi/6} and~\ref{fig:std,pi/3}].

\textbf{Remark~4:} One can have another tight certainty relation
	\begin{equation}
	\label{u2-pro,d=2,Ur} 
	u_2(p)\,u_2(q)\leq\tfrac{1}{4}\max\big\{(2-r)^2,(1+r)^2\big\}\,,
	\end{equation}
	where product of certainty measures is used.
	The relation~\eqref{u2-pro,d=2,Ur} is presented in \cite{Larsen90}
	for ${\tfrac{1}{2}\leq r}$. 
	One can verify that ${u_2(p)\,u_2(q)}$ is a convex functions on $\boldsymbol{\omega}$.
	Therefore, its absolute maximum [given in \eqref{u2-pro,d=2,Ur}] will be on the ellipse [specified by \eqref{ellipse}], and the global minimum $\tfrac{1}{4}$ will be at the center of $\boldsymbol{\omega}$.
	The product-function reaches its upper bound on the $F$-points.
	By applying the negative of the logarithm on both sides of the inequality \eqref{u2-pro,d=2,Ur}, we get the corresponding tight UR---achieved in \cite{Bosyk12}---in terms of the collision entropy (that is, the R\'{e}nyi entropy \cite{Renyi61} of order $2$).

Lastly, we pick the function 
\begin{equation}
	\label{u-max-p}
	u_{\textrm{max}}(\vec{p}\,)=
	\operatorname*{max}_{1\leq i\leq d}\;\{p_i\}
\end{equation} 
that defines a norm on $\mathbb{R}^{d}$ if we replace $p_i$ with ${|p_i|}$. Since every $p_i$ follows \eqref{p-const2}, the modulus sign is not shown in \eqref{u-max-p}.
Every norm is a convex function, so ${u_{\textrm{max}}}$ can be considered as a certainty measure on $\Omega_a$; ${u_{\textrm{max}}(\vec{p}\,)\in\big[\tfrac{1}{d}\,,1\big]}$ for every ${\vec{p}\in\Omega_a}$. 
Note that ${u_{\textrm{max}}(\vec{p}\,)}$ is not differentiable everywhere in $\Omega_a$.
Nevertheless, we can assemble a combined certainty measure with the sum
${u_{\textrm{max}}(\vec{p}\,)+u_{\textrm{max}}(\vec{q}\,)}$ on $\boldsymbol{\omega}$.

In the case of ${d=2}$, the function ${u_{\textrm{max}}(p)+u_{\textrm{max}}(q)}$ is equal to
\begin{equation}
\label{umax(p,q)}
\begin{cases}
(1-p)+(1-q) & \text{if}\ \ 
0\leq p \leq \tfrac{1}{2} \ \text{and}\ 
0\leq q \leq \tfrac{1}{2}\\
(1-p)+q & \text{if}\ \
0\leq p \leq \tfrac{1}{2} \ \text{and}\ 
\tfrac{1}{2}\leq q \leq 1\\
\quad \quad\ \, p+(1-q) & \text{if}\ \ 
\tfrac{1}{2}\leq p \leq 1 \ \text{and}\ 
0\leq q \leq \tfrac{1}{2}\\
\quad \quad\ \, p+q & \text{if}\ \
\tfrac{1}{2}\leq p \leq 1 \ \text{and}\ 
\tfrac{1}{2}\leq q \leq 1\,.
\end{cases}
\end{equation}
The limits on ${p,q}$ stated in \eqref{umax(p,q)} divide $\boldsymbol{\omega}$---that is an elliptical region [see Figs.~\ref{fig:u,pi/6} and \ref{fig:std,pi/3}]---into four 
quadrants.
The function ${u_{\textrm{max}}(p)+u_{\textrm{max}}(q)}$ is differentiable in each of the quadrants.
Furthermore, since it is a convex function on $\boldsymbol{\omega}$, its global maximum will be at the ellipse~\eqref{ellipse}.
Here we discover four critical points, one in each quadrant on the ellipse, where the combined function takes a maximum value.
In fact, these four points are the same ${F_1,\cdots,F_4}$ exhibited in Fig.~\ref{fig:std,pi/3}.

The combined measure acquires the value ${1+\sqrt{1-r}}$ at both ${F_2,F_4}$
and reaches the value $1+\sqrt{r}$ at both ${F_1,F_3}$.
Thus, like \eqref{u2,d=2,UR}, we get the tight certainty relation 
\begin{equation}
\label{umax,d=2,UR}             
u_{\textrm{max}}(p)+u_{\textrm{max}}(q)\leq 
\max\big\{1+\sqrt{1-r}\,,\,1+\sqrt{r}\,\big\}\,,
\end{equation}
for a qubit.
And, the absolute maximum (upper bound) is given by
\begin{equation}
\begin{cases}
1+\sqrt{1-r}
&\mbox{if}\ \ r\leq\tfrac{1}{2}\quad (\mbox{at } F_2,F_4 
\mbox{ in Fig.~\ref{fig:std,pi/3}})\\
1+\sqrt{r}
& \mbox{if}\ \ r\geq\tfrac{1}{2}\quad (\mbox{at } F_1,F_3 
\mbox{ in Fig.~\ref{fig:std,pi/3}})
\end{cases}
\end{equation}
analogues to \eqref{max-u2}.
Besides, ${u_{\textrm{max}}(p)+u_{\textrm{max}}(q)}$ has its
global minimum 1 at the center of $\boldsymbol{\omega}$ [exhibited by the star ${(\star)}$ in Figs.~\ref{fig:u,pi/6} and \ref{fig:std,pi/3}].

The certainty relation \eqref{umax,d=2,UR} is captured in \cite{Vicente05} using the inequality
	\small
\begin{equation}
	\label{min-TI}
	\arccos(\max_{ij} \sqrt{{r_{ij}}})\leq
	\arccos(\max_{i} \sqrt{{p_{i}}})+\arccos(\max_{j} \sqrt{{q_{j}}})\,.
\end{equation}
\normalsize
Instead of TIs~\eqref{d2-cons}, for a qubit, all the tight relation
\eqref{d2-UR}, \eqref{d2-ProUR}, \eqref{Std,d=2,UR}, \eqref{Std,d=2,UR-1}, \eqref{u2,d=2,UR}, \eqref{u2-pro,d=2,Ur}, \eqref{umax,d=2,UR}, \eqref{ur-pro,d=2,UR}, \eqref{max-min entropy UR}, and the entropy UR given in \cite{Garrett90,Sanchez-Ruiz98,Ghirardi03} can be obtained with \eqref{min-TI}.
In fact, inequality \eqref{min-TI}, that is ${\min_{ij}\theta_{ij}\leq\min_{i}\alpha_{i}+\min_{j}\beta_{j}}$,
can be produced from $d^2$ TIs~\eqref{t-ineq}, and it is weaker than the TIs: 
all those ${(\vec{p},\vec{q}\,)\in\mathbf{\Omega}}$
that are bounded by \eqref{min-TI} rather than \eqref{t-ineq} constitute a bigger combined-probability space.

\textbf{Remark~5:} One can confirm that the product 
	${u_{\textrm{max}}(p)\,u_{\textrm{max}}(q)}$ is neither a concave nor a convex function on
	$\boldsymbol{\omega}$ (for a similar observation, see \cite{Maassen88}), so it not clear to us whether or not we can
	take it as a good combined-(un)certainty measure for every qubit's state.
	It also shows that product of two convex (concave) functions is not necessarily a convex (concave) function.
	By computing the gradient of ${u_{\textrm{max}}(p)\,u_{\textrm{max}}(q)}$ in each of the four quadrants, one can realize: the function reaches its global minimum $\tfrac{1}{4}$ at the center of $\boldsymbol{\omega}$ and reaches its global maximum (on the ellipse) at the $F$-points.
Hence, we have the tight relation 
\begin{equation}
\label{ur-pro,d=2,UR}
		u_{\textrm{max}}(p)\,u_{\textrm{max}}(q)
\leq\tfrac{1}{4}\max\big\{(1+\sqrt{1-r}\,)^2,(1+\sqrt{r}\,)^2\big\}\,,\qquad
\end{equation} 
which is reported in \cite{Maassen88} (and implicitly appear in \cite{Deutsch83}).
In fact, for ${d=2}$, the ket given by Eq.~(11) in \cite{Deutsch83} 
is the ket \eqref{psi-eq-a} with ${\beta=\tfrac{\theta}{2}}$ and ${\nu=0}$, and the ket corresponds to the point $F_1$.
By applying the negative of the logarithm on both sides of the inequality \eqref{ur-pro,d=2,UR}, one can turn this relation 
in the min-entropy terms \cite{Mandayam10}.
The min-entropy ${H_{\text{min}}(q):=-\log\bm( u_{\textrm{max}}(q)\bm)}$ is the smallest in the family of R\'{e}nyi entropies \cite{Renyi61}, and it is neither concave nor convex function on the interval ${[0,1]}$.
Like above, using the min-entropy, one can have another tight relation
\begin{equation}
\label{max-min entropy UR}
-\log\bm(\max\{\,r\,,\,1-r\,\}\bm)\leq H_{\sfrac{1}{2}}(p)+H_{\text{min}}(q)\,,
\end{equation} 	
that is also given in \cite{Maassen88},
recall that ${H_{\sfrac{1}{2}}(p)=2\log\bm( u(p)\bm)}$.
The function ${H_{\sfrac{1}{2}}(p)+H_{\text{min}}(q)}$ always takes its global minimum at the endpoints $E_2$ and $E_4$ and takes its absolute maximum ${2\log2}$ at the center [shown in Fig.~\ref{fig:u,pi/6}] of $\boldsymbol{\omega}$.
In \cite{Zozor13}, a general expression for the tight lower bound of a sum of R\'{e}nyi entropies is given, which is basically the minimization of the sum on the ellipse.

\section{Conclusion and outlook}\label{sec:Conc-Out}

Taking a pure quantum state for a qudit, we present TIs~\eqref{t-ineq} and then the combined-probability space $\boldsymbol{\omega}$ for a general pair of measurement settings.
The combined space is a compact and convex set in $\mathbb{R}^{2d}$, and  
all its extreme points are represented by the $m$-parametric curves, ${1\leq m\leq d-1}$.
These curves are determined by the two settings ($\varTheta$-matrix)
and are sufficient to generate the whole
$\boldsymbol{\omega}$ as well as to provide a (un)certainty relation.

One can pick some suitable concave and convex functions on $\boldsymbol{\omega}$
to quantify the uncertainty and certainty, respectively.
Subsequently, one can establish an uncertainty (a certainty) relation by finding the absolute minimum (maximum) of a function at the parametric curves.
Due to the parametric curves, formulation of a (un)certainty relation
become a single-parameter optimization problem.

Particularly for the uncertainty measures \eqref{u-pq} and \eqref{hmax-pq}, the absolute minima 
can always be easily computed by repeating the three-step procedure
given in Sec.~\ref{sec:UM-UR} with every \textsc{m}-set, ${2\leq\textsc{m}\leq d}$, built with entries in the $\varTheta$-matrix.
And, thus, one can enjoy the corresponding URs for any pair of measurement settings.
For the other functions, one needs to find all the critical points on the curves first and then the absolute extremum at those points.
That is, still, much easier than searching the extremum on the whole space.
In each case, the extremum---that is a lower (upper) bound on an uncertainty (certainty) measure---only depends on the measurement settings, not on a quantum state. 
Every (pure or mixed) state of a qudit provides a point in $\boldsymbol{\omega}$ by the Born rule and respects every (un)certainty relation presented in this write-up.

In the case of a qubit, ${d=2}$, we show that many known tight (un)certainty relations, owing to \cite{Larsen90,Busch14,Garrett90,Sanchez-Ruiz98,Ghirardi03,Bosyk12,Vicente05,Zozor13,Deutsch83,Maassen88,Rastegin12}, 
can be derived from the TIs~\eqref{d2-cons}.
These TIs define an ellipse that represents all the parametric curves,
and each point on the ellipse (and in $\boldsymbol{\omega}$) corresponds to a qubit's state, thus we have tight relations.
The same ellipse also emerges in \cite{Lenard72,Larsen90,Kaniewski14} as a special case.
For a pair of measurement setting on a qubit, it seems that the TIs~\eqref{t-ineq} and the results in \cite{Lenard72,Larsen90,Kaniewski14,Landau61} provide more fundamental QCs than the tight (un)certainty relations.

TIs~\eqref{t-ineq} do not provide all possible QCs when the dimension
${d>2}$, hence there are still some points in $\boldsymbol{\omega}$ that correspond to no quantum state, and our URs given in Sec.~\ref{sec:UM-UR} are not tight in general.
However, all our (un)certainty relations are built on the fact that
`every point outside of $\boldsymbol{\omega}$ is, surely, not associated with any quantum state'.
One can include other QCs, namely TIs~\eqref{t-ineq-3},
then the domain $\boldsymbol{\omega}$ of a (un)certainty function will be smaller.
Consequently, better bounds and finer (un)certainty relations
can be achieved.
To get a tight bound, in the case of general settings and ${d>2}$, is a challenging task.
Tight URs are only known in some special cases: position-momentum \cite{Weyl32}, MUBs \cite{Kraus87,Maassen88,Larsen90,Sanchez-Ruiz95,Ballester07,Mandayam10}, and a qubit \cite{Larsen90,Busch14,Garrett90,Sanchez-Ruiz98,Ghirardi03,Bosyk12,Vicente05,Zozor13,Deutsch83,Maassen88,Rastegin12}.

URs have numerous applications in different strands of physics.
Recently, these are employed for certain quantum information processing tasks such as the cryptography \cite{Mandayam10} and the entanglement detection \cite{Vicente05,Hofmann03,Guhne04,Giovannetti04,Guhne04b}.
As our (un)certainty relations arise solely from TIs,
one can directly appoint TIs~\eqref{t-ineq-3} as genuine QCs for such a job.
Furthermore, in quantum state estimation \cite{Paris04}, one collects data by applying different measurement settings, thus
realizes scheme~\eqref{expt-situ} in a laboratory.
Then, $\rho_{\text{est}}$ is constructed with the data. There one needs to confirm that the estimated $\rho_{\text{est}}$ represents a legitimate quantum state.
Again TIs~\eqref{t-ineq-3} could be utilized for such a test, for instance, one can firstly check whether the estimated ${(\vec{p}_{\text{est}},\vec{q}_{\text{est}})}$
follows all the TIs or not.

\begin{acknowledgments} 
	I am very grateful to Arvind for stimulating discussions and helpful comments on the manuscript.
	I thank Arun Kumar Pati for bringing Ref.~\cite{Landau61} to my attention and J\k{e}drzej Kaniewski for explaining and making me aware about their work \cite{Kaniewski14}.
\end{acknowledgments}

\appendix

\section{Derivation of the triangle inequalities}\label{sec:Der-q-const}

Landau and Pollak obtained a single TI of the kind given in \eqref{t-ineq} for continuous-time signals. 
One can spot several similarities between their work \cite{Landau61} and the following derivation.
In this paper, the primary QCs are the TIs~\eqref{t-ineq-3}.
To derive such TIs, we consider three kets ${|\psi\rangle}$, ${|a \rangle}$, and ${|b \rangle}$ of a $d$-dimensional Hilbert space $\mathscr{H}_d$.
Their inner products are expressed in the polar form as
\begin{eqnarray}
\label{a-psi-polar}
\langle a|\psi\rangle&:=&\sqrt{p}\,e^{\text{i}\mu}=\cos\alpha\, e^{\text{i}\mu},\\
\label{b-psi-polar}
\langle b|\psi\rangle&:=&\sqrt{q}\,e^{\text{i}\nu}=\cos\beta\, e^{\text{i}\nu},\quad\mbox{and}\\
\label{ab-polar}
\langle a|b\rangle&:=&\sqrt{r}\,e^{\text{i}\delta}=\cos\theta\, e^{\text{i}\delta},
\end{eqnarray}
where the phases ${\mu,\nu,\delta\in[0,2\pi)}$.
In the main text, ${|\psi\rangle}$ is associated with a quantum state, and ${|a\rangle}$ and ${|b\rangle}$ are with the two measurement settings [see \eqref{AB-bases}].
Through the inner products, the quantum angles $\alpha$, $\beta$, and $\theta$ are related with the probabilities $p$, $q$, and $r$ [see also \eqref{pq}, \eqref{alpha-beta}, \eqref{r}, and \eqref{theta}], and ${\text{i}=\sqrt{-1}}$.
Recall that the angles lie in ${[0,\tfrac{\pi}{2}]}$, and the probabilities belong to the interval ${[0,1]}$.

It is always feasible to write one ket, say ${|\psi\rangle}$, 
as a sum of its component in the linear span of other two ${\{|a\rangle,|b\rangle\}}$ and its component in the orthogonal complement of the span [see \eqref{psi-eq}].
In general, $|a\rangle$ and $|b\rangle$ are not orthogonal to each other.
In the case of ${0<|\langle a|b\rangle|<1}$, employing the Gram-Schmidt orthogonalization process, one can convert the linearly independent set ${\{|a\rangle,|b\rangle\}}$ into an orthonormal set
${\{|b\rangle,|b^\perp\rangle\}}$ or ${\{|a\rangle,|a^\perp\rangle\}}$,
where
\begin{equation}
	\label{a-perp}
	|b^\perp\rangle=\frac{|a\rangle-\langle b|a\rangle|b\rangle}
	{\sqrt{1-|\langle a|b\rangle|^2}}\ \,\mbox{and}\ \,
	|a^\perp\rangle=\frac{|b\rangle-\langle a|b\rangle|a\rangle}
	{\sqrt{1-|\langle a|b\rangle|^2}}\,.
\end{equation}
The two sets are related by a unitary transformation:
\begin{equation}
	\label{ab-unitary}
	\begin{pmatrix}
		|b\rangle \\
		|b^\perp\rangle 
	\end{pmatrix}
	=
	\begin{pmatrix}
		\langle a|b\rangle & \sqrt{1-|\langle a|b\rangle|^2} \\
		\sqrt{1-|\langle a|b\rangle|^2} & -\langle b|a\rangle
	\end{pmatrix}
	\begin{pmatrix}
		|a\rangle \\
		|a^\perp\rangle 
	\end{pmatrix}.
\end{equation}

Now we can resolve
\begin{equation}
\label{psi-eq}
|\psi\rangle=\cos\beta\,e^{\text{i}\nu}|b\rangle+\langle b^\perp|\psi\rangle|b^\perp\rangle+\langle x|\psi\rangle|x\rangle
\end{equation}
with a suitable ket $|x\rangle$ 
that follows ${\langle b|x\rangle=0=\langle b^\perp|x\rangle}$.
If and only if ${|\psi\rangle}$ lies in the span of ${\{|a \rangle,|b\rangle\}}$, the last term in the expansion~\eqref{psi-eq} vanishes, otherwise not.
With the normalization of ${|\psi\rangle}$, 
one can recognize ${|\langle b^\perp|\psi\rangle|^2+|\langle x|\psi\rangle|^2={\sin\beta}^2}$, and subsequently 
\begin{equation}
\label{sin-alpha-gr}
0\,\leq\,|\langle x|\psi\rangle| \quad\Rightarrow\quad
|\langle b^\perp|\psi\rangle|\,\leq\,\sin\beta\,.
\end{equation}

Taking the transformation~\eqref{ab-unitary} and the polar form \eqref{ab-polar}, we realize another representation of the ket 
\begin{eqnarray}
\label{psi-eq-2}
|\psi\rangle&=&
\big(\cos\theta\cos\beta\,e^{\text{i}(\nu+\delta)}+
\sin\theta\,\langle b^\perp|\psi\rangle\big)\,|a\rangle+\nonumber\\
&&\big(\sin\theta\cos\beta\,e^{\text{i}\nu}-
\cos\theta e^{-\text{i}\delta}\langle b^\perp|\psi\rangle\big)|a^\perp\rangle+\nonumber\\
&&\langle x|\psi\rangle|x\rangle
\end{eqnarray}
from~\eqref{psi-eq}.
With the new representation~\eqref{psi-eq-2} and the polar form
\begin{equation}
\label{b-psi-perp-polar}
\langle b^\perp|\psi\rangle:=|\langle b^\perp|\psi\rangle|\,e^{\text{i}\xi}\,,\quad\qquad\xi\in[0,2\pi)\,,
\end{equation}
we attain
\begin{eqnarray}
\label{a-psi-abs-sq}
p=|\langle a|\psi\rangle|^2&=&
{\cos\theta\,}^2{\cos\beta}^2+
{\sin\theta\,}^2\,|\langle b^\perp|\psi\rangle|^2+\nonumber\\
&&2\cos\theta\sin\theta\cos\beta\,|\langle b^\perp|\psi\rangle|
\cos(\xi-(\nu+\delta))\,.\nonumber\\
\end{eqnarray}
Remember that ${\langle a|x\rangle=0=\langle a^\perp|x\rangle}$ because ${|x\rangle}$ lies in the orthogonal complement of ${\{|a\rangle,|b\rangle\}}$. Owing to
\begin{eqnarray}
\label{cos}
\cos(\xi-(\nu+\delta))\,\leq\,1\,,
\end{eqnarray}
first, we obtain the left-hand side inequality in
\begin{equation}
\label{b-psi-abs-sq-ineq-1}
p\leq
\big(\cos\theta\cos\beta+\sin\theta|\langle b^\perp|\psi\rangle|\big)^2
\leq{\cos(\theta-\beta)}^2\,,
\end{equation}
and afterwards the right-hand side inequality with the aid of \eqref{sin-alpha-gr}.
Eventually, from above, we have
\begin{equation}
\label{QC_cos_beta_sq}
p={\cos\alpha\,}^2\,\leq\,
{\cos(\theta-\beta)\,}^2
\end{equation}  
[using the polar form~\eqref{a-psi-polar}].

If there are equalities in \eqref{cos} as well as in \eqref{sin-alpha-gr}, then we reach an equality---at the place of inequality---in \eqref{QC_cos_beta_sq}:
${\xi=\nu+\delta\;(\text{mod}\,{2\pi})}$ are the solutions of equation
${\cos(\xi-(\nu+\delta))=1}$. 
And, ${|\langle x|\psi\rangle|=0}$ implies that ${|\psi\rangle}$ is contained in the subspace generated by ${\{|a \rangle,|b\rangle\}}$, thus ${|\langle b^\perp|\psi\rangle|=\sin\beta}$.
These two conditions turn \eqref{psi-eq} and \eqref{psi-eq-2} into
\begin{eqnarray}
\label{psi-eq-a}
|\psi\rangle&=&e^{\text{i}\nu}\left[\,
\cos\beta\,|b\rangle+
\sin\beta\,e^{\text{i}\delta}\,|b^\perp\rangle\,\right]\\
\label{psi-eq-b}
&=&e^{\text{i}\nu}\left[\,\cos(\theta-\beta)\,e^{\text{i}\delta}|a\rangle+ \sin(\theta-\beta)\,|a^\perp\rangle\,\right].\quad\quad
\end{eqnarray}
These $|\psi\rangle$ kets---where $\delta$ is specified by the polar form~\eqref{ab-polar}, provided ${\langle a|b\rangle\neq0}$, and the global phase $\nu$ can be any real number---are the only kets that saturate the inequality~\eqref{QC_cos_beta_sq}.
We can not straightforward use the above analysis for the next two cases ${|\langle a|b\rangle|=0,1}$, hence these are studied individually.

In the case of ${\langle a|b\rangle=0}$, 
${|b^\perp\rangle=|a\rangle}$ and ${|a^\perp\rangle=|b\rangle}$; in fact, there is no need for the orthogonalization process, and
both the representations \eqref{psi-eq} and \eqref{psi-eq-2} of
${|\psi\rangle}$ become the same.
Furthermore, $\delta$ is not determined by the polar form~\eqref{ab-polar}, whereas ${\theta=\tfrac{\pi}{2}}$.
Now the inequality~\eqref{QC_cos_beta_sq} becomes
${{\cos\alpha\,}^2+{\cos\beta\,}^2\leq1}$, which is---directly realized from \eqref{psi-eq} due to \eqref{sin-alpha-gr}---saturated by the ket~\eqref{psi-eq-a} with an arbitrary real phase $\delta$
[remember ${\cos\alpha=|\langle a|\psi\rangle|}$ due to \eqref{a-psi-polar}].

In the case of ${|\langle a|b\rangle|=1}$, ${\theta=0}$ and ${|b\rangle=e^{\text{i}\delta}|a\rangle}$ according to \eqref{ab-polar}, and the above orthogonalization process, thus ${|b^\perp\rangle}$ and ${|a^\perp\rangle}$, does not exist. 
Consequently, 
the term ${\langle b^\perp|\psi\rangle|b^\perp\rangle}$ will not then appear in the decomposition~\eqref{psi-eq} of ${|\psi\rangle}$.
At the places of \eqref{sin-alpha-gr}, \eqref{QC_cos_beta_sq}, and \eqref{psi-eq-a} we have
${0\leq|\langle x|\psi\rangle|\Rightarrow{\cos\beta\,}^2\leq1}$, ${{\cos\alpha\,}^2={\cos\beta\,}^2}$, and ${|\psi\rangle=e^{\text{i}\nu}|b\rangle}$,
respectively. In this case, there is no genuine QC, nevertheless ${{\cos\beta\,}^2\leq1}$ is saturated by the ket(s) ${|\psi\rangle=e^{\text{i}\nu}|b\rangle}$ [remember ${\cos\beta=|\langle b|\psi\rangle|}$, see \eqref{b-psi-polar}].

One can appreciate that inequality~\eqref{QC_cos_beta_sq} is a
legitimate QC, and $\alpha$ and $\beta$ must respect that for every ${\theta\in[0,\tfrac{\pi}{2}]}$. 
Applying square root to both sides of the inequality, we gain 
\begin{equation}
\label{QC_cos_beta}
\cos\alpha=|\cos\alpha|\leq
|\cos(\theta-\beta)|=\cos(\theta-\beta)\,.
\end{equation}
Since ${\alpha\in[0,\tfrac{\pi}{2}]}$ and ${(\theta-\beta)\in[-\tfrac{\pi}{2},\tfrac{\pi}{2}]}$, both
${\cos\alpha}$ and ${\cos(\theta-\beta)}$ are nonnegative numbers,
hence there is no need to use the modulus on either side of the above inequality.
As the $\arccos$ function is a strictly decreasing function and
${\arccos(\cos\varsigma)=|\varsigma|}$ for ${\varsigma\in[-\tfrac{\pi}{2},\tfrac{\pi}{2}]}$, from \eqref{QC_cos_beta}, we own an equivalent form  
\begin{equation}
\label{QC_beta-abs}
|\theta-\beta|\,\leq\,\alpha\,
\end{equation}
of \eqref{QC_cos_beta_sq}.
In fact, \eqref{QC_beta-abs} carries two TIs: ${\theta\leq\alpha+\beta}$ and ${\beta\leq\alpha+\theta}$.
${|\psi\rangle}$ of \eqref{psi-eq-a} with ${0\leq\beta\leq\theta}$ saturates the TI ${\theta\leq\alpha+\beta}$ and with ${\theta\leq\beta\leq\tfrac{\pi}{2}}$ saturates the other TI ${\beta\leq\alpha+\theta}$.
TIs such as ${\theta\leq\alpha+\beta}$ [see \eqref{t-ineq}] are used to define the combined-probability space $\boldsymbol{\omega}$ in Sec.~\ref{sec:PS-C}.

Replacing the ordered set $\{b,\beta,\nu\}$ by $\{a,\alpha,\mu\}$ in \eqref{psi-eq} and repeating the above analysis, 
one will discover 
\begin{eqnarray}
\label{QC_cos_alpha_sq}
q={\cos\beta\,}^2\,&\leq&\,{\cos(\theta-\alpha)\,}^2
\quad\mbox{and}\\
\label{QC_alpha-abs}
|\theta-\alpha|\,&\leq&\,\beta
\end{eqnarray} 
at the places of \eqref{QC_cos_beta_sq} and \eqref{QC_beta-abs},
respectively.
Jointly \eqref{QC_beta-abs} and \eqref{QC_alpha-abs} can be written as 
\begin{equation}
\label{QC_beta-alpha}
|\theta-\beta|\,\leq\,\alpha\,\leq\,\theta+\beta\,,
\end{equation}
which displays three TIs associated with 
the three angles. 
A TI says: \emph{the sum of two quantum angles must be greater than or equal to the remaining quantum angle.}

In fact, the quantum angle ``${\arccos|\langle\ |\ \rangle|}$"
is a metric (and a distinguishability measure \cite{Wootters81}) on the set $\mathcal{S}_\text{pure}$ of all pure states (${\rho=\rho^2}$).
It is because the four conditions,
\begin{enumerate}
	\item ${\arccos|\langle a|b \rangle|\geq 0}$
	\item $\arccos|\langle a|b\rangle|=0$ if and only if
	${|a\rangle\langle a|=|b\rangle\langle b|}$
	\item ${\arccos|\langle a|b \rangle|=\arccos|\langle b|a \rangle|}$
	\item ${\arccos|\langle a|b \rangle|\leq
		\arccos|\langle a|\psi \rangle|+
		\arccos|\langle \psi|b \rangle|}$\,,
\end{enumerate}
are satisfied for every ${|a\rangle\langle a|}$,
${|b\rangle\langle b|}$, and ${|\psi\rangle\langle \psi|}$ in $\mathcal{S}_\text{pure}$, where
${|\langle a|b \rangle|=
\sqrt{\text{tr}\bm(|a\rangle\langle a|\,|b\rangle\langle b|\bm)}}$.
Note that every pure state on $\mathscr{H}_d$ is made of a ket in $\mathscr{H}_d$, and two kets that are equal up to a global phase provide the same pure state.
As the $\arccos$ function is nonnegative, the first condition is valid.
The second and third are true by the virtue of 
${|\langle a|b\rangle|=1\Leftrightarrow|a\rangle\langle a|=|b\rangle\langle b|}$ and ${|\langle a|b \rangle|=|\langle b|a \rangle|}$, respectively.
The last condition is, the TI ${\theta\leq\alpha+\beta}$, already derived above.

Returning to the TIs~\eqref{QC_beta-alpha}, as ${\alpha\in[0,\tfrac{\pi}{2}]}$, $\theta+\beta$
will be a true upper bound on $\alpha$ only if it is smaller than or equal to $\tfrac{\pi}{2}$. 
Hence, we can further improve \eqref{QC_beta-alpha} as   
\begin{equation}
\label{QC_beta-alpha-imp}
|\theta-\beta|\,\leq\,\alpha\,\leq\,\min\big\{\theta+\beta\,,\tfrac{\pi}{2}\big\}\,.
\end{equation}
Taking the right-hand side inequality and applying the cosine function---that decreases monotonically on ${[0,\pi]}$---to both the terms, we get 
\begin{equation}
\label{QC_cos_beta-alpha-imp}
\max\,\{\cos(\theta+\beta)\,,0\,\}\,\leq\,\cos\alpha\,.
\end{equation}
Now, considering the Heaviside's unit step function
\begin{equation}
\label{step-fn}
\eta(\upsilon):=
\begin{cases}
0 & \text{if  }\ \upsilon < 0 \\
1 & \text{if  }\ \upsilon\geq 0
\end{cases}\,,
\end{equation}
one can rewrite \eqref{QC_cos_beta-alpha-imp} as
\begin{equation}
\label{QC_cos_beta-alpha-imp-2}
\eta\bm(\cos(\theta+\beta)\bm)\;{\cos(\theta+\beta)}
\,\leq\,\cos\alpha\,.
\end{equation}
Since the terms on either side of the above inequality are nonnegative, squaring both sides delivers 
\begin{equation}
\label{QC_cos_beta-alpha-imp-sq}
\eta\bm(\cos(\theta+\beta)\bm)\;{\cos(\theta+\beta)}^2\,\leq\,{\cos\alpha\,}^2\,.
\end{equation}

Putting \eqref{QC_cos_beta_sq} and \eqref{QC_cos_beta-alpha-imp-sq}
side by side, we accomplish
\begin{equation}
\label{QC_cos_beta-alpha-sq-imp}
\eta\bm(\cos(\theta+\beta)\bm)\,{\cos(\theta+\beta)}^2
\leq{\cos\alpha\,}^2
\leq{\cos(\theta-\beta)}^2.
\end{equation}
Furthermore, due to \eqref{a-psi-polar}--\eqref{ab-polar}, \eqref{QC_cos_beta-alpha-sq-imp} becomes
\begin{eqnarray}
\label{QC-1}
&&\eta(\tau_{-})\;{\tau_{-}}^{2}\,\leq\, p\,\leq\, {\tau_{+}}^{2}\,,
\quad\quad\qquad\mbox{where}\\
\label{tau-}
&&\tau_{-}:=\sqrt{r\,q}-\textstyle\sqrt{(1-r)(1-q)}\qquad\mbox{and}\\
\label{tau+}
&&\tau_{+}:=\sqrt{r\,q}+\textstyle\sqrt{(1-r)(1-q)}\,.
\end{eqnarray}
In essence, we obtain QCs~\eqref{QC_beta-alpha-imp} and \eqref{QC-1}
that are equivalent to each other, one is in terms of the quantum angles and the other is in terms of the probabilities.


\section{Compactness and convexity of ${\boldsymbol{\omega}\subset\mathbf{\Omega}}$}\label{sec:Compact-Convex-w}

The real vector space $\mathbb{R}^{2d}$ is also a metric space with the Euclidean distance, and both its subsets $\mathbf{\Omega}$ and $\boldsymbol{\omega}$ are closed as well as bounded, hence they are compact sets (thanks to the Heine-Borel theorem, see in \cite{Rudin76}). 
Since a convex combination of probability vectors is again a probability vector,
both $\Omega_a$ and $\Omega_b$ are convex subsets of $\mathbb{R}^{d}$. Moreover, ${\mathbf{\Omega}=\Omega_a\times\Omega_b}$ is a convex set because it is a Cartesian product of two such sets.

To prove the convexity of $\boldsymbol{\omega}$, we consider
two combined vectors ${\big(\vec{p}\,',\vec{q}\,'\big)}$ and ${\big(\vec{p}\,'',\vec{q}\,''\big)}$ that
belong to $\boldsymbol{\omega}$.
It means that their components follow the constraints~\eqref{p-const1}--\eqref{q-const2} and \eqref{rpq-2} that is
\begin{eqnarray}
\label{p-1prime}
&&p'_i+q'_j\leq r_{ij}+1+2\sqrt{r_{ij}(1-p'_i)(1-q'_j)}\;,\\
\label{p-2prime}
&&p''_i+q''_j\leq r_{ij}+1+2\sqrt{r_{ij}(1-p''_i)(1-q''_j)}
\end{eqnarray}
for every ${1\leq i,j\leq d}$.
For the proof, we need to show that a convex combination 
\begin{equation}
\label{con-comb}
\big(\vec{p},\vec{q}\,\big)=
\lambda\,\big(\vec{p}\,',\vec{q}\,'\big)
+(1-\lambda) \big(\vec{p}\,'',\vec{q}\,''\big)
\end{equation}
fulfills all the requirements~\eqref{p-const1}--\eqref{q-const2} and \eqref{rpq-2}---therefore, lies in $\boldsymbol{\omega}$---for every $\lambda\in[0,1]$.
Thanks to the convexity of $\mathbf{\Omega}$, the combination~\eqref{con-comb} belongs to $\mathbf{\Omega}$ and
$\big(\vec{p},\vec{q}\,\big)$
meets all the demands~\eqref{p-const1}--\eqref{q-const2}.

Now we demonstrate that the components $p_i$ and $q_j$ of $\big(\vec{p},\vec{q}\,\big)$ respect inequality~\eqref{rpq-2}:
\begin{eqnarray}
\label{pq-1}
p_i+q_j&=&\lambda\,(p'_i+q'_j)+(1-\lambda)(p''_i+q''_j)\\
\label{pq-2}
&\leq&r_{ij}+1+2\sqrt{r_{ij}}\ \Big[\lambda\sqrt{(1-p'_i)(1-q'_j)}+\nonumber\\
&& (1-\lambda)\sqrt{(1-p''_i)(1-q''_j)}\ \Big]\\
\label{pq-3}
&\leq&r_{ij}+1+2\sqrt{r_{ij}}\;\sqrt{1-\lambda p'_i-(1-\lambda)p''_i}\nonumber\\
&& \sqrt{1-\lambda q'_j-(1-\lambda)q''_j}\\
\label{pq-4}
&=&r_{ij}+1+2\sqrt{r_{ij}}\sqrt{(1-p_i)(1-q_j)}\;.
\end{eqnarray}
We have equality~\eqref{pq-1} due to the convex combination~\eqref{con-comb}, and then we acquire inequality~\eqref{pq-2} by employing \eqref{p-1prime} and \eqref{p-2prime}.
The next inequality~\eqref{pq-3} is attributed to the concavity of a real-valued function
\begin{equation}
\label{f}
f(p,q):=\sqrt{(1-p)(1-q)}
\end{equation}
defined on ${[0,1]\times[0,1]}$, and the last equality is again because of the combination~\eqref{con-comb}.
In conclusion, the combined-probability space $\boldsymbol{\omega}$ is a convex set in $\mathbb{R}^{2d}$.
Beside, to recognize that $f(p,q)$ is a concave function, we present the Hessian matrix
\begin{equation}
\begin{pmatrix}
\frac{\partial^2 f}{\partial p^2} &\frac{\partial^2 f}{\partial p\partial q}\\
\frac{\partial^2 f}{\partial q\partial p} &\frac{\partial^2 f}{\partial q^2}
\end{pmatrix}=
\begin{pmatrix}
\frac{{-(1-q)}^{1/2}}{{4(1-p)}^{3/2}} &\frac{1}{{4(1-p)}^{1/2}{(1-q)}^{1/2}}\\
\frac{1}{{4(1-p)}^{1/2}{(1-q)}^{1/2}}&
\frac{{-(1-p)}^{1/2}}{{4(1-q)}^{3/2}}
\end{pmatrix}
\end{equation}
that is a negative semidefinite matrix for every $p$ and $q$ in the interval $[0,1)$.
For ${p=1}$ or ${q=1}$ or both, ${f(p,q)=0}$, and the Hessian matrix is the ${2\times2}$ zero matrix.


\section{Preliminary calculations for the next appendix}\label{sec:inter-results}

With \eqref{pq}, \eqref{alpha-beta}, \eqref{r}, and \eqref{theta}, let us again acknowledge that $\text{probability}=\text{cos\,(angle)}^2$, and the quantum angles belong to the interval ${[0,\tfrac{\pi}{2}]}$.
Now we consider ${j\neq l}$ and
\begin{eqnarray}
\label{q_j+q_l=}
q_j+q_l &=& {\cos\beta_j}^2+{\cos\beta_l}^2\nonumber\\
&=& 1+\cos(\beta_j+\beta_l)\cos(\beta_j-\beta_l)\,.     
\end{eqnarray}
Since the difference between angles ${\beta_j-\beta_l\in[-\tfrac{\pi}{2},\tfrac{\pi}{2}]}$,
we have ${0\leq\cos(\beta_j-\beta_l)}$.
Hence, with~\eqref{q_j+q_l=},
one can establish
\begin{equation}
\label{q_j+q_l<1-cos}
q_j+q_l\leq1\  \Leftrightarrow\ \cos(\beta_j+\beta_l)\leq0\,,
\end{equation}
and then 
\begin{equation}
\label{q_j+q_l<1-beta}
\qquad\qquad\quad q_j+q_l\leq1\ \Leftrightarrow\  \tfrac{\pi}{2}\leq\beta_j+\beta_l
\qquad (j\neq l)
\end{equation}
due to the $\arccos$ function; note that ${\arccos(\cos\varsigma)=\varsigma}$ for ${\varsigma\in[0,\pi]}$.
One can also perceive ${\tfrac{\pi}{2}\leq\beta_j+\beta_l}$ as a TI.

Next we are going to validate a result that is applied in Appendix~\ref{sec:Extreme-Points-w}. 
\begin{equation}
\label{1<Cos^2+Cos^2-q}
\parbox{0.75\columnwidth}{%
	If ${j\neq l}$, ${0\leq\theta_{ij}-\beta_j}$, and ${0\leq\theta_{kl}-\beta_l}$, then
	${1\leq{\cos(\theta_{ij}-\beta_j)}^2+
		{\cos(\theta_{kl}-\beta_l)}^2}$.}
\end{equation}
Let us designate ${\theta_{ij}-\beta_j}$ and ${\theta_{kl}-\beta_l}$ by $\varphi_{ij}$ and $\varphi_{kl}$, respectively, and write 
\begin{equation}
\label{cos-varphi}
{\cos\varphi_{ij}}^2+{\cos\varphi_{kl}}^2=
1+\cos(\varphi_{ij}+\varphi_{kl})\cos(\varphi_{ij}-\varphi_{kl})   
\end{equation}
just like~\eqref{q_j+q_l=}.
One can show that the sum
\begin{equation}
	\label{sum-varphi}
	\varphi_{ij}+\varphi_{kl}=(\theta_{ij}+\theta_{kl})-(\beta_j+\beta_l)\leq\tfrac{\pi}{2}  
\end{equation}
due to ${\theta_{ij}+\theta_{kl}\leq\pi}$ and \eqref{q_j+q_l<1-beta}.
Clearly ${\varphi_{ij},\varphi_{kl}\leq\tfrac{\pi}{2}}$ because 
${\theta,\beta\in[0,\tfrac{\pi}{2}]}$, and if ${0\leq\varphi_{ij},\varphi_{kl}}$ [see the requirements in \eqref{1<Cos^2+Cos^2-q}] then we have ${0\leq\varphi_{ij}+\varphi_{kl}}$ and ${\varphi_{ij}-\varphi_{kl}\in[-\tfrac{\pi}{2},\tfrac{\pi}{2}]}$.
As a net result, ${0\leq\cos(\varphi_{ij}\pm\varphi_{kl})}$, the last term in \eqref{cos-varphi}
turns out to be a nonnegative function, and thus we
achieve ${1\leq{\cos\varphi_{ij}}^2+{\cos\varphi_{kl}}^2}$. It completes a proof of~\eqref{1<Cos^2+Cos^2-q}.

\begin{equation}
\label{1=Cos^2+Cos^2-q}
\parbox{0.7\columnwidth}{%
	In addition to the requirements in \eqref{1<Cos^2+Cos^2-q}, if and only if ${\theta_{ij}=\tfrac{\pi}{2}=\theta_{kl}}$ and 
	${\beta_j+\beta_l=\tfrac{\pi}{2}}$, then we acquire the
	equality ${1={\cos(\theta_{ij}-\beta_j)}^2+{\cos(\theta_{kl}-\beta_l)}^2}$
	in~\eqref{1<Cos^2+Cos^2-q}.}
\end{equation}
If ${\theta_{ij}=\tfrac{\pi}{2}=\theta_{kl}}$ and 
${\beta_j+\beta_l=\tfrac{\pi}{2}}$ then evidently we have the equality of \eqref{1=Cos^2+Cos^2-q}. Now let us prove the converse
under the requirements ${0\leq\varphi_{ij},\varphi_{kl}}$ of \eqref{1<Cos^2+Cos^2-q}.
If ${{\cos\varphi_{ij}}^2+{\cos\varphi_{kl}}^2=1}$ then the last term in \eqref{cos-varphi} must vanish, which occurs---provided ${0\leq\varphi_{ij},\varphi_{kl}}$---when the sum in~\eqref{sum-varphi} attains its upper bound $\tfrac{\pi}{2}$ or ${\varphi_{ij}-\varphi_{kl}=\pm\tfrac{\pi}{2}}$.
The case ${\varphi_{ij}-\varphi_{kl}=\tfrac{\pi}{2}}$ arises when
${\varphi_{ij}=\tfrac{\pi}{2}}$ and ${\varphi_{kl}=0}$, and 
${\varphi_{ij}-\varphi_{kl}=-\tfrac{\pi}{2}}$ happens when ${\varphi_{ij}=0}$ and ${\varphi_{kl}=\tfrac{\pi}{2}}$.
Both these cases come under ${\varphi_{ij}+\varphi_{kl}=\tfrac{\pi}{2}}$---that is when the sum in~\eqref{sum-varphi} reaches its upper bound---which materialize
if and only if ${\theta_{ij}=\tfrac{\pi}{2}=\theta_{kl}}$ and 
${\beta_j+\beta_l=\tfrac{\pi}{2}}$; it validates \eqref{1=Cos^2+Cos^2-q}.

Similar to \eqref{q_j+q_l<1-beta} we have
\begin{equation}
\label{p_i+p_k}
\qquad\qquad p_i+p_k\leq1\  \Leftrightarrow\ \tfrac{\pi}{2}\leq\alpha_i+\alpha_k
\qquad (i\neq k)\,,
\end{equation}
and to \eqref{1<Cos^2+Cos^2-q} plus \eqref{1=Cos^2+Cos^2-q} we have
\begin{equation}
\label{1<Cos^2+Cos^2-p}
\parbox{0.75\columnwidth}{%
	if ${i\neq k}$, ${0\leq\theta_{ij}-\alpha_i}$, and ${0\leq\theta_{kl}-\alpha_k}$, then
	${1\leq{\cos(\theta_{ij}-\alpha_i)}^2+
	{\cos(\theta_{kl}-\alpha_k)}^2}$. In addition,
	if and only if ${\theta_{ij}=\tfrac{\pi}{2}=\theta_{kl}}$ and 
	${\alpha_i+\alpha_k=\tfrac{\pi}{2}}$, then we own the equality
	${1={\cos(\theta_{ij}-\alpha_i)}^2+
		{\cos(\theta_{kl}-\alpha_k)}^2}$.  
}
\end{equation}


\section{Extreme points of $\boldsymbol{\omega}$}\label{sec:Extreme-Points-w}

In Appendix~\ref{sec:Compact-Convex-w}, we demonstrate that the combined-probability space $\boldsymbol{\omega}$ is a compact convex set in $\mathbb{R}^{2d}$.
According to the Krein-Milman theorem (see Theorem~${3.3.5}$ and Appendix~A.3 in \cite{Niculescu93}), every point of such a set can be decomposed into a convex combination of its extreme points.
In this appendix, starting from an arbitrary interior point of $\boldsymbol{\omega}$, we move toward its extreme points.

\subsection{\label{sec:interior-w}Interior of $\boldsymbol{\omega}$}

A point ${\big(\dot{\vec{p}},\dot{\vec{q}}\,\big)\in\boldsymbol{\omega}}$ that obeys each of the constraints \eqref{p-const2}, \eqref{q-const2}, and \eqref{t-ineq} with \emph{strict} inequality, 
\begin{equation}
\label{int}
0<\dot{p}_i, \  0<\dot{q}_j, \ \theta_{ij}<\dot{\alpha}_i+\dot{\beta}_j 
\ \mbox{for all} \ 1\leq i,j\leq d\,, 
\end{equation} 
is called an \emph{interior} point of $\boldsymbol{\omega}$.
In certain cases, such as ${d=2}$ and ${\theta\in\{0,\tfrac{\pi}{2}\}}$, there exist---no interior point---only extreme points, then the following analysis is not needed. 
However, for ${d>2}$, there is always an interior point: 
with ${\theta_{ij}\leq\tfrac{\pi}{2}<2\arccos\tfrac{1}{\sqrt{d}}}$, one can show that the center---specified by ${p_i=\tfrac{1}{d}=q_j}$ for all ${i,j}$---of $\boldsymbol{\omega}$ is an interior point when ${d>2}$.

We begin our journey from a general but fixed interior point ${\big(\dot{\vec{p}},\dot{\vec{q}}\,\big)}$ along a straight line, which is the locus of points ${\vec{P}=\big(p_1,p_2,\dot{\vec{p}}_{\mathrm{rest}},\dot{\vec{q}}\,\big)\in\mathbb{R}^{2d}}$, where $p_1,p_2$ obey the linear equation
\begin{equation}
\label{p1p2-sum}
p_1+p_2=1-\textstyle\sum\nolimits_{i=3}^{d}\dot{p}_i
=\dot{p_1}+\dot{p_2}\,\leq\,1
\end{equation} 
and ${\dot{\vec{p}}_{\mathrm{rest}}=(\dot{p}_3,\cdots,\dot{p}_d)}$.
One can acknowledge that two points on this line differ from each other only in the first two coordinates, hence $p_1,p_2$ are the only variables here. 
In \eqref{p1p2-sum}, the inequality saturates for $d=2$ and
becomes strict due to \eqref{int} when ${d>2}$.

Since we never want to move outside of the combined space, we only consider those points on the line that lie in $\boldsymbol{\omega}$. 
From Sec.~\ref{sec:PS-C} recall that a point of $\mathbb{R}^{2d}$ lies in ${\mathbf{\Omega}}$ if and only if it meets all the requirements~\eqref{p-const1}--\eqref{q-const2}, and if it also satisfies all the TIs~\eqref{t-ineq} only then it belongs to ${\boldsymbol{\omega}}$.
So a point ${\vec{P}=\big(p_1,p_2,\dot{\vec{p}}_{\mathrm{rest}},\dot{\vec{q}}\,\big)}$ on the line, defined by \eqref{p1p2-sum}, is contained in ${\mathbf{\Omega}}$ if and only if
\begin{equation}
\label{p1p2-ineq}
0\leq p_1\quad\mbox{and}\quad 0\leq p_2\,.
\end{equation}

With \eqref{p1p2-sum} and \eqref{p1p2-ineq}, one can derive
\begin{equation}
\label{p1,p2-ineq}
0\,\leq\,p_1\,,\,p_2\,\leq\,\dot{p_1}+\dot{p_2}\,.
\end{equation}
As per \eqref{pq} and \eqref{alpha-beta}, we can attach angles $\alpha_1$  and $\alpha_2$ with $p_1$ and $p_2$, correspondingly. 
If these angles comply with
\begin{equation}
\label{alpha-cons}
\theta_{1j}-\dot{\beta}_j\leq \alpha_1,\
\theta_{2k}-\dot{\beta}_k\leq \alpha_2
\ \mbox{for all} \ 1\leq j, k\leq d\,,  \
\end{equation}
only then ${\vec{P}\in\boldsymbol{\omega}}$.
Observe that the other demands for $\vec{P}$ to be in $\boldsymbol{\omega}$---\eqref{int} for ${3\leq i\leq d}$ and \eqref{q-const1}---are automatically met, because
$\dot{\vec{p}}_{\mathrm{rest}}$ and $\dot{\vec{q}}$ are also parts of the interior point ${\big(\dot{\vec{p}},\dot{\vec{q}}\,\big)\in\boldsymbol{\omega}}$.

Considering the suprema 
\begin{eqnarray}
\label{alpha1-supremum}
\theta_{1J}-\dot{\beta}_J&=&\operatorname*{max}_{1\leq j\leq d}\;\big\{\theta_{1j}-\dot{\beta}_j\big\} \quad \mbox{and} \\
\label{alpha2-supremum}
\theta_{2K}-\dot{\beta}_K&=&\operatorname*{max}_{1\leq k\leq d}\;\big\{\theta_{2k}-\dot{\beta}_k\big\}\,, 
\end{eqnarray}  
we can convert all the conditions in \eqref{alpha-cons} into two
\begin{equation}
\label{alpha-cons-supremum}
\theta_{1J}-\dot{\beta}_J\leq \alpha_1\quad \mbox{and}\quad
\theta_{2K}-\dot{\beta}_K\leq \alpha_2\,.
\end{equation}
Throughout the paper, in the subscripts of angles, capital letters are used to highlight a supremum.
A supremum, say ${\theta_{1J}-\dot{\beta}_J}$, cannot be a negative number: ${\theta_{1J}-\dot{\beta}_J<0}$ implies ${\theta_{1j}<\dot{\beta}_j}$ for every $j$ by the definition~\eqref{alpha1-supremum}. Which leads to
${r_{1j}>\dot{q}_j}$ for each $j$ by the relations~\eqref{pq}, \eqref{alpha-beta}, \eqref{r}, and \eqref{theta}, and then to the contradiction
${1=\textstyle\sum\nolimits_{j=1}^{d}r_{1j}>
	\sum\nolimits_{j=1}^{d}\dot{q}_j=1}$.
Furthermore, ${\theta_{1J}-\dot{\beta}_J=0}$ if and only if ${\theta_{1j}=\dot{\beta}_j}$ for every~$j$.
So, both suprema~\eqref{alpha1-supremum} and \eqref{alpha2-supremum} lie in ${[0,\tfrac{\pi}{2}]}$.

Since the cosine function is monotonically decreasing and nonnegative on ${[0,\tfrac{\pi}{2}]}$, we can translate the constraints~\eqref{alpha-cons-supremum} as
\begin{equation}
\label{cos-alpha-cons-supremum} 
\cos\alpha_1\leq\cos(\theta_{1J}-\dot{\beta}_J),\quad
\cos\alpha_2\leq\cos(\theta_{2K}-\dot{\beta}_K)\
\end{equation}
and then as
\begin{eqnarray}
\label{cos-alpha-sq-cons-supremum-1} 
p_1&=&{\cos\alpha_1}^2\leq{\cos(\theta_{1J}-\dot{\beta}_J)}^2,\\
\label{cos-alpha-sq-cons-supremum-2} 
p_2&=&{\cos\alpha_2}^2\leq{\cos(\theta_{2K}-\dot{\beta}_K)}^2\,.
\end{eqnarray}
By the way, inequalities~\eqref{QC_beta-abs} and
\eqref{QC_cos_beta_sq} impose stronger restrictions than
\eqref{alpha-cons}, \eqref{cos-alpha-sq-cons-supremum-1}, and  \eqref{cos-alpha-sq-cons-supremum-2}. 
Since $p_2$ follows $p_1$ with Eq.~\eqref{p1p2-sum}, 
all the restrictions~\eqref{p1,p2-ineq}, \eqref{cos-alpha-sq-cons-supremum-1}, and  \eqref{cos-alpha-sq-cons-supremum-2} 
can be put together as 
\begin{eqnarray}
\label{p1-const} 
0&\leq&\max\left\{0\,,\,\dot{p}_1+\dot{p}_2-{\cos(\theta_{2K}-\dot{\beta}_K)}^2\,\right\}
\leq p_1\qquad\qquad\nonumber\\
&\leq&
\min\left\{{\cos(\theta_{1J}-\dot{\beta}_J)}^2,\,
\dot{p}_1+\dot{p}_2\right\}\leq1\,.
\end{eqnarray}
One can witness that these bounds on $p_1$ depend on the chosen interior point ${\big(\dot{\vec{p}},\dot{\vec{q}}\,\big)}$.
In short, only those $\vec{P}$ that fulfill the requirements \eqref{p1p2-sum} and \eqref{p1-const} belong to the combined space $\boldsymbol{\omega}$.

From the interior point ${\big(\dot{\vec{p}},\dot{\vec{q}}\,\big)}$, we can travel on the line in two directions: where $p_1$ increases and where $p_1$ decreases.
While moving we pass four points ${\vec{P}_1,\cdots,\vec{P}_4}$ of $\mathbb{R}^{2d}$ that are presented in Table~\ref{tab:endpoints}.
When we proceed in the direction where $p_1$ increases, then we reach first either $\vec{P}_1$ or $\vec{P}_2$. 
It all depends on the minimum value in~\eqref{p1-const}.  
The point that we reach first belongs to $\boldsymbol{\omega}$. Whereas the other point, then, fails to satisfy \eqref{p1-const}, and thus it lies outside of $\boldsymbol{\omega}$. 
While moving in the other direction, where $p_1$ decreases,
we encounter first either $\vec{P}_3$ or $\vec{P}_4$.
Depending on the maximum value in \eqref{p1-const} one of ${\{\vec{P}_3,\vec{P}_4\}}$ will be in, other will be out of, $\boldsymbol{\omega}$ (unless both these points are the same).

All the above possibilities are
communicated through Table~\ref{tab:endpoints-cond}.
For any ${\big(\dot{\vec{p}},\dot{\vec{q}}\,\big)}$, only two of these possibilities can and will materialize, thus $\boldsymbol{\omega}$ contains only a duo of (distinct) points from Table~\ref{tab:endpoints}. 
In Table~\ref{tab:endpoint-pairs}, we present every such duo.
In fact, the interior point ${\big(\dot{\vec{p}},\dot{\vec{q}}\,\big)}$ can be expressed as a convex combination  
\begin{equation}
\label{c-comb}
\lambda\,\underbrace{\big(p'_1,p'_2,\dot{\vec{p}}_{\mathrm{rest}},\dot{\vec{q}}\;\big)}_{\textstyle\vec{P}'}
+(1-\lambda) \underbrace{\big(p''_1,p''_2,\dot{\vec{p}}_{\mathrm{rest}},\dot{\vec{q}}\;\big)}_{\textstyle\vec{P}''}\,
\end{equation}
of points of the one duo ${\vec{P}',\vec{P}''}$ that lies in $\boldsymbol{\omega}$.
For each duo, ${\lambda\in(0,1)}$ is presented in Table~\ref{tab:endpoint-pairs}.

By varying $\lambda$ from 0 to 1 in the combination~\eqref{c-comb}, one can generate the line segment from $\vec{P}''$ to $\vec{P}'$.
Recall that the line is described by~\eqref{p1p2-sum}. 
If ${\vec{P}',\vec{P}''}$ belong to the combined space, then obviously the whole segment will be in $\boldsymbol{\omega}$
thanks to its convexity.
The line segments connecting $\vec{P}_1$ with $\vec{P}_2$ (provided ${\vec{P}_1\neq\vec{P}_2}$) and connecting ${\vec{P}_3}$ with ${\vec{P}_4}$ ${(\vec{P}_3\neq\vec{P}_4)}$ remain outside of $\boldsymbol{\omega}$.
Therefore, these two duos are not listed in Table~\ref{tab:endpoint-pairs}.

 \begin{table}[]
	\centering
	\caption{A list of four points ${\vec{P}=\big(p_1,p_2,\dot{\vec{p}}_{\mathrm{rest}},\dot{\vec{q}}\;\big)\in \mathbb{R}^{2d}}$ that lie on the line characterized by~\eqref{p1p2-sum}.
		From the interior point ${\big(\dot{\vec{p}},\dot{\vec{q}}\,\big)}$,
		$\vec{P}_1,\vec{P}_2$ are in the direction where $p_1$ increases, and $\vec{P}_3,\vec{P}_4$ are in the direction where $p_1$ decreases.
		So, the value of $p_1$ for a point here is one of the four bounds [stated in \eqref{p1-const}]. 
		Once we have $p_1$---in the center column---then $p_2$ is retrieved with \eqref{p1p2-sum} and placed in the right column.}
	\label{tab:endpoints}
	\begin{tabular}{c | c | c}
		\hline\hline\rule{0pt}{3ex}  
		$\vec{P}$ & $p_1$ & $p_2$ \\
		\hline\rule{0pt}{3ex} 	
		$\vec{P}_1$ & ${{\cos(\theta_{1J}-\dot{\beta}_J)}^2}$ & ${\dot{p}_1+\dot{p}_2-{\cos(\theta_{1J}-\dot{\beta}_J)}^2}$  \\
		$\vec{P}_2$ & ${\dot{p}_1+\dot{p}_2}$  & 0  \\
		$\vec{P}_3$ & 0 & ${\dot{p}_1+\dot{p}_2}$  \\			
		$\vec{P}_4$ & ${\dot{p}_1+\dot{p}_2-{\cos(\theta_{2K}-\dot{\beta}_K)}^2}$ & ${{\cos(\theta_{2K}-\dot{\beta}_K)}^2}$  \\[1mm]			
		\hline\hline
	\end{tabular}
\end{table}

\begin{table}[ ]
	\centering
	\caption{The 
		conditions that---rely on the minimum and the maximum values in~\eqref{p1-const}---determine whether a point from Table~\ref{tab:endpoints} will be in or out of $\boldsymbol{\omega}$.  
		If a condition from the left column holds, only then the related case in the right column occurs, and vice versa.
		One can realize that at most two conditions can hold at a time.}
	\label{tab:endpoints-cond}
	\begin{tabular}{c@{\hspace{2mm}} | @{\hspace{2mm}}c}
		\hline\hline
		If and only if & Then  \\
		\hline\rule{0pt}{3.3ex}
		${\cos(\theta_{1J}-\dot{\beta}_J)}^2<\,\dot{p_1}+\dot{p_2}$ & $\vec{P}_1\in\boldsymbol{\omega}$ and $\vec{P}_2\notin\boldsymbol{\omega}$  \\[1mm]
		${\cos(\theta_{1J}-\dot{\beta}_J)}^2>\,\dot{p_1}+\dot{p_2}$ & $\vec{P}_1\notin\boldsymbol{\omega}$ and $\vec{P}_2\in\boldsymbol{\omega}$  \\[1mm]
		${\cos(\theta_{1J}-\dot{\beta}_J)}^2=\,\dot{p_1}+\dot{p_2}$ & $\vec{P}_1=\vec{P}_2\in\boldsymbol{\omega}$  \\	[1mm]
		\hline\rule{0pt}{3ex} 
		${\cos(\theta_{2K}-\dot{\beta}_K)}^2<\,\dot{p_1}+\dot{p_2}$ & $\vec{P}_3\notin\boldsymbol{\omega}$ and $\vec{P}_4\in\boldsymbol{\omega}$  \\[1mm]
		${\cos(\theta_{2K}-\dot{\beta}_K)}^2>\,\dot{p_1}+\dot{p_2}$ & $\vec{P}_3\in\boldsymbol{\omega}$ and $\vec{P}_4\notin\boldsymbol{\omega}$  \\[1mm]
		${\cos(\theta_{2K}-\dot{\beta}_K)}^2=\,\dot{p_1}+\dot{p_2}$ & $\vec{P}_3=\vec{P}_4\in\boldsymbol{\omega}$  \\	[1mm]
		\hline\hline
	\end{tabular}
\end{table}

\begin{table}[]
	\centering
	\caption{
		Duos ${\vec{P}',\vec{P}''}$ of points from Table~\ref{tab:endpoints}. Only one out of these duos---unless two or more duos are the same---lies in $\boldsymbol{\omega}$ and expresses the interior point ${\big(\dot{\vec{p}},\dot{\vec{q}}\,\big)}$ through the convex combination~\eqref{c-comb} with a real number $\lambda$. 
		Corresponding to each duo, $\lambda$ is registered in the right column. One can confirm that ${0<\lambda<1}$ by realizing  
		${0<\dot{p}_1< {\cos(\theta_{1J}-\dot{\beta}_J)}^2}$ and 
		${0<\dot{p}_2< {\cos(\theta_{2K}-\dot{\beta}_K)}^2}$.}
	\label{tab:endpoint-pairs}
	\begin{tabular}{@{\hspace{2mm}}c@{\hspace{2mm}} | c}
		\hline\hline\rule{0pt}{3ex} 
		${\vec{P}',\vec{P}''}$ & $\lambda$ \\		
		\hline\rule{0pt}{4ex} 
		${\vec{P}_1,\vec{P}_3}$ &  $\cfrac{\dot{p_1}}{{\cos(\theta_{1J}-\dot{\beta}_J)}^2}$ \\[4mm]
		${\vec{P}_1,\vec{P}_4}$&  $\cfrac{{\cos(\theta_{2K}-\dot{\beta}_K)}^2-\dot{p_2}}
		{{{\cos(\theta_{1J}-\dot{\beta}_J)}^2+\cos(\theta_{2K}-\dot{\beta}_K)}^2-\dot{p_1}-\dot{p_2}}$\\[5mm]	 
		${\vec{P}_2,\vec{P}_3}$ & 
		$1-\cfrac{\dot{p_2}}{\dot{p_1}+\dot{p_2}}$ \\[4mm]
		${\vec{P}_2,\vec{P}_4}$ & $1-\cfrac{\dot{p_2}}{{\cos(\theta_{2K}-\dot{\beta}_K)}^2}$\\[4mm]
		\hline\hline
	\end{tabular}
\end{table}

In this part, it is shown that every interior point ${\big(\dot{\vec{p}},\dot{\vec{q}}\,\big)}$ in $\boldsymbol{\omega}$ can be decomposed as a convex combination of \emph{boundary} points of $\boldsymbol{\omega}$, which are decomposed in the next part.
Note that the subsequent analysis is for ${d>2}$.
In the case of ${d=2}$, ${\dot{p_1}+\dot{p_2}=1}$, and Table~\ref{tab:endpoints}
already carries the extreme points of $\boldsymbol{\omega}$. 
In fact, for ${d=2}$, we only need $\vec{P}_1$ and $\vec{P}_4$, because $\boldsymbol{\omega}$ contains
$\vec{P}_2$ and $\vec{P}_3$ if and only if 
$\vec{P}_2=\vec{P}_1$ and $\vec{P}_3=\vec{P}_4$, respectively.

\subsection{\label{sec:boundary-w}Boundary of $\boldsymbol{\omega}$}

The boundary of $\boldsymbol{\omega}$ is made of ${2d+d^2}$ regions, where a region is characterized by equality in one of the constraints \eqref{p-const2}, \eqref{q-const2}, and \eqref{t-ineq}:
\begin{eqnarray}
\label{Pi}
\mathbf{P}_i&:=& \big\{(\vec{p},\vec{q}\,)\in\boldsymbol{\omega}\,\big|\,p_i=0\big\}\,,\\
\label{Qj}
\mathbf{Q}_j&:=& \big\{(\vec{p},\vec{q}\,)\in\boldsymbol{\omega}\,\big|\,q_j=0\big\}\,, \quad \mbox{and}\\
\label{Rij}
\mathbf{R}_{ij}&:=& \big\{(\vec{p},\vec{q}\,)\in\boldsymbol{\omega}\,\big|\,\alpha_i+\beta_j=\theta_{ij}\big\}
\end{eqnarray}
for ${1\leq i,j \leq d}$. 
A point from Table~\ref{tab:endpoints}, provided it is in $\boldsymbol{\omega}$, called a boundary point because it belongs to one of the regions~\eqref{Pi}--\eqref{Rij}.
To reveal that the boundary points of $\boldsymbol{\omega}$ can be decomposed into certain convex combinations,
let us suppose that the duo $\vec{P}_1,\vec{P}_3$ belongs to $\boldsymbol{\omega}$ and analyze first
${\vec{P}_3\in\mathbf{P}_1}$ and then ${\vec{P}_1\in\mathbf{R}_{1J}}$.
Of course, an identical treatment can be delivered in the case of other duos from Table~\ref{tab:endpoint-pairs}.

\begin{table}[H]
	\centering
	\caption{A list of four points ${\vec{P}=\big(0,p_2,p_3,  
			\dot{\vec{p}}_{\mathrm{rest}},\dot{\vec{q}}\;\big)}$ similar to Table~\ref{tab:endpoints}. The upper bounds on $p_2$ [see \eqref{p2-const}] specify the points $\vec{P}_{31}$ and $\vec{P}_{32}$, while the lower bounds determine $\vec{P}_{33}$ and $\vec{P}_{34}$. These bounds are stated in the middle column for $p_2$, and then the corresponding $p_3$ are obtained by~\eqref{p2+p3} [see the right column].}
	\label{tab:endpoints-P3}
	\begin{tabular}{c | c | c}
		\hline\hline\rule{0pt}{3ex}  
		$\vec{P}$ & $p_2$ & $p_3$ \\
		\hline\rule{0pt}{3ex} 
		$\vec{P}_{31}$ & ${{\cos(\theta_{2K}-\dot{\beta}_K)}^2}$ & ${{\textstyle\sum\nolimits_{i=1}^{3}\dot{p}_i}-{\cos(\theta_{2K}-\dot{\beta}_K)}^2}$  
		\\
		$\vec{P}_{32}$ & ${\textstyle\sum\nolimits_{i=1}^{3}\dot{p}_i}$  & 0  
		\\
		$\vec{P}_{33}$ & 0 & ${\textstyle\sum\nolimits_{i=1}^{3}\dot{p}_i}$
		\\		
		$\vec{P}_{34}$ & ${{\textstyle\sum\nolimits_{i=1}^{3}\dot{p}_i}-{\cos(\theta_{3L}-\dot{\beta}_L)}^2}$ & ${{\cos(\theta_{3L}-\dot{\beta}_L)}^2}$  \\[1mm]			
		\hline\hline
	\end{tabular}
\end{table}
\begin{table}[H]
	\centering
	\caption{
		The necessary and sufficient conditions---that arise from the restraint \eqref{p2-const}---for a point of Table~\ref{tab:endpoints-P3} to be in or out of the region ${\mathbf{P}_1\subset\boldsymbol{\omega}}$. The table is like Table~\ref{tab:endpoints-cond}.}
	\label{tab:endpoints-cond-P3}
	\begin{tabular}{ c@{\hspace{2mm}} | @{\hspace{2mm}}c}
		\hline\hline 
		If and only if & Then  \\
		\hline\rule{0pt}{3.3ex} 
		${{\cos(\theta_{2K}-\dot{\beta}_K)}^2<\,
			{\textstyle\sum\nolimits_{i=1}^{3}\dot{p}_i}}$ & $\vec{P}_{31}\in\mathbf{P}_1$ and 
		$\vec{P}_{32}\notin\mathbf{P}_1$  
		\\[1mm]
		${{\cos(\theta_{2K}-\dot{\beta}_K)}^2>\,
			{\textstyle\sum\nolimits_{i=1}^{3}\dot{p}_i}}$ & $\vec{P}_{31}\notin\mathbf{P}_1$ and 
		$\vec{P}_{32}\in\mathbf{P}_1$  
		\\[1mm]
		${{\cos(\theta_{2K}-\dot{\beta}_K)}^2=\,
			{\textstyle\sum\nolimits_{i=1}^{3}\dot{p}_i}}$ & $\vec{P}_{31}=\vec{P}_{32}\in\mathbf{P}_1$  
		\\[1mm]
		\hline\rule{0pt}{3.3ex} 
		${{\cos(\theta_{3L}-\dot{\beta}_L)}^2<\,
			{\textstyle\sum\nolimits_{i=1}^{3}\dot{p}_i}}$ & $\vec{P}_{33}\notin\mathbf{P}_1$ and 
		$\vec{P}_{34}\in\mathbf{P}_1$  
		\\[1mm]
		${{\cos(\theta_{3L}-\dot{\beta}_L)}^2>\,
			{\textstyle\sum\nolimits_{i=1}^{3}\dot{p}_i}}$ & $\vec{P}_{33}\in\mathbf{P}_1$ and 
		$\vec{P}_{34}\notin\mathbf{P}_1$  
		\\[1mm]
		${{\cos(\theta_{3L}-\dot{\beta}_L)}^2=\,
			{\textstyle\sum\nolimits_{i=1}^{3}\dot{p}_i}}$ & $\vec{P}_{33}=\vec{P}_{34}\in\mathbf{P}_1$  
		\\[1mm]
		\hline\hline
	\end{tabular}
\end{table}
\begin{table}[]
	\centering	
	\caption{
		Depending on ${\vec{P}_3}$ and the conditions in Table~\ref{tab:endpoints-cond-P3}, at most two separate points of Table~\ref{tab:endpoints-P3} can belong to $\mathbf{P}_1$. Here, the left column carries all such couples of points. To the right side of each couple ${\vec{P}',\vec{P}''}$, the value of $\lambda$ is written, which associates the couple (provided it is in $\mathbf{P}_1$) back to ${\vec{P}_3=\lambda\vec{P}'+(1-\lambda)\vec{P}''}$.
		Taking ${0<\dot{p}_3< {\cos(\theta_{3L}-\dot{\beta}_L)}^2}$ and 
		${0<\dot{p}_1+\dot{p}_2\leq{\cos(\theta_{2K}-\dot{\beta}_K)}^2}$---that determines ${\vec{P}_3\in\mathbf{P}_1}$ [see Table~\ref{tab:endpoints-cond}]---one can check that each $\lambda$ lies in the interval ${(0,1]}$.}
	\label{tab:endpoint-pairs-P3}
	\begin{tabular}{ @{\hspace{2mm}}c@{\hspace{2mm}} | c}
		\hline\hline\rule{0pt}{3ex} 
		${\vec{P}',\vec{P}''}$ & $\lambda$ \\		
		\hline\rule{0pt}{4ex} 
		${\vec{P}_{31},\vec{P}_{33}}$ & $\cfrac{\dot{p_1}+\dot{p_2}}{{\cos(\theta_{2K}-\dot{\beta}_K)}^2}$
		\\[4mm]			
		${\vec{P}_{31},\vec{P}_{34}}$ & $\cfrac{{\cos(\theta_{3L}-\dot{\beta}_L)}^2-\dot{p_3}}
		{{{\cos(\theta_{2K}-\dot{\beta}_K)}^2+\cos(\theta_{3L}-\dot{\beta}_L)}^2-{\textstyle\sum\nolimits_{i=1}^{3}\dot{p}_i}}$ 
		\\[4mm]
		${\vec{P}_{32},\vec{P}_{33}}$& 
		$1-\cfrac{\dot{p_3}}{{\textstyle\sum\nolimits_{i=1}^{3}\dot{p}_i}}$
		\\[5mm]
		${\vec{P}_{32},\vec{P}_{34}}$ &  $1-\cfrac{\dot{p_3}}{{\cos(\theta_{3L}-\dot{\beta}_L)}^2}$ \\[4mm]	 
		\hline\hline
	\end{tabular}
\end{table}

\begin{table*}[ ]
	\centering
	\caption{A set of four points ${\vec{P}=\big({\cos(\theta_{1J}-\dot{\beta}_J)}^2,p_2,p_3,\dot{\vec{p}}_{\mathrm{rest}},\dot{\vec{q}}\;\big)}$ like Tables~\ref{tab:endpoints} and \ref{tab:endpoints-P3}. Here ${\{\vec{P}_{11},\vec{P}_{12}\}}$ and ${\{\vec{P}_{13}, \vec{P}_{14}\}}$ are obtained with the upper and lower bounds in~\eqref{p2-const_P1}, correspondingly. 
		These bounds are arranged in the center column, and $p_3$ is drawn from $p_2$ with \eqref{p2+p3_P1}.}
	\label{tab:endpoints-P1}
	\begin{tabular}{c | @{\hspace{2mm}}c@{\hspace{2mm}} | @{\hspace{2mm}}c@{\hspace{2mm}}}
		\hline\hline\rule{0pt}{3ex}  
		$\vec{P}$ & $p_2$ & $p_3$ \\
		\hline\rule{0pt}{3.3ex} 	
		$\vec{P}_{11}$ & ${\cos(\theta_{2K}-\dot{\beta}_K)}^2$ & ${{\textstyle\sum\nolimits_{i=1}^{3}\dot{p}_i}-{\cos(\theta_{1J}-\dot{\beta}_J)}^2-{\cos(\theta_{2K}-\dot{\beta}_K)}^2}$  
		\\[2mm]
		$\vec{P}_{12}$ & ${\textstyle\sum\nolimits_{i=1}^{3}\dot{p}_i-
			{\cos(\theta_{1J}-\dot{\beta}_J)}^2}$  & 0  
		\\[2mm]	
		$\vec{P}_{13}$ & 0 & ${\textstyle\sum\nolimits_{i=1}^{3}\dot{p}_i-		{\cos(\theta_{1J}-\dot{\beta}_J)}^2}$  
		\\[2mm]	
		$\vec{P}_{14}$ & ${{\textstyle\sum\nolimits_{i=1}^{3}\dot{p}_i}-{\cos(\theta_{1J}-\dot{\beta}_J)}^2-{\cos(\theta_{3L}-\dot{\beta}_L)}^2}$ & ${{\cos(\theta_{3L}-\dot{\beta}_L)}^2}$ 
		\\[1mm]			
		\hline\hline
	\end{tabular}
\end{table*}

\begin{table*}[ ]
	\centering
	\caption{If there is a case from the left column, then we have the corresponding consequence in the right column. All these cases are implications of \eqref{p2-const_P1}--\eqref{cosLJ>1}.
		The table is built in the same way as Table~\ref{tab:endpoints-cond} and \ref{tab:endpoints-cond-P3}.}
	\label{tab:endpoints-cond-P1}
	\begin{tabular}{l@{\hspace{2mm}} @{\hspace{2mm}}c@{\hspace{2mm}} |@{\hspace{2mm}}c}
		\hline\hline 
		&	If  & Then  \\ 
		\hline\rule{0pt}{3.3ex} 
		${K\neq J}$ && $\vec{P}_{11}\notin\mathbf{R}_{1J}$ and $\vec{P}_{12}\in\mathbf{R}_{1J}$\\
		\hline\rule{0pt}{3.3ex} 
		\multirow{3}{*}{${K=J\quad \mbox{and}}$} 
		&
		${\cos(\theta_{1J}-\dot{\beta}_J)}^2+
		{\cos(\theta_{2K}-\dot{\beta}_K)}^2<
		{\textstyle\sum\nolimits_{i=1}^{3}\dot{p}_i}$ & $\vec{P}_{11}\in\mathbf{R}_{1J}$ and $\vec{P}_{12}\notin\mathbf{R}_{1J}$  
		\\[1mm]
		&
		${\cos(\theta_{1J}-\dot{\beta}_J)}^2+
		{\cos(\theta_{2K}-\dot{\beta}_K)}^2>
		{\textstyle\sum\nolimits_{i=1}^{3}\dot{p}_i}$ & $\vec{P}_{11}\notin\mathbf{R}_{1J}$ and $\vec{P}_{12}\in\mathbf{R}_{1J}$  
		\\[1mm]
		&
		${\cos(\theta_{1J}-\dot{\beta}_J)}^2+
		{\cos(\theta_{2K}-\dot{\beta}_K)}^2=
		{\textstyle\sum\nolimits_{i=1}^{3}\dot{p}_i}$ & $\vec{P}_{11}=\vec{P}_{12}\in\mathbf{R}_{1J}$ 
		\\[1mm]
		\hline\rule{0pt}{3.3ex}
		${L\neq J}$ && $\vec{P}_{13}\in\mathbf{R}_{1J}$ and $\vec{P}_{14}\notin\mathbf{R}_{1J}$\\
		\hline\rule{0pt}{3.3ex} 
		\multirow{3}{*}{${L=J\quad \mbox{and}}$} 
		&
		${\cos(\theta_{1J}-\dot{\beta}_J)}^2+
		{\cos(\theta_{3L}-\dot{\beta}_L)}^2<
		{\textstyle\sum\nolimits_{i=1}^{3}\dot{p}_i}$ & $\vec{P}_{13}\notin\mathbf{R}_{1J}$ and $\vec{P}_{14}\in\mathbf{R}_{1J}$  
		\\[1mm]
		&
		${\cos(\theta_{1J}-\dot{\beta}_J)}^2+
		{\cos(\theta_{3L}-\dot{\beta}_L)}^2>
		{\textstyle\sum\nolimits_{i=1}^{3}\dot{p}_i}$ & $\vec{P}_{13}\in\mathbf{R}_{1J}$ and $\vec{P}_{14}\notin\mathbf{R}_{1J}$ 
		\\[1mm]
		&
		${\cos(\theta_{1J}-\dot{\beta}_J)}^2+
		{\cos(\theta_{3L}-\dot{\beta}_L)}^2=
		{\textstyle\sum\nolimits_{i=1}^{3}\dot{p}_i}$ & $\vec{P}_{13}=\vec{P}_{14}\in\mathbf{R}_{1J}$ 
		\\[1mm]
		\hline\hline
	\end{tabular}
\end{table*}

\begin{table}[ ]
	\centering
	\caption{Taking the case ${K=J}$ and ${L=J}$, we have four duos of 
		points, and the table is arranged in the same manner as Table~\ref{tab:endpoint-pairs} and \ref{tab:endpoint-pairs-P3}. 
		Right side to each duo, we place $\lambda$ that relates the duo (when it is in $\mathbf{R}_{1J}$) to the point ${\vec{P}_1=\lambda\vec{P}'+(1-\lambda)\vec{P}''}$. 
		Having ${0<\dot{p}_i<{\cos(\theta_{iJ}-\dot{\beta}_J)}^2}$ for
		${i=1,2,3}$ and the condition ${{\cos(\theta_{1J}-\dot{\beta}_J)}^2\leq\dot{p}_1+\dot{p}_2}$ that certifies ${\vec{P}_1\in\mathbf{R}_{1J}}$ [see Table~\ref{tab:endpoints-cond}], one can show that
		$0\leq\lambda<1$ in every case.}
	\label{tab:endpoint-pairs-P1}
	\begin{tabular}{ @{\hspace{2mm}}c@{\hspace{2mm}} | @{\hspace{2mm}}c}
		\hline\hline\rule{0pt}{3ex} 
		${\vec{P}',\vec{P}''}$ & $\lambda$ \\		
		\hline\rule{0pt}{5ex} 
		${\vec{P}_{11},\vec{P}_{13}}$ & $\cfrac{\dot{p}_1+\dot{p}_2-{\cos(\theta_{1J}-\dot{\beta}_J)}^2}{{\cos(\theta_{2J}-\dot{\beta}_{J})}^2}$
		\\[4mm]			
		${\vec{P}_{11},\vec{P}_{14}}$ &
		$\cfrac{{\cos(\theta_{3J}-\dot{\beta}_J)}^2-\dot{p}_3}{\textstyle\sum\nolimits_{i=1}^{3}\left({\cos(\theta_{iJ}-\dot{\beta}_J)}^2-\dot{p}_i\right)}$
		\\[5mm]
		${\vec{P}_{12},\vec{P}_{13}}$& 
		$1-\cfrac{\dot{p}_3}{{\textstyle\sum\nolimits_{i=1}^{3}\dot{p}_i}-{\cos(\theta_{1J}-\dot{\beta}_J)}^2}$
		\\[5mm]
		${\vec{P}_{12},\vec{P}_{14}}$ &  $1-\cfrac{\dot{p}_3}{{\cos(\theta_{3J}-\dot{\beta}_J)}^2}$ \\[4mm]	 
		\hline\hline
	\end{tabular}
\end{table}

Now we start from ${\vec{P}_3}$ and travel within the region $\mathbf{P}_1$ along a new set of points
${\vec{P}=\big(0,p_2,p_3,\dot{\vec{p}}_{\mathrm{rest}},
	\dot{\vec{q}}\,\big)}$
by changing $p_2,p_3$ according to
\begin{equation}
	\label{p2+p3}
	p_2+p_3=1-\textstyle\sum\nolimits_{i=4}^{d}\dot{p}_i
	=\textstyle\sum\nolimits_{i=1}^{3}\dot{p}_i\,\leq1\,,
\end{equation}
where ${\dot{\vec{p}}_{\mathrm{rest}}=(\dot{p}_4,\cdots,\dot{p}_d)}$.
Repeating the procedure similar to Appendix~\ref{sec:interior-w}, here we have
\begin{eqnarray}
	\label{p2-const} 
	0&\leq&\max\left\{0\,,\,\textstyle\sum\nolimits_{i=1}^{3}\dot{p}_i-{\cos(\theta_{3L}-\dot{\beta}_L)}^2\,\right\}
	\leq p_2\qquad\qquad\nonumber\\
	&\leq&
	\min\left\{{\cos(\theta_{2K}-\dot{\beta}_K)}^2,\,
	\textstyle\sum\nolimits_{i=1}^{3}\dot{p}_i\right\}\leq1\,,
\end{eqnarray}
which is like~\eqref{p1-const}.
The supremum ${\theta_{2K}-\dot{\beta}_K}$ is defined by \eqref{alpha2-supremum} and
\begin{equation}
	\label{alpha3-supremum}
	\theta_{3L}-\dot{\beta}_L=\operatorname*{max}_{1\leq l\leq d}\;\big\{\theta_{3l}-\dot{\beta}_l\big\}\,.
\end{equation}
If and only if $p_2$ respects \eqref{p2-const} and
$p_3$ follows $p_2$ with \eqref{p2+p3}, then a new
${\vec{P}\in\mathbf{P}_1\subset\boldsymbol{\omega}}$.

Analogous to Tables~\ref{tab:endpoints}--\ref{tab:endpoint-pairs}, here we compose Tables~\ref{tab:endpoints-P3}--\ref{tab:endpoint-pairs-P3}, in that order. 
Table~\ref{tab:endpoints-P3} holds a collection of four points.
Table~\ref{tab:endpoints-cond-P3} has the conditions that decide
whether a point of Table~\ref{tab:endpoints-P3} is in or out of $\mathbf{P}_1$.
Table~\ref{tab:endpoint-pairs-P3} supplies all possible couples---of points from Table~\ref{tab:endpoints-P3}---out of which one belongs to $\mathbf{P}_1$, that one is determined by $\vec{P}_3$.
The line segment---connecting the one couple---carries $\vec{P}_3$ and completely occupies in the region $\mathbf{P}_1$.

Now we are going to focus on ${\vec{P}_1\in\mathbf{R}_{1J}}$. 
Let us proceed from ${\vec{P}_1}$ by altering only $p_2,p_3$ of
another new vector ${\vec{P}=\big({\cos(\theta_{1J}-\dot{\beta}_J)}^2,p_2,p_3,\dot{\vec{p}}_{\mathrm{rest}},\dot{\vec{q}}\,\big)}$ 
with respect to 
\begin{eqnarray}
\label{p2+p3_P1}
p_2+p_3&=&
1-\textstyle\sum\nolimits_{i=4}^{d}\dot{p}_i-
{\cos(\theta_{1J}-\dot{\beta}_J)}^2 \nonumber\\
&=&\textstyle\sum\nolimits_{i=1}^{3}\dot{p}_i-
{\cos(\theta_{1J}-\dot{\beta}_J)}^2\,.
\end{eqnarray}
Note that ${\dot{\vec{p}}_{\mathrm{rest}}=(\dot{p}_4,\cdots,\dot{p}_d)}$, and \eqref{p2+p3_P1} identifies a straight line, a segment of which is contained in the region ${\mathbf{R}_{1J}}$.
In addition to \eqref{p2+p3_P1}, if $p_2$ agrees to  
\begin{eqnarray}
\label{p2-const_P1} 
&&0\leq\nonumber\\
&&\max\left\{0\,,\textstyle\sum\nolimits_{i=1}^{3}\dot{p}_i-{\cos(\theta_{1J}-\dot{\beta}_J)}^2-{\cos(\theta_{3L}-\dot{\beta}_L)}^2\right\}\nonumber\\
&&\leq\, p_2\,\leq\nonumber\\
&&
\min\left\{{\cos(\theta_{2K}-\dot{\beta}_K)}^2,\,
\textstyle\sum\nolimits_{i=1}^{3}\dot{p}_i-{\cos(\theta_{1J}-\dot{\beta}_J)}^2\right\}
\nonumber\\
&&\leq1\,
\end{eqnarray}
only then the new vector ${\vec{P}\in\mathbf{R}_{1J}}$.
Like Tables~\ref{tab:endpoints} and \ref{tab:endpoints-P3}, here we assemble Table~\ref{tab:endpoints-P1} of four points using the four bounds in~\eqref{p2-const_P1}.

Due to~\eqref{1<Cos^2+Cos^2-q} and \eqref{1=Cos^2+Cos^2-q} from Appendix~\ref{sec:inter-results}, we have
\begin{eqnarray}
\label{cosKJ>1}
&&\mbox{if}\ K\neq J\ \mbox{then}\nonumber\\
&&1<{\cos(\theta_{1J}-\dot{\beta}_J)}^2+
{\cos(\theta_{2K}-\dot{\beta}_K)}^2,\ \; \mbox{and}\quad\ \\
\label{cosLJ>1}
&&\mbox{if}\ L\neq J\ \mbox{then}\nonumber\\
&&1<{\cos(\theta_{1J}-\dot{\beta}_J)}^2+ 
{\cos(\theta_{3L}-\dot{\beta}_L)}^2.
\end{eqnarray}
These inequalities are strict because a requirements in \eqref{1=Cos^2+Cos^2-q}, ${\dot{\beta}_J+\dot{\beta}_K=\tfrac{\pi}{2}}$, cannot be met since
${\dot{q}_J+\dot{q}_K<1}$ is caused by \eqref{int}.
Now taking \eqref{p2-const_P1}--\eqref{cosLJ>1} with 
${\textstyle\sum\nolimits_{i=1}^{3}\dot{p}_i\leq1}$, one can deduce that the vectors $\vec{P}_{11}$ and $\vec{P}_{14}$ of Table~\ref{tab:endpoints-P1} can not belong to ${\mathbf{R}_{1J}}$ unless ${K=J}$ and ${L=J}$, respectively.
This fact is recorded in Table~\ref{tab:endpoints-cond-P1} with some other conditions, together they tell when a point of Table~\ref{tab:endpoints-P1} 
will be in or out of the region ${\mathbf{R}_{1J}}$.

A duo, out of the four listed in Table~\ref{tab:endpoint-pairs-P1}, resides in ${\mathbf{R}_{1J}}$ and
expresses $\vec{P}_1$ through a convex combination.
As Tables~\ref{tab:endpoints}--\ref{tab:endpoint-pairs} are linked with the interior point ${\big(\dot{\vec{p}},\dot{\vec{q}}\,\big)\in\boldsymbol{\omega}}$ and Tables~\ref{tab:endpoints-P3}--\ref{tab:endpoint-pairs-P3} are attached to ${\vec{P}_3\in\mathbf{P}_{1}}$, Tables~\ref{tab:endpoints-P1}--\ref{tab:endpoint-pairs-P1} are associated with ${\vec{P}_1\in\mathbf{R}_{1J}}$.
Tables~\ref{tab:endpoints}, \ref{tab:endpoints-P3}, and
\ref{tab:endpoints-P1} carry the boundary points of $\boldsymbol{\omega}$,
$\mathbf{P}_1$, and $\mathbf{R}_{1J}$, respectively.

\subsection{\label{sec:extreme-w}Extreme of $\boldsymbol{\omega}$}

In the above parts, it is demonstrated that
every interior point ${\big(\dot{\vec{p}},\dot{\vec{q}}\,\big)\in\boldsymbol{\omega}}$ can be decomposed into a convex combination of the boundary points of $\boldsymbol{\omega}$,  
which can further be decomposed into convex combinations of the boundary points of regions~\eqref{Pi}--\eqref{Rij}. 
Continuing this decomposition process, we reach at a point
${\big(\mathring{\vec{p}}\,,\dot{\vec{q}}\,\big)}$, where
\begin{eqnarray}
\label{vec-p-not}
&&\mathring{\vec{p}}=
\big({\cos\mathring{\alpha}_1}^2,\cdots,{\cos\mathring{\alpha}_m}^2,
\mathbf{0}\,,\,\mathring{p}_s\,,\,\mathbf{0}\big)\,,\qquad \\
\label{alpha_i-not}	                
&&\mathring{\alpha}_i=\theta_{iJ}-\dot{\beta}_J\qquad\qquad\quad
(\mbox{for all }i=1,\cdots,m)\,,\\
\label{p_s-not}
&&\mathring{p}_s=1-{\textstyle\sum\nolimits_{i=1}^{m}
	                {\cos\mathring{\alpha}_i}^2}\quad\qquad
	                (m+1\leq s\leq d)\,,\qquad\qquad \\
\label{bf-0}
&&\mathbf{0}\equiv 0,\cdots,0\,,\quad\mbox{and}\\
\label{m}
&&1\leq m \leq d-1\,.
\end{eqnarray}
Since every ${\mathring{\alpha}_i}$ of \eqref{alpha_i-not} is a supremum, ${0\leq\mathring{\alpha}_i}$ [see the explanation below \eqref{alpha-cons-supremum}] and ${\mathring{\alpha}_i<\dot{\alpha}_i<\tfrac{\pi}{2}}$ due to~\eqref{int}, we deduce that
\begin{equation}
\label{alpha_i limits}
\qquad\qquad
0\leq\mathring{\alpha}_i<\tfrac{\pi}{2}\quad\quad
(\mbox{for all }i=1,\cdots,m)\,.
\end{equation}

The point ${\big(\mathring{\vec{p}}\,,\dot{\vec{q}}\,\big)}$, 
designated by \eqref{vec-p-not}--\eqref{m},
satisfies $m$ and ${d-(m+1)}$ number of \emph{equality} constraints of type~\eqref{t-ineq} and \eqref{p-const2}, respectively.
If ${\mathring{p}_s}$ of \eqref{p_s-not} follows
\begin{equation}
	\label{pk-not-cons}
    0\,\leq\,\mathring{p}_s\,\leq\,{\cos(\theta_{sZ}-\dot{\beta}_Z)}^2
\end{equation}
then ${\big(\mathring{\vec{p}}\,,\dot{\vec{q}}\,\big)
	\in\boldsymbol{\omega}}$, where
\begin{equation}
	\label{Sup-s}
\theta_{sZ}-\dot{\beta}_Z=\operatorname*{max}_{1\leq z\leq d}\;\big\{\theta_{sz}-\dot{\beta}_z\big\}
\end{equation}
is a supremum like~\eqref{alpha1-supremum}, \eqref{alpha2-supremum}, \eqref{alpha3-supremum}, and \eqref{alpha_i-not}.
One can check that points in Table~\ref{tab:endpoints} for ${d=2}$
and in Tables~\ref{tab:endpoints-P3} as well as \ref{tab:endpoints-P1}---provided ${K=J}$ and ${L=J}$---for ${d=3}$ are like ${\big(\mathring{\vec{p}}\,,\dot{\vec{q}}\,\big)}$; remember that ${\textstyle\sum\nolimits_{i=1}^{d}\dot{p}_i=1}$ due to \eqref{p-const1}.
Furthermore, one can easily recognize $\mathring{p}_s$ in each of these points. Then, one can see through Table~\ref{tab:endpoints-cond}, \ref{tab:endpoints-cond-P3}, and \ref{tab:endpoints-cond-P1} that one of the two inequalities in~\eqref{pk-not-cons} is required for a point 
to be in $\boldsymbol{\omega}$. The other inequality is automatically obeyed due to \eqref{int} and the conditions appeared in the earlier decompositions.

If we start our journey from a point ${\big(\dot{\vec{p}}\,,\dot{\vec{q}}\,\big)}$, where
\begin{equation}
\label{ext-pt_m=1,0=p q}
\dot{\vec{q}}=(r_{11},\cdots,r_{1d})\,,
\end{equation}
then we will arrive at the point ${\big(\mathring{\vec{p}}\,,\dot{\vec{q}}\,\big)}$, where
\begin{equation}
\label{ext-pt_m=1,0=p p}
\mathring{\vec{p}}=\big(1,\mathbf{0}\big)
\end{equation}
[for \textbf{0}, see \eqref{bf-0}].
This point represents an extreme point of $\boldsymbol{\omega}$ and a special case
\begin{equation}
\label{case m=1,0=p}
m=1\quad\mbox{with}\quad 0=\mathring{p}_s
\end{equation}
of \eqref{m} and \eqref{p_s-not}.
In the case~\eqref{case m=1,0=p}, the supremum 
${\mathring{\alpha}_1=\theta_{1J}-\dot{\beta}_J=0}$ that is possible if and only if ${\theta_{1j}=\dot{\beta}_j}$, means ${r_{1j}=\dot{q}_j}$, for every $j$. Indeed, it is so [see \eqref{ext-pt_m=1,0=p q}].
In all other cases, ${0<\mathring{\alpha}_i}$ for every ${1\leq i\leq m}$ [see the limits~\eqref{alpha_i limits} on ${\mathring{\alpha}_i}$ of \eqref{alpha_i-not}], and ${\big(\mathring{\vec{p}}\,,\dot{\vec{q}}\,\big)}$
can be decomposed further by adopting the same procedure as before.

Without loss of generality, let us suppose ${J=1}$ for the subsequent analysis.
Here we begin with ${\vec{Q}=\big(\mathring{\vec{p}}\,,\dot{q}_1,q_2,q_3,\dot{\vec{q}}_{\mathrm{rest}}\,\big)}$, where
\begin{equation}
\label{q_2+q_3=}
q_2+q_3=1-\textstyle\sum\nolimits_{i=4}^{d}\dot{q}_j-\underbrace{{\cos\dot{\beta}_1}^2}_{\textstyle\dot{q}_1}
=\dot{q}_2+\dot{q}_3
\end{equation}
and ${\dot{\vec{q}}_{\mathrm{rest}}=(\dot{q}_4,\cdots,\dot{q}_d)}$.
One can acknowledge that $\vec{Q}$ represents all those points, including ${\big(\mathring{\vec{p}}\,,\dot{\vec{q}}\,\big)}$, that fall on the straight line characterized by~\eqref{q_2+q_3=}.

If $q_3$ stays on the line with $q_2$, which follows
\begin{eqnarray}
\label{q2-const} 
0&\leq&\max\left\{0\,,\,\dot{q}_2+\dot{q}_3-
{\cos(\theta_{L3}-\mathring{\alpha}_L)}^2\,\right\}
\leq q_2\qquad\qquad\nonumber\\
&\leq&
\min\left\{{\cos(\theta_{K2}-\mathring{\alpha}_K)}^2,\,
\dot{q}_2+\dot{q}_3\right\}\leq1\,,
\end{eqnarray}
then ${\vec{Q}\in\boldsymbol{\omega}}$. 
Here
\begin{eqnarray}
\label{beta2-supremum}
&& \theta_{K2}-\mathring{\alpha}_K=\operatorname*{max}_{1\leq k\leq d}\;\big\{\theta_{k2}-\mathring{\alpha}_k\big\} \quad \mbox{and} \\
\label{beta3-supremum}
&& \theta_{L3}-\mathring{\alpha}_L=\operatorname*{max}_{1\leq l\leq d}\;\big\{\theta_{l3}-\mathring{\alpha}_l\big\} 
\end{eqnarray} 
are suprema, and the angles $\mathring{\alpha}$
are related to the components of $\mathring{\vec{p}}$ through \eqref{pq} and \eqref{alpha-beta} [see also \eqref{vec-p-not} and \eqref{alpha_i-not}].
The constraints~\eqref{q2-const} look alike \eqref{p1-const} and \eqref{p2-const}.
Identical to Tables~\ref{tab:endpoints}, \ref{tab:endpoints-P3}, and \ref{tab:endpoints-P1}, we enter a list of four points in  Table~\ref{tab:endpoints-Q}, where the points are drawn from the four bounds on $q_2$ given in~\eqref{q2-const}.

\begin{table}[ ]
	\centering
	\caption{Four points
		${\vec{Q}=\big(\mathring{\vec{p}}\,,\dot{q}_1,q_2,q_3,\dot{\vec{q}}_{\mathrm{rest}}\,\big)\in\mathbb{R}^{2d}}$ that rest on the
		line specified by~\eqref{q_2+q_3=}.
		From the point ${\big(\mathring{\vec{p}}\,,\dot{\vec{q}}\,\big)}$,
		the coordinate $q_2$ increases towards ${\{\vec{Q}_1,\vec{Q}_2\}}$, while it decreases towards ${\{\vec{Q}_3,\vec{Q}_4\}}$.
		The middle column carries the four bounds given in \eqref{q2-const}, and then $q_3$ is obtained with \eqref{q_2+q_3=}.  
		The table is prepared in the same fashion as Tables~\ref{tab:endpoints}, \ref{tab:endpoints-P3}, and \ref{tab:endpoints-P1}. }
	\label{tab:endpoints-Q}
	\begin{tabular}{c | c | c}
		\hline\hline\rule{0pt}{1ex}  
		$\vec{Q}$ & $q_2$ & $q_3$ \\
		\hline\rule{0pt}{3ex} 	
		$\vec{Q}_1$ & ${{\cos(\theta_{K2}-\mathring{\alpha}_K)}^2}$ & ${\dot{q}_2+\dot{q}_3-{\cos(\theta_{K2}-\mathring{\alpha}_K)}^2}$  \\	
		$\vec{Q}_2$ & ${\dot{q}_2+\dot{q}_3}$  & 0  \\	
		$\vec{Q}_3$ & 0 & ${\dot{q}_2+\dot{q}_3}$  \\	
		$\vec{Q}_4$ & ${\dot{q}_2+\dot{q}_3-{\cos(\theta_{L3}-\mathring{\alpha}_L)}^2}$ & ${{\cos(\theta_{L3}-\mathring{\alpha}_L)}^2}$  \\[1mm]			
		\hline\hline
	\end{tabular}
\end{table}

Now, to establish criteria for a point of Table~\ref{tab:endpoints-Q} to be in or out of $\boldsymbol{\omega}$, we are going to address
the two cases
\begin{eqnarray}
\label{case m=1,0<p}
m=1\quad&\mbox{with}&\quad 0<\mathring{p}_s\quad\mbox{and}\\
\label{case m>1}
m>1\quad&\mbox{with}&\quad 0<\mathring{p}_s
\end{eqnarray}
individually [see Eq.~\eqref{p_s-not} for $\mathring{p}_s$ and the range~\eqref{m} of $m$].
Let us first take the case~\eqref{case m>1}: whatever the suprema~\eqref{beta2-supremum} and \eqref{beta3-supremum} are, we have
\begin{eqnarray}
\label{cos1K>1}
&&1<{\cos\dot{\beta}_1}^2+{\cos(\theta_{K2}-\mathring{\alpha}_K)}^2 \quad\mbox{and}\quad\ \\
\label{cos1L>1}
&&1<{\cos\dot{\beta}_1}^2+{\cos(\theta_{L3}-\mathring{\alpha}_L)}^2. 
\end{eqnarray}
To demonstrate this, we consider ${m=2}$, the cases with ${m>2}$ can be handled likewise.
For ${m=2}$, we have ${\dot{\beta}_1=\theta_{i1}-\mathring{\alpha}_i}$ (where ${i=1,2}$) due to \eqref{alpha_i-not}.
If $K$ associated with the supremum~\eqref{beta2-supremum} is 1, then by taking ${\dot{\beta}_1=\theta_{21}-\mathring{\alpha}_2}$ we can validate the strict inequality \eqref{cos1K>1} thanks to~\eqref{1<Cos^2+Cos^2-p}.
If ${K\neq1}$, we can do the same by now considering ${\dot{\beta}_1=\theta_{11}-\mathring{\alpha}_1}$.
In a similar fashion, we can establish the other inequality~\eqref{cos1L>1}.

We draw the following inferences from inequalities \eqref{cos1K>1} and \eqref{cos1L>1}.
\begin{eqnarray}
\label{cos1K>1-q1}
{\cos(\theta_{K2}-\mathring{\alpha}_K)}^2&>&
1-\dot{q}_1=\textstyle\sum\nolimits_{j=2}^{d}\dot{q}_j\geq
\dot{q}_2+\dot{q}_3\,,\qquad\quad  \\
\label{cos1L>1-q1}
{\cos(\theta_{L3}-\mathring{\alpha}_L)}^2&>&
1-\dot{q}_1=\textstyle\sum\nolimits_{j=2}^{d}\dot{q}_j\geq
\dot{q}_2+\dot{q}_3\,
\end{eqnarray}
implies that the maximum and the minimum values in \eqref{q2-const} are 0 and ${\dot{q}_2+\dot{q}_3}$, respectively.
Consequently, the points $\vec{Q}_1$ and $\vec{Q}_4$ of Table~\ref{tab:endpoints-Q} never, whereas $\vec{Q}_2$ and $\vec{Q}_3$ always,
belong to $\boldsymbol{\omega}$ in the case~\eqref{case m>1}.
Moreover, ${\big(\mathring{\vec{p}}\,,\dot{\vec{q}}\,\big)}$
can be broken into the convex combination
${\lambda\,\vec{Q}_2+(1-\lambda)\,\vec{Q}_3}$, where ${\lambda=\tfrac{\dot{q}_2}{\dot{q}_2+\dot{q}_3}}$ [see Table~\ref{tab:endpoint-pairs-Q}].

Next, it is not difficult to realize that both $\vec{Q}_2$ and $\vec{Q}_3$ can be decomposed further and further until 
we arrive at a point $\big(\mathring{\vec{p}}\,,\mathring{\vec{q}}\,\big)$, where
\begin{equation}
\label{vec-q-not}
\mathring{\vec{q}}=
\big(\dot{q}_1\,,\mathbf{0}\,,\,\mathring{q}_t\,,\,\mathbf{0}\big)
\quad\mbox{with}\quad\mathring{q}_t=1-\dot{q}_1\quad
(2\leq t\leq d)\,.\quad
\end{equation}
In the decomposition process one will encounter inequalities, such as 
\eqref{cos1K>1} and \eqref{cos1L>1}, that can be tacked like the above.
For ${m>1}$, a point $\big(\mathring{\vec{p}}\,,\mathring{\vec{q}}\,\big)$
defined by \eqref{vec-p-not}--\eqref{bf-0} and \eqref{vec-q-not}
is an extreme point of $\boldsymbol{\omega}$, because
it cannot be written into a convex combination of other points of $\boldsymbol{\omega}$.
Furthermore,
$\big(\mathring{\vec{p}}\,,\mathring{\vec{q}}\,\big)$
is a vector-valued function of $\dot{\beta}_1$ since $\theta$-angles
are fixed by \eqref{theta} once the measurement settings 
are selected in~\eqref{AB-bases}.

Let us now turn to the case~\eqref{case m=1,0<p},
where ${\dot{\beta}_1=\theta_{11}-\mathring{\alpha}_1}$ according to~\eqref{alpha_i-not},
\begin{eqnarray}
\label{Q, m=1}
&&\vec{Q}=\big(\mathring{\vec{p}}\,,\,
{\cos(\theta_{11}-\mathring{\alpha}_{1})}^2\,,\,
q_2\,,\,q_3\,,\,\dot{\vec{q}}_{\mathrm{rest}}\,\big)\,,\quad\mbox{and}\\
\label{vec-p-not m=1}
&&\mathring{\vec{p}}=
\big(\mathring{p}_{1}\,,
\mathbf{0}\,,\,\mathring{p}_s\,,\,\mathbf{0}\big)
\ \mbox{with}\ \;
1-\mathring{p}_s=\mathring{p}_{1}
={\cos\mathring{\alpha}_1}^2\,.\quad\qquad
\end{eqnarray}
Since supremum~\eqref{beta2-supremum} is a nonnegative number, $K$ can either be $s$ or 1 here.
It is due to ${\theta_{i2}-\mathring{\alpha}_i\leq0}$ when ${i\neq s}$ and ${i\neq 1}$, because then ${\mathring{\alpha}_i=\tfrac{\pi}{2}}$ and every ${\theta\leq\tfrac{\pi}{2}}$.
Similarly, $L$ related to the supremum~\eqref{beta3-supremum} can either be $s$ or 1 here.

When ${K=s}$ or ${L=s}$ or both, we encounter situation similar to the case~\eqref{case m>1}: 
When ${K=s}$ then---due to~\eqref{1<Cos^2+Cos^2-p}---we have 
\begin{eqnarray}
\label{cos1s>=1}
&&{\cos(\theta_{K2}-\mathring{\alpha}_K)}^2+
{\cos(\theta_{11}-\mathring{\alpha}_1)}^2\geq1 \quad\mbox{and thus}\quad\quad \\
\label{cos1s>=1-q1}
&&{\cos(\theta_{K2}-\mathring{\alpha}_K)}^2\geq
1-\dot{q}_1=\textstyle\sum\nolimits_{j=2}^{d}\dot{q}_j\geq
\dot{q}_2+\dot{q}_3\,. \qquad\quad
\end{eqnarray}
One can perceive that \eqref{cos1s>=1} and \eqref{cos1s>=1-q1} are analogues to \eqref{cos1K>1} and \eqref{cos1K>1-q1}, respectively. 
The inequalities in~\eqref{cos1s>=1-q1} suggest that ${\dot{q}_2+\dot{q}_3}$ is the minimum value in \eqref{q2-const}. Therefore, without exception $\vec{Q}_2$ lies in $\boldsymbol{\omega}$,
if ${\vec{Q}_1=\vec{Q}_2}$ then ${\vec{Q}_1\in\boldsymbol{\omega}}$.
Identically, for ${L=s}$, always $\vec{Q}_3\in\boldsymbol{\omega}$,
and $\vec{Q}_4$ belongs to $\boldsymbol{\omega}$ only when it is $\vec{Q}_3$.

\begin{table}[]
	\centering
	\caption{Group of conditions for the case~\eqref{case m=1,0<p}, where ${\mathring{\alpha}_1=\theta_{11}-\dot{\beta_1}}$.
		A condition from the left column delivers what is on its right side.
		These conditions originate from \eqref{q2-const} and the discussion around \eqref{cos1s>=1-q1}.
		At most two conditions can hold simultaneously, thus more than two distinct points of Table~\ref{tab:endpoints-Q} cannot be a part of $\boldsymbol{\omega}$. The table looks like Table~\ref{tab:endpoints-cond-P1}.}
	\label{tab:endpoints-cond-Q}
	\begin{tabular}{l@{\hspace{2mm}} @{\hspace{2mm}}c@{\hspace{2mm}} | @{\hspace{2mm}}c}
		\hline\hline 
		&	If  & Then  \\ 
		\hline\rule{0pt}{3.3ex} 
		${K=s}$ && $\vec{Q}_{2}\in\boldsymbol{\omega}$\\
		\hline\rule{0pt}{3.3ex} 
		\multirow{3}{*}{${K=1,}$} 
		&
		${\cos(\theta_{K2}-\mathring{\alpha}_K)}^2<\,\dot{q_2}+\dot{q_3}$ & $\vec{Q}_1\in\boldsymbol{\omega}$ and $\vec{Q}_2\notin\boldsymbol{\omega}$  
		\\[1mm]
		&
		${\cos(\theta_{K2}-\mathring{\alpha}_K)}^2>\,\dot{q_2}+\dot{q_3}$ & $\vec{Q}_1\notin\boldsymbol{\omega}$ and $\vec{Q}_2\in\boldsymbol{\omega}$ 
		\\[1mm]
		&
		${\cos(\theta_{K2}-\mathring{\alpha}_K)}^2=\,\dot{q_2}+\dot{q_3}$ & ${\vec{Q}_1=\vec{Q}_2\in\boldsymbol{\omega}}$ 
		\\[1mm]
		\hline\rule{0pt}{3.3ex}
		${L=s}$ && $\vec{Q}_{3}\in\boldsymbol{\omega}$\\
		\hline\rule{0pt}{3.3ex} 
		\multirow{3}{*}{${L=1,}$} 
		&
		${\cos(\theta_{L3}-\mathring{\alpha}_L)}^2<\,\dot{q_2}+\dot{q_3}$ & $\vec{Q}_3\notin\boldsymbol{\omega}$ and $\vec{Q}_4\in\boldsymbol{\omega}$ 
		\\[1mm]
		&
		${\cos(\theta_{L3}-\mathring{\alpha}_L)}^2>\,\dot{q_2}+\dot{q_3}$ & $\vec{Q}_3\in\boldsymbol{\omega}$ and $\vec{Q}_4\notin\boldsymbol{\omega}$ 
		\\[1mm]
		&
		${\cos(\theta_{L3}-\mathring{\alpha}_L)}^2=\,\dot{q_2}+\dot{q_3}$ & ${\vec{Q}_3=\vec{Q}_4\in\boldsymbol{\omega}}$ 
		\\[1mm]
		\hline\hline
	\end{tabular}
\end{table}

\begin{table}[]
	\centering
	\caption{
		Collection of duplets ${\vec{Q}',\vec{Q}''}$ of points from Table~\ref{tab:endpoints-Q}.
		Only one of these duplets---except if two or more are the same---belongs to $\boldsymbol{\omega}$ and represents the point ${\big(\mathring{\vec{p}}\,,\dot{\vec{q}}\,\big)}$ with the convex combination	${\lambda\,\vec{Q}'+(1-\lambda)\,\vec{Q}''}$.
		Here we assume ${K=1}$ and ${L=1}$, otherwise $\vec{Q}_1$
		and $\vec{Q}_4$ can not belong to $\boldsymbol{\omega}$ without being equal to $\vec{Q}_2$ and $\vec{Q}_3$, respectively [see Table~\ref{tab:endpoints-cond-Q}].
		The right column has the values of $\lambda$ for each duplet, provided the duplet lies in $\boldsymbol{\omega}$. 
		One can check that ${\lambda\in[0,1]}$ with 
		${0<\dot{q}_2\leq {\cos(\theta_{K2}-\mathring{\alpha}_K)}^2}$ and 
		${0<\dot{q}_3\leq{\cos(\theta_{L3}-\mathring{\alpha}_L)}^2}$ [see \eqref{q2-const}].
	}
	\label{tab:endpoint-pairs-Q}
	\begin{tabular}{@{\hspace{2mm}}c@{\hspace{2mm}} | c}
		\hline\hline\rule{0pt}{3ex} 
		${\vec{Q}',\vec{Q}''}$ & $\lambda$ \\		
		\hline\rule{0pt}{4ex} 
		${\vec{Q}_1,\vec{Q}_3}$ &  $\cfrac{\dot{q}_2}{{\cos(\theta_{12}-\mathring{\alpha}_1)}^2}$ \\[4mm]
		${\vec{Q}_1,\vec{Q}_4}$&  $\cfrac{{\cos(\theta_{13}-\mathring{\alpha}_1)}^2-\dot{q}_3}
		{{{\cos(\theta_{12}-\mathring{\alpha}_1)}^2+
				{\cos(\theta_{13}-\mathring{\alpha}_1)}^2-\dot{q}_2-\dot{q}_3}}$\\[5mm]	 
		${\vec{Q}_2,\vec{Q}_3}$ & 
		$1-\cfrac{\dot{q}_3}{\dot{q}_2+\dot{q}_3}$ \\[4mm]
		${\vec{Q}_2,\vec{Q}_4}$ & $1-\cfrac{\dot{q}_3}{{\cos(\theta_{13}-\mathring{\alpha}_1)}^2}$\\[4mm]
		\hline\hline
	\end{tabular}
\end{table}

When ${K=1}$ and ${L=1}$ only then ${\vec{Q}_1}$
and ${\vec{Q}_4}$ can be in $\boldsymbol{\omega}$
without being equal to $\vec{Q}_2$ and $\vec{Q}_3$, respectively [see Table~\ref{tab:endpoints-cond-Q}].
With Table~\ref{tab:endpoints-cond-Q}, for the case~\eqref{case m=1,0<p}, one can find out whether or not a duplet of points from Table~\ref{tab:endpoints-Q} lies in $\boldsymbol{\omega}$. 
All such duplets are gathered in Table~\ref{tab:endpoint-pairs-Q}, which
reveals that the point ${\big(\mathring{\vec{p}}\,,\dot{\vec{q}}\,\big)}$
can be split into a convex combination.
As before, we can break the points of Table~\ref{tab:endpoints-Q} further and further until we reach extreme points of $\boldsymbol{\omega}$.

In the case~\eqref{case m=1,0<p}, the decomposition process leads to 
\begin{eqnarray}
\label{vec-q-not m=1}
&&\mathring{\vec{q}}=
\big({\cos\mathring{\beta}_1}^2,\cdots,{\cos\mathring{\beta}_n}^2,
\mathbf{0}\,,\,\mathring{q}_t\,,\,\mathbf{0}\big)\,,\quad\mbox{where} \\
\label{bera_j-not}	                
&&\mathring{\beta}_j=\theta_{1j}-\mathring{\alpha}_1\qquad\qquad
(\mbox{for all }j=1,\cdots,n)\,,\\
\label{q_t-not}
&&\mathring{q}_t=1-{\textstyle\sum\nolimits_{j=1}^{n}
	{\cos\mathring{\beta}_j}^2}\quad
(n+1\leq t\leq d)\,,\ \mbox{and}\qquad \\
\label{n}
&&1\leq n \leq d-1\,.
\end{eqnarray}
If $\mathring{q}_t$ of \eqref{q_t-not} obeys
\begin{eqnarray}
\label{qt-not-cons}
&&0\,\leq\,\mathring{q}_t\,\leq\,{\cos(\theta_{Zt}-\mathring{\alpha}_Z)}^2,
\quad\mbox{where}\\
\label{Sup-z}
&&\theta_{Zt}-\mathring{\alpha}_Z=\operatorname*{max}_{1\leq z\leq d}\;\big\{\theta_{zt}-\mathring{\alpha}_z\big\}\,,
\end{eqnarray} 
then the point ${\big(\mathring{\vec{p}}\,,\mathring{\vec{q}}\,\big)}$
stated by \eqref{vec-p-not m=1} and \eqref{vec-q-not m=1} belongs to
$\boldsymbol{\omega}$.
It is an extreme point of $\boldsymbol{\omega}$ in the case~\eqref{case m=1,0<p}.
One can also realize that both there $\mathring{\vec{p}}$ and 
$\mathring{\vec{q}}$ are functions of $\dot{\beta}_1$ by noticing ${\mathring{\beta}_j=\theta_{1j}-\theta_{11}+\dot{\beta}_1}$ in \eqref{bera_j-not} with ${\mathring{\alpha}_1=\theta_{11}-\dot{\beta}_1}$.
In fact, the extreme point identified by \eqref{ext-pt_m=1,0=p p} and \eqref{ext-pt_m=1,0=p q} in the case~\eqref{case m=1,0=p}
can also be represented
with these $\mathring{\vec{p}}$ and $\mathring{\vec{q}}$ of \eqref{vec-p-not m=1} and \eqref{vec-q-not m=1} by taking ${\mathring{\alpha}_1=0}$, which make it as an endpoint of the parametric curve
${\big(\mathring{\vec{p}}{\scriptstyle(\mathring{\alpha}_1)}\,,\mathring{\vec{q}}{\scriptstyle(\mathring{\alpha}_1)}\,\big)}$.
In conclusion, we realize the \emph{structure} of extreme points of $\boldsymbol{\omega}$:

\begin{equation}
\label{Ext-pts}
\parbox{0.85\columnwidth}{%
	The point ${\big(\mathring{\vec{p}}\,,\mathring{\vec{q}}\,\big)}$, where ``${\mathring{\vec{p}}}$ is specified by \eqref{vec-p-not}--\eqref{bf-0} and ${\mathring{\vec{q}}}$ is given by \eqref{vec-q-not}" when ${m>1}$ and ``${\mathring{\vec{p}}}$ is describe by \eqref{vec-p-not m=1} and ${\mathring{\vec{q}}}$ is presented by 
    \eqref{vec-q-not m=1}--\eqref{n}" when ${m=1}$, represents an extreme point of $\boldsymbol{\omega}$ provided $\dot{\beta}_1$ is within suitable limits presented in the next part.
    For every ${1\leq m\leq(d-1)}$, ${\big(\mathring{\vec{p}}{\scriptstyle(\dot{\beta}_1)}\,,\mathring{\vec{q}}{\scriptstyle(\dot{\beta}_1)}\,\big)}$ is a vector-valued function of a real parameter $\dot{\beta}_1$,
    thus it characterizes an $m$-parametric curve in $\boldsymbol{\omega}$. Such curves are presented in Sec.~\ref{sec:PS-C}. 
}
\end{equation}

\subsection{\label{sec:limit-beta-1}Limits on $\beta_1$}

We start with the $m$-parametric curve
${\big(\vec{p}{\scriptstyle(\beta_1)}\,,\vec{q}{\scriptstyle(\beta_1)}\,\big)}$ identified by \eqref{m-t-eq}--\eqref{bf-0-}.
According to \eqref{Ext-pts}, a part of the curve that lies in $\boldsymbol{\omega}$ represents its extreme points.
This part is specified by the upper and lower limits of $\beta_1$.
To compute these limits, here, we only need to consider
\begin{eqnarray}
\label{0<ps}
0&\leq& p_s\,,\\
\label{t-ineq_it}
\theta_{it}&\leq&\alpha_i+\beta_t\quad(\mbox{for } i=1,\cdots,m\,,s)\,,
\quad\mbox{and}\qquad\\
\label{t-ineq_sj}
\theta_{sj}&\leq&\alpha_s+\beta_j\quad(\mbox{for } j=1,t)\,.
\end{eqnarray}
When ${i>m}$ and ${i\neq s}$ then ${\alpha_i=\tfrac{\pi}{2}}$, and when ${j\neq1}$ and ${j\neq t}$, then ${\beta_j=\tfrac{\pi}{2}}$.
So one can easily perceive that the points ${\big(\vec{p}{\scriptstyle(\beta_1)}\,,\vec{q}{\scriptstyle(\beta_1)}\,\big)}$ fulfill rest of the requirements~\eqref{t-ineq} as well as \eqref{p-const1}--\eqref{q-const2} to be in $\boldsymbol{\omega}$.

For ${i=s}$ in \eqref{t-ineq_it} or ${j=t}$ in \eqref{t-ineq_sj}, the TI is always obeyed: due to
\begin{eqnarray}  
\label{alpha_s+beta_t-1} 
\tfrac{\pi}{2} &\leq&\alpha_s+\alpha_1 \\ 
\label{alpha_s+beta_t-2}      
&=&\alpha_s+\theta_{11}-\beta_1\\
\label{alpha_s+beta_t-3}  
&=&\alpha_s+\theta_{11}-\tfrac{\pi}{2}+\beta_t\,,
\quad\mbox{we have}\qquad\quad \\
\label{alpha_s+beta_t-4} 
\tfrac{\pi}{2}\leq\pi-\theta_{11}&\leq&\alpha_s+\beta_t\,.
\end{eqnarray}
With \eqref{p_i+p_k}, \eqref{m-t-eq}, and \eqref{q_j+q_l<1-beta}
one can sequentially go through the steps~\eqref{alpha_s+beta_t-1}--\eqref{alpha_s+beta_t-3}, and the left-hand side inequality in
\eqref{alpha_s+beta_t-4} is a consequence of ${\theta\leq\tfrac{\pi}{2}}$.
Since $\alpha_s$ and $\beta_t$ obey
${\tfrac{\pi}{2}\leq\alpha_s+\beta_t}$,
they certainly follow the TI ${\theta_{st}\leq\alpha_s+\beta_t}$ as every ${\theta\leq\tfrac{\pi}{2}}$.

If we decrease $\beta_1$ then ${\alpha_s+\beta_1}$ decreases,
and $\beta_1$ reaches its lower limit $\beta'$ when the inequality \eqref{t-ineq_sj}, for ${j=1}$, gets 
saturated.
It means that $\beta'$ is a solution of the equation ${\theta_{s1}-\beta'=\alpha_s}$ and thus of
\begin{equation}
\label{beta'}
{\cos(\theta_{s1}-\beta')}^2=p_s=
1-{\textstyle\sum\nolimits_{i=1}^{m}{\cos(\theta_{i1}-\beta')}^2}
\end{equation}
[by \eqref{m-t-eq} and \eqref{p_s}].
If we increase $\beta_1$ then $p_s$ and ${\alpha_i+\beta_t}$ ${(i=1,\cdots,m)}$ decrease, and $\beta_1$ attains its upper limit $\beta''$ as soon as one of the inequalities~\eqref{0<ps} and \eqref{t-ineq_it} gets saturated.
Using \eqref{m-t-eq}, \eqref{p_s}, and ${\beta_t=\tfrac{\pi}{2}-\beta_1}$ [owing to \eqref{q_j+q_l<1-beta}], these inequalities can be expressed as
\begin{eqnarray}
\label{0<ps 2} 
0&\leq&
1-{\textstyle\sum\nolimits_{i=1}^{m}{\cos(\theta_{i1}-\beta_1)}^2}
\qquad\mbox{and}\\
\label{t-ineq_it 2}
\beta_1&\leq&\tfrac{\theta_{i1}-\theta_{it}}{2}+\tfrac{\pi}{4}
\qquad(\mbox{for } i=1,\cdots,m)\,.
\end{eqnarray}
Now we need to investigate the two cases, ${m=1}$ and ${1< m\leq(d-1)}$ listed in \eqref{Ext-pts}, separately for $\beta''$.

In the case ${m=1}$, \eqref{0<ps 2} clearly holds, and
the upper limit
\begin{equation}
\label{beta'' m=1}
\beta''=\tfrac{\theta_{11}-\theta_{1t}}{2}+\tfrac{\pi}{4}
\end{equation}
is obtained when \eqref{t-ineq_it 2} is saturated.
Corresponding to $\beta''$ of \eqref{beta'' m=1}, we have 
\begin{equation}
\label{alpha1 m=1}
\alpha_1=\theta_{11}-\beta''=\tfrac{\theta_{11}+\theta_{1t}}{2}-\tfrac{\pi}{4}
\end{equation}
which is a root of the equation
\begin{equation}
\label{alpha1 m=1 eq}
{\cos(\theta_{11}-\alpha_1)}^2+{\cos(\theta_{1t}-\alpha_1)}^2=1\,.
\end{equation}

In the case ${1< m\leq(d-1)}$, when we increase $\beta_1$ then the inequality \eqref{0<ps 2}, rather than \eqref{t-ineq_it 2}, gets saturated first. Hence, $\beta''$ is now a solution of 
\begin{equation}
\label{beta'' m>1}
{\textstyle\sum\nolimits_{i=1}^{m}{\cos(\theta_{i1}-\beta'')}^2}=1\,.
\end{equation}
One can justify these statements by proving
\begin{equation}
\label{beta'<beta''}
\beta''\,\leq\,
\underbrace{\tfrac{\theta_{i1}+\theta_{i'1}}{2}-\tfrac{\pi}{4}}_{\widetilde{\beta}}
\,\leq\,
\tfrac{\theta_{i1}-\theta_{it}}{2}+\tfrac{\pi}{4}\,,
\end{equation}
where ${1\leq i,i'\leq m}$.
As $\beta''$ is a root of Eq.~\eqref{beta'' m>1},
$\widetilde{\beta}$ is a root of
\begin{equation}
\label{eq-beta-tilde}
{\cos\big(\theta_{i1}-\widetilde{\beta}\,\big)}^2+
{\cos\big(\theta_{i'1}-\widetilde{\beta}\,\big)}^2=1\,.
\end{equation}
Equations~\eqref{beta'}, \eqref{beta'' m>1}, and \eqref{eq-beta-tilde} are of the form
\begin{equation}
\label{eq-beta-1}
\textstyle\sum\nolimits_{i=1}^{\textsc{m}}
{\cos(\theta_{i1}-\beta_1)}^2=1\,,
\end{equation}
where $\textsc{m}$ angles---the $\textsc{m}$-set ${\{\theta_{11},\cdots,\theta_{\textsc{m}1}\}}$---are taken from the first column of $\varTheta$ matrix [given in \eqref{r-theta-matrices}]. Always, we must choose the root of Eq.~\eqref{eq-beta-1} that respects 
${0\leq\beta_1\leq\theta_{i1}}$ for every ${i=1,\cdots,\textsc{m}}$.
Furthermore, as we add more angles from the first column to the $\textsc{m}$-set, the number of nonnegative terms increases on the left-hand side of Eq.~\eqref{eq-beta-1}.
Then $\beta_1$ of smaller value will satisfy Eq.~\eqref{eq-beta-1}.
So, by comparing Eqs.~\eqref{beta'' m>1} and \eqref{eq-beta-tilde} in this way, we can certify the left-hand side inequality in \eqref{beta'<beta''}.
Whereas, after a simplification, the right-hand side inequality turns into ${\theta_{i'1}+\theta_{it}\leq\pi}$, which is true as every ${\theta\leq\tfrac{\pi}{2}}$.

\begin{equation}
	\label{lim-bete}
	\parbox{0.85\columnwidth}{%
In conclusion, the lower limit $\beta'$ is the root of Eq.~\eqref{beta'} for every ${1\leq m\leq(d-1)}$. The upper limit $\beta''$, for ${m=1}$, is given by \eqref{beta'' m=1} and can be derived from Eq.~\eqref{alpha1 m=1 eq}.
For ${1<m}$, $\beta''$ is the solution of Eq.~\eqref{beta'' m>1}.
	}
\end{equation}

In fact, Eq.~\eqref{alpha1 m=1 eq}---where two angles are taken from the first column of $\varTheta$---is also like Eq.~\eqref{eq-beta-1}.
Basically, one needs to solve equation such as \eqref{eq-beta-1}---where ${2\leq \textsc{m}\leq d}$ angles are picked from a row or a column of $\varTheta$---to get a limit and then an endpoint of an $m$-parametric curve. When ${m=1}$ then \textsc{m} can only be 2 [see \eqref{beta'} and \eqref{alpha1 m=1 eq}]. And, when ${1< m\leq(d-1)}$ then \textsc{m} can either be $m$ or ${m+1}$ [see \eqref{beta'' m>1} and \eqref{beta'}].

To solve Eq.~\eqref{eq-beta-1} for $\beta_1$, we transform it into
\begin{eqnarray}
\label{eq-beta-1 xyz}
&&\textbf{x}\,{\cos\beta_1}^2+\textbf{y}\,\sin\beta_1\,\cos\beta_1+
\textbf{z}=0\,,\quad\mbox{where}\qquad\\
\label{x}
&&\textbf{x}:=\textstyle\sum\nolimits_{i=1}^{\textsc{m}}\cos 2\theta_{i1}=
2\textstyle\sum\nolimits_{i=1}^{\textsc{m}}r_{i1}-\textsc{m}
\,,\\
\label{y}
&&\textbf{y}:=\textstyle\sum\nolimits_{i=1}^{\textsc{m}}\sin 2\theta_{i1}=
2\textstyle\sum\nolimits_{i=1}^{\textsc{m}}\sqrt{r_{i1}\,(1-r_{i1})}\,,
\quad\mbox{and}\qquad\quad\\
\label{z}
&&\textbf{z}:=\textstyle\sum\nolimits_{i=1}^{\textsc{m}}{\sin\theta_{i1}}^2-1=
\textsc{m}-\textstyle\sum\nolimits_{i=1}^{\textsc{m}}r_{i1}-1
\,.
\end{eqnarray}
Calling ${{\cos\beta_1}^2=q_1}$ by the relations \eqref{pq} and \eqref{alpha-beta}, we can write Eq.~\eqref{eq-beta-1 xyz} as 
\begin{equation}
\label{eq-beta-3}
\textbf{x}\,q_1+\textbf{y}\,\sqrt{q_1\,(1-q_1)}+
\textbf{z}=0\,.
\end{equation}
The two roots of Eq.~\eqref{eq-beta-3} are 
\begin{equation}
\label{eq-beta-roots}
{\cos\beta_1}^2=q_1=\frac{(\textbf{y}^2-2\,\textbf{x}\,\textbf{z})\pm
	\textbf{y}\sqrt{\textbf{y}^2-4\,\textbf{z}\,(\textbf{x}+\textbf{z})}}
{2\,(\textbf{x}^2+\textbf{y}^2)}\,,
\end{equation}
which only depend on the $\textsc{m}$-set ${\{\theta_{11},\cdots,\theta_{\textsc{m}1}\}}$ associated with Eq.~\eqref{eq-beta-1}.

We pick the root~\eqref{eq-beta-roots} with +~sign due to the following reasons.
First, for ${\textsc{m}=2}$, we have equation such as \eqref{eq-beta-tilde}, and its root $\widetilde{\beta}$---given in \eqref{beta'<beta''}---corresponds to the +~sign solution [see also \eqref{alpha1 m=1} with \eqref{alpha1 m=1 eq}].
Second, for ${\textsc{m}=d}$, ${\beta_1=0}$ is the only permissible solution
of Eq.~\eqref{eq-beta-1}.
It is because angles $\theta_{i1}$ are not random real numbers, they follow ${\textstyle\sum\nolimits_{i=1}^{d}{\cos\theta_{i1}}^2=1}$.
When ${\textsc{m}=d}$, ${\textbf{z}=d-2=-\textbf{x}}$  [see \eqref{x} and \eqref{z}], and always the solution~\eqref{eq-beta-roots} with +~sign offers ${\beta_1=0}$.
Third reason, for a pair of MUBs \cite{Durt10}, where every ${\theta}$ is the same ${\arccos\tfrac{1}{\sqrt{d}}}$, one can directly solve Eq.~\eqref{eq-beta-1}.
For every $\textsc{m}$-set, we get the same $\beta_1$ [see $\chi$ in~\eqref{beta_end-MUB}], which corresponds to
\begin{equation}
\label{eq-beta-roots-MUB}
{\cos\beta_1}^2=\frac{\left[1+
	\sqrt{(d-1)(\textsc{m}-1)}\right]^2}{d\,\textsc{m}}
\end{equation}
that is clearly the root~\eqref{eq-beta-roots} with +~sign.



\begin{thebibliography}{99}
		
	\bibitem{Heisenberg27} 
	W. Heisenberg, 
	Z. Phys. \textbf{43}, 172 (1927); English translation in \cite{Wheeler83}. 
	
	
	\bibitem{Wheeler83}
	J.~A. Wheeler and W.~H. Zurek, eds., \textit{Quantum Theory and Measurement}
	(Princeton University Press, Princeton, New Jersey, 1983), pp. 62--84.
	
	
	\bibitem{Weyl32}
	H. Weyl, \textit{The Theory of Groups and Quantum Mechanics}, 
	English translated by H. P. Robertson
	(E.P. Dutton, New York, 1932), Chapter~2, Section~7 and Appendix~1.
	
	
	
	\bibitem{Busch07}
	P. Busch, T. Heinonen, and P. Lahti,
	Phys. Rep. \textbf{452}, 155 (2007).

	
	\bibitem{Robertson29} 
	H. P. Robertson, 
	Phys. Rev. \textbf{34}, 163 (1929).
	
	
	\bibitem{Deutsch83} 
	D. Deutsch, 
	Phys. Rev. Lett. \textbf{50}, 631 (1983).
	
	
	\bibitem{Kraus87} 
	K. Kraus, 
	Phys. Rev. D \textbf{35}, 3070 (1987).
	
	
	\bibitem{Maassen88} 
	H. Maassen and J. B. M. Uffink, 
	Phys. Rev. Lett. \textbf{60}, 1103 (1988).
	
	
	
	\bibitem{Wehner10} 
	S. Wehner and A. Winter, 
	New J. Phys. \textbf{12}, 025009 (2010).
	
	
	\bibitem{Bialynicki11}
	I. Bialynicki-Birula and \L{}. Rudnicki,
	\textit{Entropic Uncertainty Relations in Quantum Physics}, in ``Statistical Complexity: Applications in Electronic Structure'', edited by K. D. Sen (Springer, Netherlands, 2011), 
	pp. 1--34; e-print arXiv:1001.4668 [quant-ph] (2011).
	
	
	\bibitem{Coles17}
    P. J. Coles, M. Berta, M. Tomamichel, and S. Wehner,
    Rev. Mod. Phys. \textbf{89}, 015002 (2017).


	
	 	 
	\bibitem{Wootters81}
	W. K. Wootters, 
	Phys. Rev. D \textbf{23}, 357 (1981).
	


	
    \bibitem{Landau61}
	H. J. Landau and H. O. Pollak, 
	Bell System Tech. J. \textbf{40}, 65 (1961).
	
	
	\bibitem{Folland97}
	G. B. Folland and A. Sitaram, J. Fourier Anal. Appl. \textbf{3}, 207 (1997).
	
	
	\bibitem{Lenard72}
	A. Lenard,
	J. Functional Analysis \textbf{10}, 410 (1972).
	
	
	\bibitem{Larsen90}
	U. Larsen, 
	J. Phys. A: Math. Gen. \textbf{23}, 1041 (1990).
	
	
	\bibitem{Kaniewski14}
	J. Kaniewski, M. Tomamichel, and S. Wehner,
	Phys. Rev. A \textbf{90}, 012332 (2014).


	
	
   \bibitem{Durt10}
   T. Durt, B.-G. Englert, I. Bengtsson, and K. \.{Z}yczkowski, Int. J. Quantum. Inform. \textbf{8}, 535 (2010).
   
  
   
   
   \bibitem{Ivanovic92}
   I. D. Ivanovic, J. Phys. A: Math. Gen. \textbf{25}, 363 (1992).
   
   
   \bibitem{Sanchez-Ruiz95}
   J. S\'{a}nchez-Ruiz, Phys. Lett. A \textbf{201}, 125 (1995).
   
   
   \bibitem{Ballester07} 
   M. A. Ballester and S. Wehner, 
   Phys. Rev. A \textbf{75}, 022319 (2007).
   
   
   \bibitem{Wu09}
   S. Wu, S. Yu, and K. M\o{}lmer, Phys. Rev. A
   \textbf{79}, 022104 (2009).
   
   
   
   \bibitem{Mandayam10}
   P. Mandayam, S. Wehner, and N. Balachandran,
   J. Math. Phys. \textbf{51}, 082201 (2010).
   
	





    \bibitem{Rastegin12}
    A. E. Rastegin, Int. J. Theor.
    Phys. \textbf{51}, 1300 (2012).
     
    
    
    \bibitem{Busch14}
    P. Busch, P. Lahti, and R. F. Werner,
    Phys. Rev. A \textbf{89}, 012129 (2014).
    
    
    
    \bibitem{Garrett90}
    A.J.M. Garrett and S.F. Gull,
    Phys. Lett. A \textbf{151}, 453 (1990).
    
 
   
	\bibitem{Sanchez-Ruiz98} 
    J. S\'{a}nches-Ruiz, Phys. Lett. A \textbf{244}, 189 (1998).
   
   
	\bibitem{Ghirardi03}
    G.C. Ghirardi, L. Marinatto, and R. Romano,
    Phys. Lett. A \textbf{317}, 32 (2003).
    
    
    
    \bibitem{Bosyk12} 
    G. M. Bosyk, M. Portesi, and A. Plastino,
    Phys. Rev. A \textbf{85}, 012108 (2012).
    
    
    \bibitem{Vicente05}
     J. I. de Vicente and J. S\'{a}nchez-Ruiz,
     Phys. Rev. A \textbf{71}, 052325 (2005).
     
     
     \bibitem{Zozor13}
     S. Zozor, G. M. Bosyk, and M. Portesi,
     J. Phys. A: Math. Theor. \textbf{46}, 465301 (2013).

   

	 
   
  	

   \bibitem{Maccone14} 
   L. Maccone and A. K. Pati,
   Phys. Rev. Lett. \textbf{113}, 260401 (2014).



	\bibitem{Luis11} 
    A. Luis,
    Phys. Rev. A \textbf{84}, 034101 (2011).	





   
   
   	\bibitem{Hofmann03}
   	H. F. Hofmann and S. Takeuch,
   	Phys. Rev. A \textbf{68}, 032103 (2003).
   	
   	\bibitem{Guhne04}
   	O. G\"{u}hne,
   	Phys. Rev. Lett. \textbf{92}, 117903 (2004).
   	
   	
   	\bibitem{Guhne04b}
   	O. G\"{u}hne and M. Lewenstein, 
   	Phys. Rev. A \textbf{70}, 022316 (2004).
   		
   	\bibitem{Giovannetti04}
   	V. Giovannetti,
   	Phys. Rev. A \textbf{70}, 012102 (2004).
  
 	
		




    \bibitem{Niculescu93}
    C. Niculescu and L.-E. Persson,
    \textit{Convex Functions and their Applications: A Contemporary Approach,} (Springer-Verlag, New York, 2006).



    \bibitem{Peres93}
    A. Peres, 
    \textit{Quantum Theory: Concepts and Methods} 
    (Kluwer Academic Publishers, 1995).

  
   
    \bibitem{Tsallis88}
   C. Tsallis, J. Stat. Phys. \textbf{52}, 479 (1988).

    
    
    \bibitem{Rockafellar70}
     R. T. Rockafellar, \textit{Convex Analysis} 
     (Princeton University Press, Princeton, New Jersey, 1970).



    \bibitem{Sehrawat16}
    A. Sehrawat, e-print arXiv:1611.09760 [quant-ph] (2016).
      
    
    \bibitem{Shannon48}
    C. E. Shannon, Bell Syst. Tech. J. \textbf{27}, 379 (1948).
   
   
   
    \bibitem{Renyi61}
    A. R\'{e}nyi, \textit{On measures of information and entropy}, in Proceedings of the 4th Berkeley Symposium on Mathematical Statistics and Probability, Vol.~1 (University of California Press, Berkeley, CA, 1961), pp. 547--561.
    
  
	
	\bibitem{Paris04}
	M.~Paris and J.~\v{R}eh\'a\v{c}ek, eds., \textit{Quantum State Estimation} (Springer-Verlag, Heidelberg, 2004).
	
	
	
	\bibitem{Rudin76} 
	W. Rudin, 
	\textit{Principles of mathematical analysis} 
	(McGraw-Hill, 1976), Chapter~2.
	
    \bibitem{prep} Evidently, this paper only talks about preparation (un)certainty relations.
	
	\bibitem{diff-routes}
	In \cite{Lenard72}, the region of allowed expectation values of a couple of orthogonal projectors ${(Q,P)}$ is obtained as the convex hull of two ellipses.
	For ${Q=|a\rangle\langle a|}$ and ${P=|b\rangle\langle b|}$, the two ellipses become the one, which is same as ours.
	A pair of ellipses in \cite{Larsen90}---is associated with the two ellipses in \cite{Lenard72}---specifies an allowed region for a pair of purities related to a pair of projective measurements.
    In \cite{Kaniewski14}, by taking $M$ binary observables an ellipsoid, inside a hypercube, is presented as an allowed region for the expectation values of the $M$ observables. For ${M=2}$, the ellipsoid turns into our ellipse.
	

	
		
\end{thebibliography}
\end{document}